\documentclass[11pt,a4paper,final]{article}
\usepackage{amsmath,amssymb,epsfig,latexsym,enumerate}
\usepackage{graphicx}
\usepackage{cite}
\usepackage{slashed}
\usepackage{color}
\usepackage{pifont}
\usepackage{stackrel}
\usepackage[T1]{fontenc}
\usepackage{hyperref}
\usepackage[affil-it]{authblk}
\usepackage{geometry}
\numberwithin{equation}{section}

\newcommand{\eps}{\epsilon}

\def\dim{{d}}

\def\CF{{C_F}}

\def\MS{{\overline{\text{\scshape{ms}}}}}
\def\DR{{\overline{\text{\scshape{dr}}}}}

\def\NNLO{{\scshape nnlo}}
\def\DS{{\scshape ds}}

\def\FDH{{\scshape fdh}}

\def\DRED{{\scshape dred}}
\def\CDR{{\scshape cdr}}
\def\DREG{{\scshape dreg}}

\def\IReg{{\scshape ireg}}
\def\FDR{{\scshape fdr}}
\def\FDU{{\scshape fdu}}

\def\UV{{\scshape uv}}
\def\IR{{\scshape ir}}
\def\LTD{{\scshape ltd}}
\def\FTT{{\scshape ftt}}

\def\mFDH{{\rm \scriptscriptstyle FDH}}

\newcommand{\be}{\begin{equation}}
\newcommand{\ee}{\end{equation}}
\newcommand{\ba}{\begin{array}}
\newcommand{\ea}{\end{array}}
\newcommand{\bq}{\begin{eqnarray}}
\newcommand{\eq}{\end{eqnarray} }
\newcommand{\bal}{\begin{align}}
\newcommand{\eal}{\end{align}}

\newcommand{\beq}{\begin{eqnarray}}
\newcommand{\eeq}{\end{eqnarray}}
\newcommand{\mc}{\mathcal}
\newcommand{\nn}{\nonumber}
\newcommand{\bS}[1]{{\bf S}_{#1}}
\newcommand{\bC}[1]{{\bf C}_{#1}}
\newcommand{\bSC}[1]{{\bf SC}_{#1}}

\newcommand{\bbS}[1]{\overline{\bf S}_{#1}}
\newcommand{\bbC}[1]{\overline{\bf C}_{#1}}
\newcommand{\bbSC}[1]{\overline{\bf SC}_{#1}}

\newcommand{\W}[1]{\mc{W}_{#1}}
\newcommand{\bW}[1]{\overline{\mc{W}}_{#1}}

\newcommand{\Norm}{\mc{N}_1}

\newcommand{\Bn}{B}

\newcommand{\kk}[2]{\bar k_{#1}^{(#2)}}
\newcommand{\sk}[2]{\bar s_{#1}^{(#2)}}

\newcommand{\npo}{{n+1}}
\newcommand{\npt}{{n+2}}

\newcommand\sss{\scriptscriptstyle}
\newcommand{\bqa}{\begin{eqnarray}}
\newcommand{\eqa}{\end{eqnarray}}

\def\NNLO{{\scshape nnlo}}

\def\mFDH{{\rm \scriptscriptstyle FDH}}

\def\mFDR{{\rm \scriptscriptstyle FDR}}

\def\Feyncalc{{{\sc FeynCalc}}}
\def\PackageX{{{\sc Package-X}}}
\graphicspath{{fdf.files/figs/}}

\newcommand{\Neps}{n_\epsilon}

\def\nn{\nonumber}

\def\Eq#1{Eq.~(\ref{#1})}
\def\beqa{\begin{eqnarray}}
\def\eeqa{\end{eqnarray}}

\def\g{g_{s}}

\def\as{\alpha_{s}}

\newcommand\asopi{\frac{\as}{\pi}}

\newcommand\dsigdqtz{\frac{\mathd\hat\sigma_{ab}\!\lp\qt,z\rp}{d\qt^2}}

\newcommand\qt{q_{\sss\rm T}}

\newcommand{\qtcut}{q_{\sss \rm T}^{\sss\rm cut}}
\newcommand{\qtmax}{q_{\sss \rm T}^{\sss\rm max}}

\def\beeq{\begin{eqnarray}} 
\def\eeeq{\end{eqnarray}} 
\def\ito{\leftarrow}
 
\def\nn{\nonumber} 

\def\lq{\left[} 
\def\rq{\right]} 
\def\rg{\right\}} 
\def\lg{\left\{} 
\def\lp{\left(} 
\def\rp{\right)}

\newcommand{\mathd}{\mathrm{d}}

\newcommand{\footnotew}[1]{\footnote{#1}}

\allowdisplaybreaks

\newcommand{\munich}{Max-Planck-Institut f{\"u}r Physik, Werner-Heisenberg-Institut, 80805 M{\"u}nchen, Germany.}
\newcommand{\valencia}{Instituto de F\'{\i}sica Corpuscular, UVEG--CSIC, 46980 Paterna, Spain.}
\newcommand{\culiacan}{Facultad de Ciencias F\'isico-Matem\'aticas, Universidad Aut\'onoma de Sinaloa, 80000 Culiac\'an, Mexico.}
\newcommand{\andre}{CCNH, Universidade Federal do ABC, 09210-580 , Santo Andr\'e, Brazil.}
\newcommand{\coimbra}{CFisUC, Department of Physics, University of Coimbra, 3004-516 Coimbra, Portugal.}
\newcommand{\psiz}{Paul Scherrer Institut, 5232 Villigen, PSI, Switzerland.}
\newcommand{\uzh}{Physik-Institut, Universit\"at Z\"urich,  8057 Z\"urich, Switzerland.}
\newcommand{\wurz}{Universit\"at W\"urzburg, Institut f\"ur Theoretische Physik und Astrophysik, 97074 W\"urzburg, Germany.}
\newcommand{\turin}{Dipartimento di Fisica and Arnold-Regge Center, Universit\`{a} di Torino and INFN, 10125 Torino, Italy.}
\newcommand{\granada}{Dep. de F\'isica Te\'orica y del Cosmos and CAFPE,
Universidad de Granada, 18071 Granada, Spain.}
\newcommand{\kit}{Institut fur Theoretische Teilchenphysik,
Karlsruher Institut fur Technologie,
76128 Karlsruhe, Germany.}
\newcommand{\firenze}{INFN, Sezione di Firenze, 50019 Sesto Fiorentino, Italy.}
\newcommand{\genova}{INFN, Sezione di Genova, 16146, Genova, Italy.}
\newcommand{\desy}{Deutsches Elektronensynchrotron DESY, 15738 Zeuthen, Germany.}
\newcommand{\milan}{Dipartimento di Fisica, Universit\`a di Milano and INFN, Sezione di Milano, 20133 Milan, Italy.}
\newcommand{\bicocca}{Universit\`a di Milano\,-\,Bicocca and INFN, Sezione di
  Milano\,-\,Bicocca, 20126 Milano, Italy.}
\newcommand{\lisboa}{Laboratório de Instrumentação e Física de Partículas LIP, 1649-003 Lisboa, Portugal.}
\newcommand{\dresden}{Institut f\"ur Kern- und Teilchenphysik, TU Dresden, 01062 Dresden, Germany.}
\newcommand{\durham}{Institute for Particle Physics Phenomenology, DH1 3LE, Durham, UK.}
\newcommand{\napoli}{Universit\`a di Napoli and INFN, Sezione di
  Napoli, 80126 Napoli, Italy.}

\begin{document}
%\maketitle

\interfootnotelinepenalty=10000
\begin{flushright}
IFIC/20-54; MPP-2020-218\\
TTP20-043; P3H-20-079
\end{flushright}

\vspace{0.5cm}
\begin{center}
{\Large\bf May the four be with you:\\
Novel IR-subtraction methods to tackle NNLO calculations}
\bigskip\vspace{1.cm}{
{\large\sc 
W.J.~Torres~Bobadilla${}^{a,b}$\footnote{e-mail: torres@mpp.mpg.de},
{G.F.R.~Sborlini}${}^{b,c}$,\\
{P. Banerjee}${}^{d}$,
{S. Catani}${}^{e}$,
{A.L. Cherchiglia}${}^{f}$,
{L. Cieri}${}^{e}$,
{P.K. Dhani}${}^{e,g}$,
{F.~Driencourt-Mangin}${}^{b}$,
{T.~Engel}${}^{d,h}$,
{G.~Ferrera}${}^{i}$,
{C.~Gnendiger}${}^{d}$,
{R.J.~Hern\'andez-Pinto}${}^{j}$,
{B.~Hiller}${}^{k}$,
{G.~Pelliccioli}${}^{l}$,
{J.~Pires}${}^{m}$,
{R.~Pittau}${}^{n}$,
{M.~Rocco}${}^{o}$,
{G.~Rodrigo}${}^{b}$,
{M.~Sampaio}${}^{f}$,
{A.~Signer}${}^{d,h}$,
{C.~Signorile-Signorile}${}^{p,q}$,
{D.~St\"o{}ckinger}${}^{r}$,
{F.~Tramontano}${}^{s}$,
{and~Y.~Ulrich}${}^{d,h,t}$
} \\[7mm]
}
{\it \footnotesize
${}^a${\munich}\\
${}^b${\valencia}\\
${}^c${\desy}\\
${}^d${\psiz}\\
${}^e${\firenze}\\
${}^f${\andre}\\
${}^g${\genova}\\
${}^h${\uzh}\\
${}^i${\milan}\\
${}^j${\culiacan}\\
${}^k${\coimbra}\\
${}^l${\wurz}\\
${}^m${\lisboa}\\
${}^n${\granada}\\
${}^o${\bicocca}\\
${}^p${\kit}\\
${}^q${\turin}\\
${}^r${\dresden}\\
${}^s${\napoli}\\
${}^t${\durham}\\
{}
}
\end{center}
\bigskip
\bigskip

\thispagestyle{empty}
%\vspace{2em}
\begin{abstract}
In this report, we present a discussion about different frameworks 
to perform precise higher-order computations for high-energy physics. 
These approaches implement novel strategies to deal with 
infrared and ultraviolet singularities in quantum field theories. 
A special emphasis is devoted to the local cancellation of these singularities,
which can enhance the efficiency of computations and lead to discover novel mathematical
properties in quantum field theories.
\end{abstract}
\newpage
\setcounter{page}{1}

 \noindent\hrulefill
 \tableofcontents
 \noindent\hrulefill

\setcounter{footnote}{0}
\renewcommand{\thefootnote}{\arabic{footnote}}

%%%%%%%%%%%%%%%%%%%%%%%%%%%%%%%%%%%%%%%%%%%%%%%%%%%
%%%%%%%%%%%%%%%%%%%%%%%%%%%%%%%%%%%%%%%%%%%%%%%%%%%

\section{Introduction}

Nowadays the calculation of observables relevant for 
high-energy physics (HEP) colliders is extremely important, in
view of the future experiments that will provide new data with accuracy not
reached so far. Therefore, insight and support from the theory
side is necessary to understand the new findings. Clearly, the HEP
community has to be ready to tackle this kind of problems and various
approaches have to be implemented or reformulated. In particular,
the perturbative framework applied to Quantum Field Theories (QFTs) 
has shown to be very important for providing highly precise predictions. 
The so-called Next-to-Leading Order (NLO) revolution was possible due 
to the emergence of novel techniques inspired by clever ideas.
Likewise, predictions at NNLO are currently
being calculated, but a fully established framework as at NLO is not yet complete.
There are indeed several ideas and already working methods 
that can, for some processes, produce NNLO predictions
to be compared with available experimental measurements.

In the spirit of providing observables at NNLO and beyond, one encounters several
obstacles that do not allow to easily perform an evaluation in the physical 
four space-time dimensions. For instance, the calculation of multi-loop Feynman
integrals constitutes a challenge due to the presence of singularities. 
Hence, proper procedures to deal with infrared (IR) and ultraviolet (UV) 
divergences and with physical threshold singularities have to be devised and implemented. 
As seen in many applications, starting with an integrand free of singularities
makes the evaluation more stable and leads to reliable numerical results. 
Such an integrand-level representation of physical observables, 
characterised by a point-by-point, or \emph{local} cancellation of singularities, 
is one of the main topics of this report and a valuable item in the HEP community wish-list.

This report is one of the outcomes of the discussions and activities
of the workshop ``WorkStop/ThinkStart~3.0: paving the way to alternative NNLO strategies'', 
which took place on 4.-6. November 2019 at the
Galileo Galilei Institute for Theoretical Physics (GGI) in Florence.
In this report, we compare several strategies to locally cancel IR singularities 
and, thus, providing local integrand-level representations of physical 
observables in four space-time dimensions.
In order to analyse their features, we consider the NNLO correction 
to the scattering process $e^{+}e^{-}\to\gamma^{*}\to q\bar{q}$.
The adopted techniques extend the ones summarised in Ref.~\cite{Gnendiger:2017pys},
where thoughtful and complete descriptions of the NLO 
calculations for this process are provided.
Although the full calculation of NNLO predictions
requires several ingredients, we elucidate among the different frameworks
their main features to perform such computations.
Special emphasis is put on comparing the advantages 
and limitations of each strategy, 
in order to provide the reader with a better understanding 
of the techniques that are currently available.

It is clear that having a fully local representation of any physical observable 
allows for a smooth numerical evaluation and,
thus, a more direct calculation of highly accurate predictions 
to be compared with the experimental data.
In the context of this kind of calculations, taking care of the regularisation 
techniques applied to reach well-defined results is very important. 
Speaking in a wide sense, in this report we consider two kinds of techniques, 
depending on the underlying dimensionality of the integration space: 
four vs. $d$ dimensional implementations. 
Both alternatives are constrained to fulfil several conditions.
In the following, we list a few of them. 
\begin{itemize}

\item 
A comparison between various versions of regularisation 
schemes that regulate singularities by treating the integration momenta in 
$d$ dimensional space-time
(dimensional schemes)
and schemes that do not alter the space-time dimension
(non-dimensional schemes)
is carried out at NNLO, 
showing transition rules between both approaches at
intermediate steps of the calculation.

\item The ultraviolet renormalisation, preserving all the symmetry properties of the
amplitude, is under study in the different approaches.
Some of these methods aim at an integr\textit{and}-level renormalisation, 
which differs from the traditional integr\textit{al}-level framework.
In other words, one is interested in extracting the usual UV counterterms
directly from the bare amplitude 
rather than subtracting integrated counterterms. 
Once again, the main focus is put on the locality: 
subtracting the UV singularities directly from the 
amplitude leads to a local cancellation of 
non-integrable contributions in the UV region.

\item It is clear that multi-loop level calculations are, in general,
 contaminated not only by UV singularities. 
In fact, the presence of IR ones makes a direct integration more involved. This
is because the standard IR singularities need to be canceled by the
corresponding real corrections unless an IR counterterm is encountered.

%\item The treatment of $\gamma_{5}$ has to be elucidated in order to properly
%find agreement among the different schemes. Interesting results starting
%at one- and promoted to multi-loop level 
%have been recently considered, 
%showing that purely four dimensional 
%representations simplify some issues 
%related to the formal properties of the former. 

\end{itemize}
A clear understanding of the aforementioned points will pave an avenue
to  provide a systematic procedure to generate NNLO calculations.
On top of the $d$- or not to $d$-dimensional techniques, 
the treatment of $\gamma_{5}$ might be elucidated to, thus, 
find agreement between the different schemes.
Although this topic was not considered on of main the targets of this workshop,
interesting discussions took place. 
In particular, at the closing discussion, which we summarise at the end
of this report.

Nevertheless, with the very interesting developments at the HEP 
colliders proposed for the near future, it is currently mandatory
to consider higher-order predictions. Therefore, NNLO is no longer
the ultimate goal and all methods need to overcome the obstacles
of providing observables at N$^3$LO and N$^4$LO accuracy. 
Hence, for these reasons, 
presenting a collection of different methods has the intention of illustrating
where we are and what we can do next. 
To this end, in the present report, we comment on the features of the
following regularisation/renormalisation schemes as well as methods
that are only focused on the local cancellation of IR singularities: 
\begin{itemize}
\item Dimensional schemes: four dimensional helicity scheme
(\FDH) and dimensional reduction~(\DRED)
\item Non-dimensional schemes: four-dimensional
unsubtracted scheme~(\FDU), four
dimensional regularisation/renormalisation (\FDR),
and implicit regularisation~(\IReg).
\item Subtraction methods: the $\qt$-subtraction method, the antenna
 and the local analytic sector subtraction. 
\end{itemize}

In the last section of this report, we summarise the advantages and disadvantages 
of the above-mentioned methods. 
Furthermore, we provide a very brief summary of issues that were mentioned
during the closing discussion session at the Workshop.

\graphicspath{{fdrfdh/}}
\section{NNLO processes in FDH/DRED and FDR}

The vast majority of higher-order calculations in QFT are done using
conventional dimensional regularisation (\CDR) to deal with ultraviolet
(UV) and infrared (IR) singularities in intermediate expressions. As
discussed in~\cite{Gnendiger:2017pys}, there are several alternative
approaches, trying to reduce or in fact even eliminate the need to
shift from four to $d\!=\!4-\!2\epsilon$ dimensions.

In this contribution we elaborate further along these lines. In a
first step, we corroborate the relation between individual parts
(double-virtual, real-virtual, double-real) of NNLO cross sections
computed in different variants of dimensional regularisation such as
the four-dimensional helicity scheme (\FDH) and dimensional reduction
(\DRED). In particular, we compute the individual parts of
$H\!\to\! b\bar{b}$ at NNLO in \DRED\ and \FDH, and reproduce the decay rate
obtained in \CDR. As for the process $\gamma\!\to\!q \bar q$~%
\cite{Gnendiger:2019vnp}, the double-real corrections in \DRED\
are simply obtained by integrating the four-dimensional matrix element
squared over the phase space.

The differences that occur by dropping the $\mathcal{O}(\epsilon)$
terms in the real matrix element are compensated by similar
modifications in the real-virtual corrections and adapted UV
renormalisation, such that the physical cross section is scheme
independent. This cancellation of the scheme dependence is best
understood by treating the $\epsilon$-scalars that need to be
introduced to consistently define \FDH\ and \DRED\ as spurious
physical particles. The UV and IR singularities of processes involving
$\epsilon$-scalars cancel for physical processes after consistent UV
renormalisation and combining double-virtual, real-virtual, and
double-real parts. This leaves us with a finite contribution
multiplied by $\Neps=2\epsilon$, the multiplicity of the $\epsilon$-scalars.
Setting $\epsilon\to 0$ in the final result the contribution
of the $\epsilon$-scalars drops out or, equivalently, the scheme
dependence cancels.

Since the contribution of $\epsilon$-scalars drops out for physical
observables it is, of course, possible to leave them out from the very
beginning. This is nothing but computing in \CDR. However, including
$\epsilon$-scalars sometimes offers advantages, as it is (from a
technical point of view) equivalent to performing the algebra in four
dimensions. We reiterate the statement that introducing $\epsilon$-scalars
in diagrams and counterterms is nothing but a consistent procedure
(also beyond leading order) to technically implement the often made
instruction to ``perform the algebra in four dimensions''.\footnotew{
Alternative studies that have a four-dimensional representation 
of the $d$-dimensional space-time have been studied, at one-loop level,
in Ref.~\cite{Fazio:2014xea}, displaying interesting features in 
formal~\cite{Mastrolia:2015maa,Primo:2016omk}
and in phenomenological applications~\cite{Gnendiger:2017rfh,Bruque:2018bmy}. 
 }

Once we know how to transform from \CDR\ to dimensional schemes where
some degrees of freedom are kept in four dimensions, we ask the question
whether the latter can be related to an entirely four-dimensional
calculation using four-dimensional regularisation
(\FDR)~\cite{Pittau:2012zd, Page:2015zca, Page:2018ljf}.\footnotew{
A similar approach that does not alter the space-time dimension is considered 
in Section~\ref{sec:ireg}, where a rewriting of the Feynman propagator,
as shall be described, allows to explicitly extract the UV dependence 
from the propagators.  
}
It is clear that physical results must not depend on the regularisation scheme and,
therefore, \FDR\ (and \FDH\ and \DRED) has to reproduce the results
obtained in \CDR, as long as the same renormalisation scheme
(typically $\MS$) is used. For that reason we investigate the possibility
to relate  \textit{individual parts} of the calculation obtained in \FDR\
to the corresponding ones in \FDH\ and \DRED.  Such a relation would deepen
our understanding of the alternative schemes and tighten the argument
that they are consistent (at least to the order investigated). On a
more practical level, it opens up the possibility to compute
individual parts in different schemes and consistently combine these
results.

We start in Section~\ref{sec:dimschemes} by presenting the new results
for $H\!\to\! b\bar{b}$ in \DRED\ and \FDH\ and by discussing the
results for $\gamma\!\to\! q \bar q$.  The corresponding NLO results
and the $N_F$ part of the NNLO results in \FDR\ are presented in
Section~\ref{sec:fdr}.  In Section~\ref{sec:relate} we study the
relation of the individual parts between the dimensional schemes and
\FDR.

\subsection{FDH and DRED}\label{sec:dimschemes}

The relation between \CDR\ and other dimensional schemes such as
\FDH\ and \DRED\ has been investigated thoroughly in the
literature~\cite{Broggio:2015dga, Gnendiger:2016cpg}.
Before we consider the processes $H\!\to\!b\bar{b}$ and
$\gamma^*\!\to\!q\bar{q}$ we collect the renormalisation constants that
are required. As is well known \cite{Jack:2007ni}, in \FDH\ and \DRED\
the evanescent coupling of an $\epsilon$-scalar to a fermion,
$a_e\!=\!\alpha_e/(4\pi)$, has to be distinguished from the gauge coupling
$a_s\!=\!\alpha_s/(4\pi)$. Identifying the renormalisation and regularisation
scales, the relation between the bare couplings $a_s^0$ and $a_e^0$ to the
$\MS$-renormalised ones is given as
\begin{subequations}
\label{eq:ren1}
\begin{align}
a_{s}^{0}  &= Z_{a_s}^{\MS} \, a_{s} =
    a_{s}\,S_{\epsilon}^{-1}\,
    \Big\{1- 
    \frac{a_{s}}{\epsilon}
    \Big[
	 C_A\Big(
	    \frac{11}{3}
	    -\frac{\Neps}{6}
	    \Big)
	-\frac{2}{3}N_F
	\Big]
  +\mathcal{O}(a^2) \Big\}\,,
   \label{eq:ren1a}
\\
a_{e}^{0}  &= Z_{a_e}^{\MS} \, a_{e} =
    a_{e}\,S_{\epsilon}^{-1}\,
    \Big\{1-
      \frac{a_{s}}{\epsilon}\Big[ 
        6\,C_F
        \Big] 
      -\frac{a_{e}}{\epsilon}\Big[
    	C_A(2\!-\!\Neps)
    	-C_F(4\!-\!\Neps)
    	-N_F
    	\Big]
  + \mathcal{O}(a^2) \Big\}\,,
   \label{eq:ren1b}
\end{align}
 \end{subequations}
with $S_{\epsilon}\!=\!e^{-\epsilon\gamma_{E}}(4\pi)^{\epsilon}$.
In \CDR\ we only need \eqref{eq:ren1a} with $\Neps\!\to\!0$,
whereas in \FDH\ and \DRED\ \mbox{$\Neps\!=\!2\epsilon$}.
The corresponding values in the $\DR$ scheme are simply obtained
by setting $\Neps\!=\!0$ in \eqref{eq:ren1}. As the $N_F$ part
does not depend on $\Neps$ at the one-loop level, it is the same
both in $\MS$ and $\DR$. For later purpose we further need the
difference between \eqref{eq:ren1b} and \eqref{eq:ren1a} which
is given by
\begin{align}
    \delta_{Z}^{\MS} \equiv S_{\epsilon}
    \Big[Z_{a_e}^{\MS}\!-\!Z_{a_s}^{\MS} \Big]_{a_e=a_s}
    =\frac{a_s^2}{\epsilon}\Big[
        C_F\Big(\!-\!2\!-\!\Neps\Big)
        \!+\!C_A\Big(\frac{5}{3}\!+\!\frac{5}{6}\Neps\Big)
        \!+\!\frac{1}{3}N_F\Big]
    + \mathcal{O}(a^3) \,.
    \label{eq:delta}
\end{align}
For $H\!\to\!b \bar{b}$ we also need the renormalisation of the Yukawa
coupling $y_{b}^{0} $ which is defined as the ratio of the bottom quark mass
and the Higgs vacuum expectation value, see e.\,g.\
\cite{Anastasiou:2011qx,Gehrmann:2014vha}.
Apart from its appearance through the Yukawa coupling we will set the
bottom quark mass to zero.  While the renormalisation of $y_{b}^{0}$
is well known in \CDR, for \FDH/\DRED\ this constitutes an additional
calculational step. Using the $\MS$ scheme we get
\begin{align}
y_{b}^{0} =  y_{b} \Big[ 1 &+ a_s\, S_{\epsilon}^{-1}\, C_F\,\Big(
        \!-\!\frac{3}{\epsilon}
        \!-\!\frac{\Neps}{2\,\epsilon}\Big) \nonumber \\
    &
    + a^2_s\, S_{\epsilon}^{-2}\,C^2_F\,\Big(
      \frac{9}{2\,\epsilon^2}
    - \frac{3}{4\,\epsilon}
    + \frac{2\,\Neps}{\epsilon^2}
    - \frac{\Neps}{\epsilon}
    + \frac{3\,\Neps^2}{8\,\epsilon^2}
    - \frac{9\,\Neps^2}{16\,\epsilon} \Big) \nonumber \\
    &
    + a^2_s\,  S_{\epsilon}^{-2}\,C_A C_F\,\Big(
      \frac{11}{2\,\epsilon^2}
    - \frac{97}{12\,\epsilon}
    + \frac{\Neps}{4\,\epsilon^2}
    - \frac{19\,\Neps}{24\,\epsilon}
    - \frac{\Neps^2}{4\,\epsilon^2}
    + \frac{\Neps^2}{4\,\epsilon} \Big) \nonumber \\
    & 
    + a^2_s\,  S_{\epsilon}^{-2}\, C_F N_F\,\Big(
     - \frac{1}{\epsilon^2}
    + \frac{5}{6\,\epsilon}
    - \frac{\Neps}{4\,\epsilon^2}
    + \frac{\Neps}{8\,\epsilon} \Big) 
    \label{eq:yuk_ren}
\Big]\,.
\end{align}
Similar to \eqref{eq:delta} we again have set equal the renormalised(!)
couplings $a_e\!=\!a_s$.  The Yukawa renormalisation can be obtained from
the UV divergences of an off-shell computation of the $H\!\to\!b\bar{b}$
Green functions; however a technically simpler determination is also
possible \cite{GnendigerPhD} by taking the on-shell form factor and
subtracting the IR divergences obtained from the known general
structure \cite{Broggio:2015dga}.

\subsubsection{$H\to b \bar b$ in FDH/DRED}
\label{sec:fdhdred}
\begin{figure}[t]
\begin{center}
\includegraphics[width=1.8in]{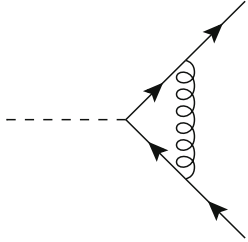}
\end{center}
\vskip -0.2cm
\caption{\label{fig:FDR2} The one-loop vertex correction to the
    $H\!\to\!b\bar b$ amplitude. The diagram only contains a
    'singular' gluon field. In \DRED\ and \FDH\ there is an additional
    diagram with the gluon replaced by an $\epsilon$-scalar.}
\end{figure}

In the following we consider one- and two-loop corrections to
$H\!\to\!b\bar b$ in \FDH\ and \DRED. Our discussion follows closely
\cite{Gnendiger:2017pys} (for NLO) and
\cite{Gnendiger:2019vnp} (for NNLO)
where the corresponding calculations for the process
$e^+e^-\!\to\!\gamma^*\!\to\! q\bar{q}$ are discussed.
To start, we notice that disentangling the $\epsilon$-scalar
contributions is actually simpler for $H\!\to\!b\bar b$ than for
$\gamma^*\!\to\!q \bar q$.  This is due to the fact that the tree-level
interaction is not mediated by a gauge boson and, accordingly,
we therefore only have to split the gluon field in the one-loop
contributions.  Moreover, the major difference between \FDH\ and
\DRED\ is in the treatment of so-called 'regular' vector fields, see
e.\,g.\ Table~1 of \cite{Gnendiger:2017pys}.  As in the present case
only 'singular' vector fields contribute, the virtual correction is the
same in both schemes, i.\,e.
\begin{align}
    \mathcal{A}_{\text{\FDH}}(H\!\to\! b\bar{b})\ \equiv\
    \mathcal{A}_{\text{\DRED}}(H\!\to\! b\bar{b})\ \equiv\
    \mathcal{A},
\end{align}
see also Figure~\ref{fig:FDR2}. Using form factor coefficients
$F^{(n)}$ of loop-order $n$, the bare amplitude for
$H\!\to\! b \bar{b}$ can be written as
\begin{align}
\mathcal{A}_{\text{bare}}%(H\to b\bar{b})
&= \mathcal{A}^{(0)}_{\text{bare}} \Big[ 1 +
  \sum_n \Big(\frac{\mu_0^2}{\!-\!M_H^2}\Big)^{n\,\epsilon}\,
    S_{\epsilon}^{n}\,F^{(n)}_{\text{bare}} \Big]\,,
\end{align}
including the regularisation scale $\mu_0$ and the Higgs mass
$M_H$. The tree-level amplitude
$\mathcal{A}^{(0)}_{\text{bare}}\!=\! i\,y_{b}^{0}\, \bar{u}(p_b)v(p_{\bar{b}})$
results in the tree-level decay width $\Gamma^{(0)}\!=\!2\,M_H^2 y_b^2/(2 M_H)$.

\subsubsection*{NLO}
The NLO virtual corrections are directly related the one-loop form
factor for which we get in \FDH/\DRED
\begin{align}
F^{(1)}_{\text{bare}}
= %a_{s}^{0}\,
a_{s}^{0}\,C_F\,
\Big[
  \!-\! \frac{2}{\epsilon^2}
  \!-\! 2
    \!+\! \frac{\pi^2}{6} 
    \!+\! \mathcal{O}(\epsilon)
    \Big]
+
a_{e}^{0}\,C_F\,\Neps\,
\Big[
  \frac{1}{\epsilon}
  \!+\! 2
    \!+\! \mathcal{O}(\epsilon)
    \Big]
\,.
\label{fhbare:oneloop}
\end{align}
The well-known \CDR\ result which is given e.\,g.\ in
\cite{Anastasiou:2011qx,Gehrmann:2014vha} is obtained by setting
$\Neps\!\to\!0$. Identifying the (bare) couplings in
\eqref{fhbare:oneloop} {\it before} making replacement \eqref{eq:ren1}
corresponds to so-called 'naive' \FDH\ and it has been shown that this
leads to inconsistent results at higher orders~\cite{Kilgore:2012tb,
Gnendiger:2014nxa, Broggio:2015ata}. Instead, we first have to
renormalise $a_{s}^{0}$ and $a_{e}^{0}$ and only then we are allowed
to identify the renormalised couplings $a_e\!=\!a_s$.

Integrating over the phase space and using the conventions of
\cite{Gnendiger:2017pys} we then obtain for the UV-renormalised
virtual cross section 
\begin{align}
\Gamma^{(v)}_{\text{\DS}%\mFDH/\mDRED
} = 
\Gamma^{(0)}\,\CF\,
  \Phi_2(\epsilon)\,c_\Gamma(\epsilon)\,
  s^{-\epsilon}
  \Big\{a_s
    \Big[
        \!-\!\frac{4}{\epsilon^2}
        \!-\!\frac{6}{\epsilon}
        \!-\!4
        \!+\!2\,\pi^2
        \!+\!\mathcal{O}(\epsilon)
    \Big] + a_e
    \Big[
        \!+\!\frac{\Neps}{\epsilon}
        \!+\!\mathcal{O}(\epsilon^0)
    \Big]
  \Big\}\,,
\label{sig_bbv_four}
\end{align}
where we introduce the $\epsilon$-dependent prefactors
\begin{subequations}
  \label{eq:prefactors}
\begin{align}
c_\Gamma(\epsilon)
&=(4\pi)^{\epsilon}
  \frac{\Gamma(1+\epsilon)\,\Gamma^2(1-\epsilon)}{\Gamma(1-2\epsilon)}
  =1+\mathcal{O}(\epsilon)\,,
  \label{cgammadef2}\\
\Phi_2(\epsilon)
&=\Big(\frac{4\pi}{s}\Big)^\epsilon
  \frac{\Gamma(1-\epsilon)}{\Gamma(2-2\epsilon)}
=1+{\cal O}(\epsilon)\,,
\label{eq:Phi2}\\
  \Phi_3(\epsilon)
&=\Big(\frac{4\pi}{s}\Big)^{2\epsilon}\frac{1}{\Gamma(2-2\epsilon)}
= 1 + {\cal O}(\epsilon)\, .
\label{eq:Phi3}
\end{align}
\end{subequations}
The subscript \DS\ in \eqref{sig_bbv_four} and in what follows indicates
that the results for all dimensional schemes can be obtained from this
expression.  For the evaluation of the real contribution we use the setup
and notation of \cite{Gnendiger:2017pys} and arrive at
\begin{align}
\Gamma^{(r)}_{\text{\DS}}
& = 
\Gamma^{(0)}\,\CF\,\Phi_3(\epsilon)\,
\Big\{a_s\Big[
        \!+\!\frac{4}{\epsilon^2}
        \!+\!\frac{6}{\epsilon}
        \!+\!21
        \!-\!2\,\pi^2
        \!+\!\mathcal{O}(\epsilon)
        \Big]
    + a_e\Big[
        \!-\!\frac{\Neps}{\epsilon}
        \!+\!\mathcal{O}(\epsilon^0)
    \Big]
    \Big\}\,.
\label{sig_bbr_four}
\end{align}
As for the virtual cross section, the result is the same in \FDH\ and
\DRED\ which is due to the absence of 'regular' gauge bosons at NLO for
the considered process. The presence of a 'singular' gluon, however,
leads to $a_s$ contributions (which stem from the $\dim$-dimensional gluon)
and $a_e$ contributions (which stem from the associated $\epsilon$-scalar gluon).
Again, at NLO such a distinction is of course not strictly necessary;
at NNLO, however, the different renormalisation of $a_s$ and $a_e$
has to be taken into account, see also Sec.~\ref{sec:relatevv}.

Finally, since $c_\Gamma\,\Phi_2 s^{-\epsilon}/\Phi_3 =1+{\cal O}(\epsilon^3)$,
the cancellation of singularities between \eqref{sig_bbv_four} and~%
\eqref{sig_bbr_four} takes place and leaves us with the well known
finite answer
\begin{align}
  \label{hbb:oneloop}
\Gamma^{(1)}
=\Gamma^{(0)} + \Gamma^{(v)}_{\text{\DS}} + \Gamma^{(r)}_{\text{\DS}}\Big|_{
    \dim\to4
    }
=\Gamma^{(0)}\Big[
    1 + a_s\,17\,\CF  \Big]\,.
\end{align}
In particular, the evanescent contributions $\propto a_e\Neps $ drop out.

\subsubsection*{NNLO}

In the following we provide separately the double-virtual,
double-real, and real-virtual contributions to $\Gamma$ in \FDH/\DRED.
All of these results are new and have not been published elsewhere
before.  For brevity reasons we keep the dependence on $\Neps$
explicit in the divergent terms, but drop finite $\Neps$ terms.
Setting $\Neps\!\to\!0\ (2\epsilon) $ then yields the
\CDR\ (\FDH/\DRED) result. As before, we give the UV-renormalised
results after having set $a_e\!\equiv\!a_s$.

Writing the double virtual corrections as
\begin{align}\label{dec}
  \Gamma^{(vv)}_{\text{\DS}}
  &= \Gamma^{(0)}\,\Phi_2(\epsilon)\,a^2_s\,\CF\,\Big[
    C_F\, \Gamma^{(vv)}_{\text{\DS}}(C_F)
    + C_A\, \Gamma^{(vv)}_{\text{\DS}}(C_A)
    + N_F\, \Gamma^{(vv)}_{\text{\DS}}(N_F) \Big]\, ,
\end{align}
we then obtain for the individual parts
\begin{subequations}
\label{eq:vv_ren}
\begin{align}
\Gamma^{(vv)}_{\text{\DS}}(C_F)
&=
     + \frac{8}{\epsilon^4}
     + \frac{24-4\Neps}{\epsilon^3}
     + \frac{1}{\epsilon^2} \Big(
      34-\frac{28\pi^2}{3}-23\Neps\Big)\\*
     &+ \frac{1}{\epsilon} \Big(
        \frac{109}{2}
        \!-\!12\pi^2
        \!-\!\frac{184\zeta_3}{3}
        \!-\!62 \Neps
        \!+\!\frac{20\pi^2 \Neps}{3}
        \!+\! \frac{31\Neps^2}{8}
        \Big)
     \!+\!128
     \!-\! \frac{40\pi^2}{3}
     \!+\! \frac{137\pi^4}{45}
     \!-\! 116\zeta_3\, ,
     \nonumber \\
 \Gamma^{(vv)}_{\text{\DS}}(C_A)
 &=
     + \frac{22-\Neps}{2\,\epsilon^3}
     + \frac{1}{\epsilon^2} \Big(
        \frac{32}{9}
        \!+\!\frac{\pi^2}{3}
        \!-\!\frac{19\Neps}{18}
        \!+\!\frac{\Neps^2}{2}
        \Big) \\*
     &+\frac{1}{\epsilon} \Big(
        \!-\!\frac{961}{54}
        \!-\!\frac{11\pi^2}{6}
        \!+\!26\zeta_3
        \!+\!\frac{761\Neps}{108}
        \!+\!\frac{ \pi^2 \Neps}{12}
        \Big)
     \!-\!\frac{934}{81}
     \!+\!\frac{701\pi^2}{54}
     \!-\!\frac{8\pi^4}{45}
     \!+\!\frac{302\zeta_3}{9}\, ,
   \nonumber  \\
 \Gamma^{(vv)}_{\text{\DS}}(N_F)
 &=
     - \frac{2}{\epsilon^3}
     + \frac{1}{\epsilon^2} \Big(
        \!-\! \frac{8}{9}
        \!+\!\frac{\Neps}{2}
        \Big)
     + \frac{1}{\epsilon} \Big(
        \frac{65}{27}
        \!+\!\frac{\pi^2}{3}
        \!-\!\frac{3\Neps}{4}
        \Big)
     + \frac{400}{81}
     \!-\! \frac{55\pi^2}{27}
     \!+\! \frac{4\zeta_3}{9}\, .
     \label{eq:vv_nf_ren}
\end{align}
\end{subequations}
We note that in \eqref{dec} we could have pulled out additional
prefactors such as $c_\Gamma^2$, in analogy to \eqref{sig_bbv_four}.
This would of course modify the subleading poles and finite terms of
\eqref{eq:vv_ren} (as well as \eqref{eq:rv_ren} and \eqref{eq:rr}
below). The conclusions, however, we will draw in Section~\ref{sec:relate}
are not affected by the choice of the prefactor.

As mentioned before, the one- and two-loop renormalisation of the
Yukawa coupling is included in \eqref{eq:vv_ren} in order to get
consistent results. We would like to stress, that in \FDH\ and \DRED\
there are finite terms associated with the $\MS$ renormalisation factors. 
The terms $\propto\Neps^m/\epsilon^n$ ($m,n\!>\!0$) that are potentially
finite when setting $\Neps\!=\!2\epsilon$ cancel when combining the
double-virtual with the real-virtual and double-real contribution,
as will be shown below. However, if the double-real (and real-virtual)
corrections are computed by doing the algebra in four dimensions, the
$\Neps$ terms are not disentangled any longer and, therefore, all these
terms need to be included to obtain a consistent result. 
Moreover, as no $\epsilon$-scalars are present at the tree level,
\eqref{eq:ren1} is sufficient for the renormalisation of $a_s^0$ and $a_e^0$. 

Splitting the real-virtual and double-real contributions in a similar way
as before, we then get
\begin{subequations}
\label{eq:rv_ren}
\begin{align}
\Gamma^{(rv)}_{\text{\DS}}(C_F)
&=
     - \frac{16}{\epsilon^4}
     - \frac{48\!-\!8\Neps}{\epsilon^3}
     + \frac{1}{\epsilon^2} \Big(
        \!-\!146
        \!+\!\frac{64\pi^2}{3}
        \!+\!42\Neps
        \!-\!\frac{\Neps^2}{2}
        \Big) \\*&\quad
    +\frac{1}{\epsilon} \Big(
        \!-\!524
        \!+\!46\pi^2
        \!+\!\frac{848\zeta_3}{3}
        \!+\!147\Neps
        \!-\!\frac{37\pi^2\Neps}{3}
        \!-\!\frac{13\Neps^2}{2}
        \Big)  
     \!-\!1879
     \!+\!170\pi^2
     \!-\!\frac{38\pi^4}{9}
     \!+\!624\zeta_3\, ,
     \nonumber \\
 \Gamma^{(rv)}_{\text{\DS}}(C_A)
 &=
     -\frac{2}{\epsilon^4}
     -\frac{62\!-\!5\Neps}{3\,\epsilon^3}
     +\frac{1}{\epsilon^2} \Big(
        \!-\!52
        \!+\!\frac{7\pi^2}{3}
        \!+\!\Neps
        \!-\!\frac{\Neps^2}{2}
        \Big)  \\*&\quad
    + \frac{1}{\epsilon} \Big(
        \!-\!209
        \!+\!\frac{158\pi^2}{9}
        \!+\!\frac{16\zeta_3}{3}
        \!+\!\frac{25\Neps}{2}
        \!-\!\frac{17\pi^2\Neps}{9}
        \Big)
     \!-\! \frac{4769}{6}
     \!+\! \frac{355\pi^2}{6}
     \!-\! \frac{47\pi^4}{36}
     \!+\! \frac{2000\zeta_3}{9}\, , \nonumber\\
 \Gamma^{(rv)}_{\text{\DS}}(N_F)
 &=
     + \frac{8}{3\,\epsilon^3}
     + \frac{1}{\epsilon^2} \Big(4-\Neps \Big)
     + \frac{1}{\epsilon} \Big(
        14
        \!-\!\frac{14\pi^2}{9}
        \!-\!\frac{5\Neps}{2}
        \Big)
     \!+\! \frac{127}{3}
     \!-\! \frac{7\pi^2}{3}
     \!-\! \frac{200\zeta_3}{9}\,
\end{align}
\end{subequations}
and
\begin{subequations}\label{eq:rr}
\begin{align}
\Gamma^{(rr)}_{\text{\DS}}(C_F)
&=
     + \frac{8}{\epsilon^4}
     + \frac{24\!-\!4\Neps}{\epsilon^3}
     + \frac{1}{\epsilon^2} \Big(
        112
        \!-\!12\pi^2
        \!-\!19\Neps
        \!+\!\frac{\Neps^2}{2}
        \Big) \\*&\quad
    +\frac{1}{\epsilon} \Big(
        \frac{939}{2}
        \!-\! 34\pi^2
        \!-\!\frac{664\zeta_3}{3}
        \!-\!85\Neps
        \!+\!\frac{17\pi^2 \Neps}{3}
        \!+\!\frac{21\Neps^2}{8}
       \Big) 
     \!+\! \frac{7695}{4}
     \!-\! \frac{488\pi^2}{3}
     \!+\! \frac{53\pi^4}{45}
     \!-\! 544\zeta_3\, ,   \nonumber  \\
 \Gamma^{(rr)}_{\text{\DS}}(C_A)
 &=
     + \frac{2}{\epsilon^4}
     + \frac{58\!-\!7\Neps}{6\epsilon^3}
     + \frac{1}{\epsilon^2} \Big(
        \frac{436}{9}
        \!-\!\frac{8\pi^2}{3}
        \!-\!\frac{89\Neps}{18}
          \Big) \\*&\quad
    + \frac{1}{\epsilon} \Big(
        \frac{12247}{54}
        \!-\! \frac{283\pi^2}{18}
        \!-\! \frac{94\zeta_3}{3}
        \!-\! \frac{2111\Neps}{108}
        \!+\! \frac{65 \pi^2 \Neps}{36}
        \Big) 
     \!+\! \frac{333595}{324}
     \!-\! \frac{2047\pi^2}{27}
     \!+\! \frac{89\pi^4}{69}
     \!-\!\frac{2860\zeta_3}{9}\, , \nonumber\\
 \Gamma^{(rr)}_{\text{\DS}}(N_F)
 &=
     - \frac{2}{3\epsilon^3}
     + \frac{1}{\epsilon^2} \Big(
       \!-\!\frac{28}{9}
       \!+\!\frac{\Neps}{2}
       \Big)
     + \frac{1}{\epsilon} \Big(
        \!-\!\frac{443}{27}
        \!+\!\frac{11\pi^2}{9}
        \!+\!\frac{13\Neps}{4}
        \Big)
     \!-\! \frac{12923}{162}
     \!+\! \frac{136\pi^2}{27}
     \!+\! \frac{268\zeta_3}{9}\, .
\end{align}
\end{subequations}
To get the double-real contribution we used the integrals
of~\cite{Gehrmann-DeRidder:2003pne} and the \verb"FORM" code
of~\cite{GehrmannDeRidder:2004tv}. As for the process
$\gamma^*\!\to\!q\bar{q}$~\cite{Gnendiger:2019vnp}, the
double-real corrections in \DRED/\FDH\ are simply obtained
by integrating the four-dimensional matrix element squared
over the phase space. Their determination is therefore
significantly simplified compared to the case of \CDR.

Finally, combining these results, we can extend
\eqref{hbb:oneloop} to NNLO 
as\footnotew{Let us remark that this cancellation is obtained after individually 
integrating contributions that are divergent in four dimensions and 
combining them to obtain a finite remainder. In Section~\ref{sec:fdu}, we provide
the main ingredients towards a local cancellation at integrand level. 
}
\begin{subequations}
\begin{align}
  \label{hbb:twoloop}
\Gamma^{(2)}
&=\Gamma^{(1)}
+ \Gamma^{(vv)}_{\text{\DS}}
+ \Gamma^{(rv)}_{\text{\DS}}
+ \Gamma^{(rr)}_{\text{\DS}}
\Big|_{\dim\to4} \\*
&=\Gamma^{(0)}\Big[ 
  1 + a_s\,17\,\CF
     + a^2_s\, C^2_F \Big(
        \frac{691}{4}
        \!-\!6\pi^2
        \!-\!36\zeta_3
        \Big)
     \nonumber \\*
     & \qquad \qquad
     + a^2_s\, C_F C_A \Big(
        \frac{893}{4}
        \!-\!\frac{11\pi^2}{3}
        \!-\!62\zeta_3
        \Big)
     + a^2_s\, C_F N_F \Big(
        \!-\!\frac{65}{2}
        \!+\!\frac{2\pi^2}{3}
        \!+\! 8\zeta_3
        \Big)
     \Big]\,
\end{align}
\end{subequations}
in agreement with~\cite{Baikov:2005rw}.
As expected, the divergent $\Neps$ parts cancel in the final result
which is nothing but the scheme-independence of the physical result.

\subsubsection{$\gamma \to q \bar q$ in FDH/DRED}

The scheme dependence of the cross section
$e^+e^-\!\to\!\gamma^*\!\to\!q\bar{q}$ at NNLO is discussed in detail
in~\cite{Gnendiger:2019vnp}. For this particular process the individual
results in \DRED\ and \FDH\ differ, as in \DRED\ there are
$\epsilon$-scalar photons present in the tree-level process.
While \cite{Gnendiger:2019vnp} mainly deals with \DRED, here we only
recall a few points that are relevant for the comparison of \FDH\ with \FDR.
In particular, we want to extend to NNLO the investigation of the interplay
between \FDH\ and \FDR\ presented at NLO in~\cite{Gnendiger:2017pys}. 

\subsubsection*{NLO}

Copying the results given in Section 2.3 of \cite{Gnendiger:2017pys},
we write the virtual and real cross sections as 
\begin{align}
\sigma^{(v)}_\mFDH & =
\sigma^{(0)}\,\CF\,
  \Phi_2(\epsilon)\,c_\Gamma(\epsilon)\,
  s^{-\epsilon}
  \Big\{a_s
      \Big[
        \!-\!\frac{4}{\epsilon^2}
        \!-\!\frac{6}{\epsilon}
        \!-\!16
        \!+\!2\pi^2
        \!+\!\mathcal{O}(\epsilon)
      \Big] \!+\! a_e
      \Big[
        \!+\!\frac{\Neps}{\epsilon}
        \!+\!\mathcal{O}(\epsilon^0)
      \Big]
    \Big\}\,,
\label{sig_eev_four} \\
\sigma^{(r)}_\mFDH  &= 
\sigma^{(0)}\,\CF\,
  \Phi_3(\epsilon)\,
  \Big\{
      a_s\Big[
          \!+\!\frac{4}{\epsilon^2}
          \!+\!\frac{6}{\epsilon}
          \!+\!19
          \!-\!2\pi^2
          \!+\!\mathcal{O}(\epsilon)
      \Big] \!+\! a_e\Big[
        \!-\!\frac{\Neps}{\epsilon}
        \!+\!\mathcal{O}(\epsilon^0)    
      \Big]
  \Big\}  \, ,
  \phantom{\bigg|}
  \label{sig_eer_fdh}
\end{align}
where $\sigma^{(0)}\!=\!e^4/(4\pi)\,Q_q N_c/(3s)$ includes the electric
charge and the colour number of the quark as well as the flux factor
1/(2s). Combining these two contributions we find the well-known 
regularisation-scheme independent physical cross section
\begin{align}
  \label{yqq:oneloop}
\sigma^{(1)}
=\sigma^{(0)}
+ \sigma^{(v)}_{\text{\FDH}}
+ \sigma^{(r)}_{\text{\FDH}}\Big|_{\dim\to4}
=\sigma^{(0)}\Big[
    1 + a_s\,3\,\CF  \Big]\,.
\end{align}

\subsubsection*{NNLO}
Moving on to NNLO, we first split the cross section into a double-virtual,
a double-real, and a real-virtual part, i.\,e.\
\begin{align}
        \sigma^{\text{\NNLO}}_{\text{\FDH}}=
        \sigma^{(vv)}_{\text{\FDH}}+
        \sigma^{(rr)}_{\text{\FDH}}+
        \sigma^{(rv)}_{\text{\FDH}} \, .
\end{align}
For the comparison with \FDR\ in Section~\ref{sec:relate}, we here focus on the
$N_F$ part of the respective contributions.
The double-virtual part can be extracted from the \DRED\ result given in (3.8b) of \cite{Gnendiger:2019vnp} as
%\begin{subequations}
\begin{align}
     \sigma^{(vv)}_{\text{\FDH}}(N_F)
     &=\sigma^{(0)}\,\Phi_2(\epsilon)\,
        a_{s}^2\,C_F N_F    \Big[
        \!-\!\frac{2}{\epsilon^3}
        \!-\!\frac{8}{9 \epsilon^2}
        \!+\!\frac{1}{\epsilon} \Big(\frac{92}{27} 
        \!+\!\frac{\pi^2}{3}\Big)
        \!+\!\frac{1921}{81}
        \!-\!\frac{91\pi^2}{27}
        \!+\!\frac{4}{9}\zeta_3
        \Big]\, .
        \label{eq:resFDHvv}
\end{align}
The other two contributions are given by
\begin{align}
     \sigma^{(rv)}_{\text{\FDH}}(N_F)
     &=\sigma^{(0)}\,\Phi_2(\epsilon)\,
        a_{s}^2\,C_F N_F    \Big[
        \frac{8}{3\epsilon^3}
        \!+\!\frac{4}{\epsilon^2}
        \!+\!\frac{1}{\epsilon} \Big(\frac{32}{3} 
        \!-\!\frac{14\pi^2}{9}\Big)
        \!+\!\frac{94}{3}
        \!-\!\frac{7\pi^2}{3}
        \!-\!\frac{200}{9}\zeta_3
        \Big]\, ,
        \label{eq:resFDHrv} \\
     \sigma^{(rr)}_{\text{\FDH}}(N_F)
     &=\sigma^{(0)}\,\Phi_2(\epsilon)\,
        a_{s}^2\,C_F N_F    \Big[
        \!-\!\frac{2}{3\epsilon^3}
        \!-\!\frac{28}{9 \epsilon^2}
        \!+\!\frac{1}{\epsilon} \Big(\!-\!\frac{380}{27} 
        \!+\!\frac{11\pi^2}{9}\Big)
        \!-\!\frac{5350}{81}
        \!+\!\frac{154\pi^2}{27}
        \!+\!\frac{268}{9}\zeta_3
        \Big]\,.
        \label{eq:resFDHrr} 
    \end{align}
%\end{subequations}
The combination of all three parts results in
\begin{align}\label{ed:fdhNF}
\sigma^{(vv)}_{\text{\FDH}}(N_F) + 
\sigma^{(rv)}_{\text{\FDH}}(N_F) +
\sigma^{(rr)}_{\text{\FDH}}(N_F)
&= 
\sigma^{(0)}\,a_{s}^2\,C_F N_F 
\big[\!-\!11\!+\!8 \zeta_3 \big]
\end{align}
in agreement with the literature~\cite{Celmaster:1979xr, Chetyrkin:1979bj}.   
Note that the constant terms of \eqref{eq:resFDHrv} and
\eqref{eq:resFDHrr} differ from the corresponding results in \DRED, as
given in (3.20) and (3.17b) of \cite{Gnendiger:2019vnp}.

\subsection{FDR: Four-dimensional regularisation/renormalisation}
\label{sec:fdr}

\subsubsection{$H\to b \bar b$ in FDR}

\subsubsection*{NLO}

\begin{figure}[t]
\begin{center}
\includegraphics[width=2.in]{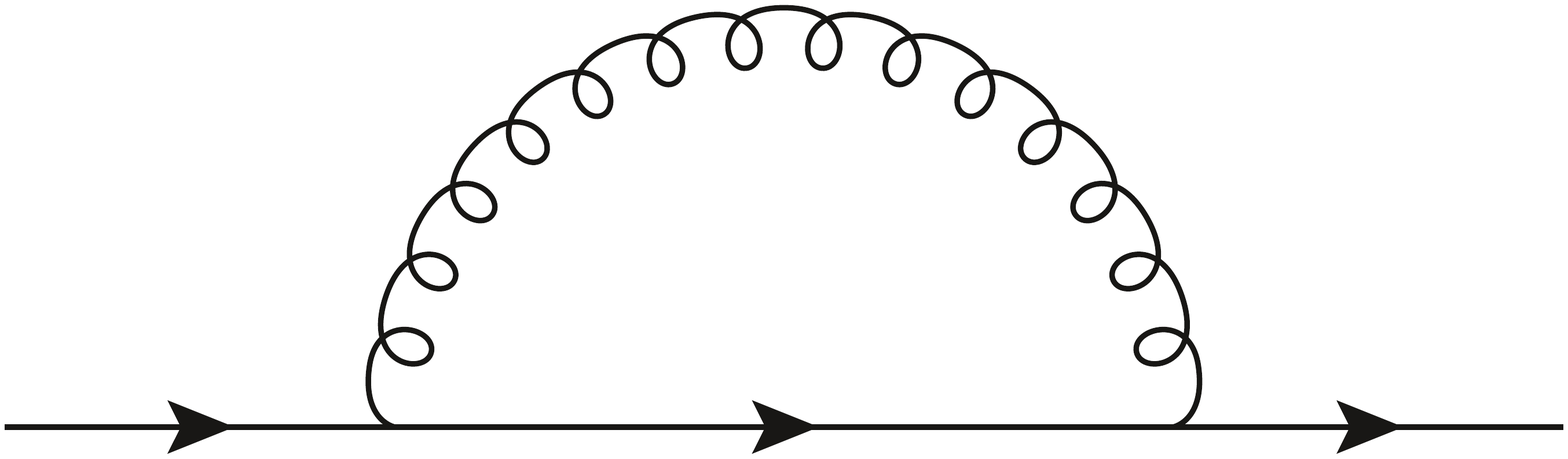}
\end{center}
\vskip -1.5cm
\caption{\label{fig:FDR1} The one-loop correction to the bottom mass.}
\end{figure}

Here we describe the \FDR\ NLO calculation of the decay rate
$\Gamma_{H\to b \bar b(g)}$.  The strong coupling constant does not
appear at the tree-level and as in Section~\ref{sec:dimschemes} we use
the $\MS$ value with $a=\alpha_s/{4 \pi}$.  As for the
unrenormalised bottom mass, it is denoted by $m_0$ and again it is
taken to be different from zero only in the Yukawa coupling.  The
one-loop relation between $m_0$ and the physical pole mass $m$,
defined as the value of the four-momentum at which the bottom quark
propagator develops a pole, is obtained by evaluating the diagram of
Figure \ref{fig:FDR1},
\begin{eqnarray}
\label{eq:FDR1}
m_0  = m (1+a \delta m), \quad
\delta m = - \CF (3 L^{\prime\prime}+5), \quad
L^{\prime\prime} :=\ln \mu^2-\ln m^2.
\end{eqnarray}

\begin{figure}[t]
\begin{center}
\includegraphics[width=2.5in]{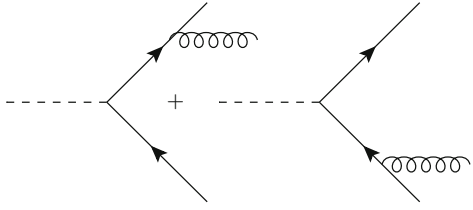}
\end{center}
\vskip -1.2cm
\caption{\label{fig:FDR3} The two diagrams contributing to the $H\to b \bar b g$
    amplitude. In \DRED\ and \FDH\ there are additional diagrams with gluons
    replaced by $\epsilon$-scalars. }
\end{figure}

The unphysical mass $\mu^2$ in \eqref{eq:FDR1} is the \FDR\ UV
regulator, which in the present calculation is taken to coincide with
the \FDR\ IR regulator.\footnotew{This is done in such a way that one- and two-loop
scaleless Feynman integrals are still set to zero in \FDR.}
Once the decay amplitude is renormalised
(namely expressed in terms of physical quantities only) all the
$\mu^2$s of UV origin get replaced by physical scales, and the
remaining ones are IR regulators which cancel in the sum of virtual
and real contributions.  As a matter of fact, we will encounter two
additional combinations besides $L^{\prime\prime}$,
\begin{eqnarray}\label{defL}
  L^\prime := \ln \mu^2-\ln (-s-i 0^+)\quad {\rm and}\quad
  L := \ln\mu^2-\ln s,  
\end{eqnarray}
where for $H\to b \bar{b}$ we have $s=M_H^2$, the Higgs mass squared. 
The amplitude contributing to $H\to b \bar b$ can be written as
\begin{eqnarray}
\label{eq:FDR2}
A_2 = m_0 A^{(0)}_2 (1+ a A_2^{(v)}),
\end{eqnarray}
where $A^{(0)}_2$ is the tree level result and the one-loop correction of
Figure \ref{fig:FDR2} reads
\begin{eqnarray}
A_2^{(v)} = -\CF (L^\prime)^2.
\end{eqnarray}
Inserting \eqref{eq:FDR1} in  \eqref{eq:FDR2} produces the
renormalised one-loop amplitude
\begin{eqnarray}
\label{eq:FDR3}
A_2^{(1)}= m A^{(0)}_2
\left[
1-a \CF \big(5+3 L^{\prime \prime }+(L^\prime)^2 \big)
\right].
\end{eqnarray}
Upon integration over the 2-body phase-space, the square of
\eqref{eq:FDR3} gives the virtual part of the NLO corrections,
\begin{eqnarray}
\label{eq:FDR4}
\Gamma^{(v)}_{\mFDR}(m^2) =
-a \CF \Gamma^{(0)}(m^2)\, {\cal R}e
\left[2 (L^\prime)^2+6 L^{\prime \prime}+10
\right].
\end{eqnarray}
This can be translated to the $\MS$ scheme by expressing
$m^2$ in terms of the running $\MS$ bottom mass
$\underline{m}^2(M^2_H)$ \cite{Bednyakov:2016onn}
\begin{eqnarray}
\label{massMS}
m^2= \underline{m}^2(M^2_H) 
\left[
1+a\CF \big(8-6 \ln \frac{m^2}{M^2_H}\big) 
\right]= \underline{m}^2(M^2_H) 
\left[
1+a\CF \big(8-6 (L - L^{\prime\prime})\big) 
\right] \, .
\end{eqnarray}
With the corresponding change in the Yukawa we obtain
\begin{eqnarray}
\label{eq:FDRv}
\Gamma^{(v)}_{\mFDR}=
-a \CF \Gamma^{(0)}\, {\cal R}e
\left[2 (L^\prime)^2+6 L+2
\right]\, ,
\end{eqnarray}
which can be directly compared with \eqref{sig_bbv_four}.  The real
counterpart is obtained by squaring the diagrams of Figure
\ref{fig:FDR3} and integrating over a 3-body phase-space in which all
final-state particles acquire a small mass $\mu$
\cite{Pittau:2013qla},
\begin{eqnarray}
\label{eq:FDR5}
\Gamma^{(r)}_{\mFDR} = \Gamma^{\rm \scriptscriptstyle R}_{H\to b \bar b g}=
a \CF \Gamma^{(0)}_{H\to b \bar b}(m^2)
\left[
2 L^2+6 L+19-2 \pi^2
\right].
\end{eqnarray}
Replacing ${\cal R}e(L')^2 = L^2+\pi^2$, the sum of \eqref{eq:FDRv} and
\eqref{eq:FDR5} gives the UV and IR finite decay rate up to the NLO accuracy
\begin{eqnarray}
\Gamma^{(1)}_{\mFDR}= \Gamma^{(0)}\, \left[ 1+ 17 a \CF \right]\, ,
\end{eqnarray}
which agrees with \eqref{hbb:oneloop}.

\subsubsection*{NNLO, $N_F$ part}

The $N_F$ parts contributing to the NNLO cross section in \FDR\ are
given in (4.19) of \cite{Page:2018ljf} in terms of the Yukawa coupling
defined through the pole mass of the bottom quark. If we express these
results in terms of the $\MS$ Yukawa couplings, as done for
\eqref{eq:FDRv}, we get
%\begin{subequations}
\begin{align}
     \Gamma^{(vv)}_{\text{\FDR}}(N_F)
     &=\Gamma^{(0)}\,
        a_{s}^2\,C_F N_F    \Big[
        \!-\!\frac{8}{9} L^3
        \!+\!\frac{2}{9} L^2
        \!+\! L \Big(\frac{278}{27}\!+\!\frac{4}{9}\pi^2\Big)
        \!+\!\frac{3425}{162}
        \!-\!\frac{40}{27}\pi^2
        \!-\!\frac{16}{3}\zeta_3
        \Big]\,,
        \label{eq:hresFDR} \\
     \Gamma^{(rv)}_{\text{\FDR}}(N_F)
     &=\Gamma^{(0)}\,
        a_{s}^2\,C_F N_F    \Big[
        \!+\!\frac{4}{3} L^3
        \!+\! 4 L^2
        \!+\! L \Big(\frac{38}{3}\!-\!\frac{4}{3}\pi^2\Big)
        \Big]\,,
        \label{eq:hresFDRrv} \\
     \Gamma^{(rr)}_{\text{\FDR}}(N_F)
     &=\Gamma^{(0)}\,
        a_{s}^2\,C_F N_F    \Big[
        \!-\!\frac{4}{9} L^3
        \!-\!\frac{38}{9} L^2
        \!-\! L \Big(\frac{620}{27}\!-\!\frac{8}{9}\pi^2\Big)
        \!-\!\frac{4345}{81}
        \!+\!\frac{58}{27}\pi^2
        \!+\!\frac{40}{3}\zeta_3
        \Big]\,.
        \label{eq:hresFDRrr} 
\end{align}
%\end{subequations}

\subsubsection{$\gamma \to q \bar q$ in FDR}

\subsubsection*{NLO}

The computation of $\gamma \to q \bar q$ in \FDR\ at NLO has been
discussed in~\cite{Gnendiger:2017pys}. Here we just list the results
for the virtual and real corrections, as given in (5.2) and (5.35) of
\cite{Gnendiger:2017pys}. They read
%\begin{subequations}
\begin{align}
\sigma^{(v)}_\mFDR &=
\sigma^{(0)}\,a_s\,\CF\,
    \Big[
    \!-\!2 L^2
    \!-\! 6 L
    \!-\!14
    \!+\!2\pi^2
\Big]\,,
    \phantom{\bigg|}
\label{sig_eev_fdr} \\
\sigma^{(r)}_\mFDR &=
\sigma^{(0)}\,a_s\,\CF\,
  \Big[
  \!+\! 2 L^2
  \!+\!6 L
  \!+\!17
  \!-\!2\pi^2
  \Big]\,,
  \phantom{\bigg|}
  \label{sig_eer_fdr}
\end{align}
%\end{subequations}
where $L$ is defined in \eqref{defL}, with $s$ now the centre-of-mass energy squared.

\subsubsection*{NNLO, $N_F$ part}

The $N_F$ parts contributing to the NNLO cross section in \FDR\ are
given in (5.13) of \cite{Page:2018ljf} and read
%\begin{subequations}
\begin{align}
     \sigma^{(vv)}_{\text{\FDR}}(N_F)
     &=\sigma^{(0)}\,
        a_{s}^2\,C_F N_F \Big[
        \!-\!\frac{8}{9} L^3
        \!+\!\frac{2}{9} L^2
        \!+\! L \Big(\frac{278}{27}\!+\!\frac{4}{9}\pi^2\Big)
        \!+\!\frac{3355}{81}
        \!-\!\frac{76}{27}\pi^2
        \!-\!\frac{16}{3}\zeta_3
        \Big]\,,
        \label{eq:resFDR} \\
     \sigma^{(rv)}_{\text{\FDR}}(N_F)
     &=\sigma^{(0)}\,
        a_{s}^2\,C_F N_F   \Big[
        \!+\!\frac{4}{3} L^3
        \!+\! 4 L^2
        \!+\! L \Big(\frac{34}{3}\!-\!\frac{4}{3}\pi^2\Big)
        \Big]\,,
        \label{eq:resFDRrv} \\
     \sigma^{(rr)}_{\text{\FDR}}(N_F)
     &=\sigma^{(0)}\,
        a_{s}^2\,C_F N_F    \Big[
        \!-\!\frac{4}{9} L^3
        \!-\!\frac{38}{9} L^2
        \!+\! L \Big(
            \!-\!\frac{584}{27}
            \!+\!\frac{8}{9}\pi^2
            \Big)
        \!-\!\frac{4246}{81}
        \!+\!\frac{76}{27}\pi^2
        \!+\!\frac{40}{3}\zeta_3
        \Big]\,.
        \label{eq:resFDRrr} 
\end{align}
%\end{subequations}

\subsection{Relations between FDH and FDR}
\label{sec:relate}

\subsubsection*{NLO}

The relation between IR divergences of virtual (and therefore real)
one-loop results obtained in \FDH\ and \FDR\ has been established
in (5.41) of \cite{Gnendiger:2017pys} for the process
$\gamma^*\!\to\! q \bar{q}$ and reads
\begin{align} 
\label{relNLO}
    \frac{1}{\epsilon^2} \leftrightarrow \frac{1}{2} L^2_{}\, ,\qquad
        \frac{1}{\epsilon} \leftrightarrow L_{}\, . 
\end{align}
These relations are here confirmed by the results of $H\!\to\!b \bar{b}$.
In fact, \eqref{sig_bbr_four} and \eqref{eq:FDR5} are related through
\eqref{relNLO} and so are \eqref{sig_bbv_four} and \eqref{eq:FDRv} if
the $\MS$ mass \eqref{massMS} is used. Moreover, if $\epsilon\!=\!0$
in prefactors that are related to integration in $\dim$ dimensions,
i.\,e.\ $c_{\Gamma}(\epsilon\!=\!0)$, $\Phi_{2}(\epsilon\!=\!0)$, and
$\Phi_{3}(\epsilon\!=\!0)$ in~\eqref{eq:prefactors}, the finite terms
of the individual parts are the same in \FDH\ and \FDR, including the
$\pi^2$ terms.

In order to find similar transition rules at NNLO, we also follow a
second approach for the comparison between \FDH\ and \FDR\ which is
slightly different.
To start, we multiply a generic virtual one-loop result obtained in
\FDH\ by an $\epsilon$-dependent function $\phi_1(\epsilon)$ and demand
equality with the corresponding \FDR\ result, i.\,e.\
\begin{align}
\label{eq:matchgv}
\phi_1(\epsilon)\,\Gamma^{(v)}_{\text \FDH}
\,\equiv\,  \Gamma^{(v)}_{\text \FDR}\,, \qquad
\phi_1(\epsilon)\,\sigma^{(v)}_{\text \FDH}
\,\equiv\,  \sigma^{(v)}_{\text \FDR}\,
\end{align}
with
\begin{align}
\label{eq:phi1}
\phi_1(\epsilon)
= 1
    +\epsilon\,   b_{11}
    +\epsilon^2\, b_{12}\,.
\end{align}
The coefficients $b_{11}$ and $b_{12}$ are so far unknown and can
be obtained by inserting known \FDH\ and \FDR\ results. Before we
can do this, however, we have to rescale powers of $1/\epsilon$ by
using
\begin{align}
    \Big(\frac{1}{\epsilon}\Big)^{k_1}
    \to(\lambda_1\,L)^{k_1}\,,\qquad
    k_1\in\{1,2\}\,.
    \label{eq:rescale1}
\end{align}
The scale factor $\lambda_1$ is so far unknown and sets the
size of the \FDH\ $1/\epsilon$-pole with respect to the \FDR\
logarithm $L$ at NLO. Evaluating \eqref{eq:matchgv} by
using for instance \eqref{sig_eev_four} and \eqref{sig_eev_fdr}
we find after expanding, applying shift \eqref{eq:rescale1},
and dropping $\mathcal{O}(\epsilon)$ terms
\begin{subequations}
\label{eq:bijres1}
\begin{align}
    b_{11} & \ = \ -\frac{3}{2}+\frac{3}{2\,\lambda_1}\,,\\
    b_{12} & \ = \ +\frac{9}{4}-\frac{9}{4\,\lambda_1}\,,\\
    (\lambda_1)^2 & \ = \ \frac{1}{2}\,.
    \label{eq:lamnda1res}
\end{align}
\end{subequations}
Eqs.~\eqref{eq:phi1}--\eqref{eq:bijres1} are equivalent
to \eqref{relNLO} in that they allow to translate virtual
(and therefore real) one-loop results obtained in \FDH\ to \FDR\
and vice versa, including the finite terms. The validity of the rules
has been checked explicitly for NLO corrections to $H\!\to\!b\bar{b}$
and $\gamma^*\!\to\!q\bar{q}$.

\subsubsection*{NNLO, double-virtual}
\label{sec:relatevv}

We start our NNLO comparisons with the IR divergences in the
$N_F$ part of the double-virtual contributions.
As mentioned repeatedly, to get correct and unitary NNLO results
in \FDH\ it is crucial to distinguish evanescent couplings at
NLO, see for instance \eqref{fhbare:oneloop} and the text below
as well as \eqref{sig_eev_four}. In \FDR, on the other hand,
unitarity is restored by treating UV-divergent one-loop subdiagrams
appearing in two-loop amplitudes in the same way they are treated at NLO.
This can be achieved by introducing 'extra-integrals', as explained
in Appendix A of~\cite{Page:2018ljf}. After taking this contribution
into account, it is possible to compare the IR divergences of the
double-virtual results obtained in \FDH\ and \FDR.

In order to find transition rules between the two schemes we follow
the approach described in the previous section and adjust the
involved quantities accordingly, i.\,e. we demand
\begin{align}
\label{eq:matchgv2}
\phi_2(\epsilon)\,\Gamma^{(vv)}_{\text \FDH}(N_F)
\ \equiv\  \Gamma^{(vv)}_{\text \FDR}(N_F)\,,\qquad
\phi_2(\epsilon)\,\sigma^{(vv)}_{\text \FDH}(N_F)
\ \equiv\  \sigma^{(vv)}_{\text \FDR}(N_F)\,
\end{align}
with
\begin{align}
\label{eq:phi2}
\phi_2(\epsilon)
= 1
    +\epsilon\,   b_{21}
    +\epsilon^2\, b_{22}
    +\epsilon^3\, b_{23}\,
\end{align}
and get after the rescaling
\begin{align}
    \Big(\frac{1}{\epsilon}\Big)^{k_2}
    \to(\lambda_2\,L)^{k_2}\,,\qquad
    k_2\in\{1,2,3\}\,
    \label{eq:rescale2}
\end{align}
and the use of \eqref{eq:vv_nf_ren} and \eqref{eq:resFDHvv}
the two-loop coefficients
\begin{subequations}
\label{eq:bijres2}
 \begin{align}
 \label{eq:b21res}
 b_{21} &=
    -\frac{4}{9}
    \!-\!\frac{1}{9\,(\lambda_{2})^2}\,, \\
\label{eq:b22res}
 b_{22} &=
    \frac{154}{81}
    \!-\!\frac{139}{27\,\lambda_{2}}
    \!+\!\frac{4}{81\,(\lambda_{2})^2}
    \!+\!\pi^2 \Big(
        \frac{1}{6}
        \!-\!\frac{2}{9\,\lambda_{2}}
        \Big)\,, \\
\label{eq:b23res}
 b_{23} &=
    -\frac{7621}{729}
    \!+\!\frac{556}{243\,\lambda_{2}}
    \!-\!\frac{154}{729\,(\lambda_{2})^2}
    \!-\!\pi^2 \Big(
        \frac{23}{54}
        \!-\!\frac{8}{81\,\lambda_{2}}
        \!+\!\frac{1}{54\,(\lambda_{2})^2}
        \Big)
    \!+\!\frac{26}{9}\zeta_3
        \,,\\
    (\lambda_2)^3 & \ = \ \frac{4}{9}\,.
    \label{eq:lamnda2res}
 \end{align}
 \end{subequations}
Similar to the one-loop case, we have reabsorbed the difference between
the two schemes in the prefactor $\phi_2(\epsilon)$, including the finite
terms. Restricting ourselves to the $N_F$ part of the double-virtual
corrections for a single process, \eqref{eq:phi2}--\eqref{eq:bijres2}
can be seen as using four parameters $\{\lambda_2, b_{21}, b_{22}, b_{23}\}$
to enforce four equalities between the $L^n$ and $\epsilon^{-n}$ terms for
$n\in\{3,2,1,0\}$. What is remarkable is that the same relations hold for
$H\!\to\!b\bar{b}$ and $\gamma^*\!\to\!q\bar{q}$. Despite their apparent
similarity, these two processes have a completely different behaviour
regarding UV renormalisation.  The question whether these rules also apply
to the $C_F^2$ and $C_A\,C_F$ part of the aforementioned processes or
to other processes can not be answered at the moment.
Of course, the precise form of the subleading poles depends on whether
or not prefactors like $c_\Gamma^2(\epsilon)$ are factored out in the
\FDH\ results. This explains the $\pi^2$ and $\zeta_3$ terms in 
\eqref{eq:b22res} and \eqref{eq:b23res}. Moreover, note that the
two-loop scaling factor \eqref{eq:lamnda2res} is different from the
one-loop scaling \eqref{eq:lamnda1res}.

\subsubsection*{NNLO, real-virtual}

Regarding the real-virtual corrections we first notice that for the
considered processes the $N_F$ part solely stems from the
(sub)renormalisation of the bare couplings in the real one-loop result.
In other words, it is given by the product of two
one-loop quantities: the real NLO contribution (which contains double
and single IR divergences) times a one-loop renormalisation constant
(which contains a single UV divergence).

In \FDH, the corresponding terms originate from inserting \eqref{eq:ren1a}
and \eqref{eq:ren1b} in \eqref{sig_bbr_four} and \eqref{sig_eer_fdh}
as the Yukawa coupling $y_b^0$ does not depend on $N_F$ at NLO.
Similar to the double-virtual contributions it is crucial to distinguish
gauge and evanescent couplings in order to avoid the wrong UV-renormalisation
of the $a_e$ terms. Ignoring this distinction, however, leads to results
in 'naive' \FDH\ which are different from \FDH:
\begin{align}
\label{eq:rv1a}
\text{\phantom{'naive' }\FDH:}\qquad
    \Gamma^{(rv)}_{\text{\FDH}}(N_F)&=
     \frac{8}{3\,\epsilon^3}
     + \frac{4}{\epsilon^2}
     + \frac{1}{\epsilon} \Big(
        12
        \!-\!\frac{14\pi^2}{9}
        \Big)
    +\mathcal{O}(\epsilon^0)\, ,
   \\*
 \label{eq:rv2a}
    \sigma^{(rv)}_{\text{\FDH}}(N_F)&\propto
     \frac{8}{3\,\epsilon^3}
     + \frac{4}{\epsilon^2}
     + \frac{1}{\epsilon} \Big(
        \frac{32}{3}
        \!-\!\frac{14\pi^2}{9}
        \Big)
    +\mathcal{O}(\epsilon^0)\,,\\
\label{eq:rv1b}
\text{'naive' \FDH:}\qquad
     \Gamma^{(rv)}_{\text{\FDH'}}(N_F)&=
     \frac{8}{3\,\epsilon^3}
     + \frac{4}{\epsilon^2}
     + \frac{1}{\epsilon} \Big(
        \frac{38}{3}
        \!-\!\frac{14\pi^2}{9}
        \Big)
 +\mathcal{O}(\epsilon^0) \, ,
    \\*
\label{eq:rv2b}
    \sigma^{(rv)}_{\text{\FDH'}}(N_F)&\propto
     \frac{8}{3\,\epsilon^3}
     + \frac{4}{\epsilon^2}
     + \frac{1}{\epsilon} \Big(
        \frac{34}{3}
        \!-\!\frac{14\pi^2}{9}
        \Big)
    +\mathcal{O}(\epsilon^0)\,.
\end{align}
In \FDR, the corresponding results are given in \eqref{eq:hresFDRrv}
and \eqref{eq:resFDRrr} and read
\begin{align}
\label{eq:rv1c}
\text{\phantom{'naive' }\FDR:}\qquad
    \Gamma^{(rv)}_{\text{\FDH}}(N_F)&=
        \frac{4}{3} L^3
        \!+\! 4 L^2
        \!+\! L \Big(\frac{38}{3}\!-\!\frac{4}{3}\pi^2\Big)\,,\\
\label{eq:rv2c}
    \sigma^{(rv)}_{\text{\FDH}}(N_F)&=
        \frac{4}{3} L^3
        \!+\! 4 L^2
        \!+\! L \Big(\frac{34}{3}\!-\!\frac{4}{3}\pi^2\Big)\,.
        \qquad\qquad\
\end{align}
Similar to \FDH, they stem from the $a_s$-renormalisation in \eqref{eq:FDR5}
and \eqref{sig_eer_fdr}, see also (3.1) and (3.2) of \cite{Page:2018ljf}.
In contrast to the double-virtual contributions, however, the different
renormalisation of $a_e$ and $a_s$ is \textit{not} taken into account
via 'extra-integrals'. Accordingly, \eqref{eq:rv1c} and \eqref{eq:rv2c}
correspond to 'naive' \FDH, i.\,e.\ \eqref{eq:rv1b} and \eqref{eq:rv2b},
rather than \FDH.
Since only one-loop quantities are involved, the transition between
'naive' \FDH\ and \FDR\ is already known and given by e.\,g.\ \eqref{relNLO}
times the transition $1/\epsilon_{\text{\UV}}\leftrightarrow L$ for
the UV divergence, i.\,e.%
\begin{align} 
\label{relNLO2}
    \frac{1}{\epsilon^2_{\text{\IR}}}\times\frac{1}{\epsilon_{\text{\UV}}}
        \leftrightarrow \frac{1}{2} L^2 \times L\, ,\qquad
   \frac{1}{\epsilon_{\text{\IR}}}\times\frac{1}{\epsilon_{\text{\UV}}}
        \leftrightarrow L \times L\,, \qquad
   (\epsilon_{\text{\IR}})^0\times\frac{1}{\epsilon_{\text{\UV}}}
        \leftrightarrow 1 \times L\,.
\end{align}
Note that the transition of the divergent $\pi^2$ terms depends on the
$\dim$-dependent prefactors that have been factored out in the \FDH\ result,
similar to the double-virtual contributions.

Regarding the finite terms it is clear that a transition between ('naive')
\FDH\ and \FDR\ can not exist. The reason is that for the considered processes
the $N_F$ part of the real-virtual contribution in \FDR\ is obtained via
multiplying \eqref{eq:FDR5} and \eqref{sig_eer_fdr} by a pure UV divergence.
As a consequence, \eqref{eq:rv1c} and \eqref{eq:rv2c} only contain pure
divergences (which are parametrized as powers of $L$) and no finite terms.
Therefore, the $N_F$ part of the real-virtual contribution alone does not
contain a finite part which is different from any dimensional scheme.

\subsubsection*{NNLO, double-real}
In the previous section we have seen that a transition rule for
real-virtual contributions between \FDH\ and \FDR\ does not exist.
As the physical result has to be scheme independent, this is
also the case for the double-real contributions.
%\subsubsection*{Combining NNLO real-virtual and double-real}
Given the fact that a transition rule exists for the
double-virtual part, however, %one expects
it is clear that \eqref{eq:phi2} can also be used to translate
the sum of the real-virtual and double-real components from \FDH\
to \FDR. This is due to the fact that the divergences are the same
(apart from their sign) and that the finite term is given by the
difference between between the physical result and the finite part
of the double-virtual contribution. We have checked this explicitly
and find indeed
\begin{align}
\label{eq:matchrest}
\phi_2(\epsilon) \Big[
 \Gamma^{(rv)}_{\text{\FDH}}(N_F)
+\Gamma^{(rr)}_{\text{\FDH}}(N_F)
\Big]&=
 \Gamma^{(rv)}_{\text{\FDR}}(N_F)
+\Gamma^{(rr)}_{\text{\FDR}}(N_F)\,,
\\
 \phi_2(\epsilon) \Big[
  \sigma^{(rv)}_{\text{\FDH}}(N_F)
 +\sigma^{(rr)}_{\text{\FDH}}(N_F)
\Big] &=
 \sigma^{(rv)}_{\text{\FDR}}(N_F)
 +\sigma^{(rr)}_{\text{\FDR}}(N_F)\,.
\end{align}

\subsubsection*{Unitarity restoration in FDH and FDR}
As we have commented many times, \FDH\ and \FDR\ use different strategies
to restore unitarity. In this paragraph, we report on an attempt towards
a comparison of the two unitarity restoration methods.

Our starting point is 'naive' \FDH\, in which no distinction is made
between gauge and evanescent couplings, and we try to extract from \FDR\ the
contribution of the \FDH\ evanescent $a_e$ terms.
This is achieved by interpreting the \FDR\ 'extra-integrals' as UV-divergent
dimensionally regulated integrals subtracted at the integrand level, as
explained in Section~6 of~\cite{Page:2015zca}.\footnotew{
Although it is not studied in this report, it would be very interesting
to establish a comparison between the \FDR\ 'extra-integrals' 
and the integrals one obtains after applying the 
local UV renormalisation summarised in Section~\ref{ssec:FDUrenormalisation}.}
By dropping the subtraction
term, one obtains the so called 'extra-extra-integrals of type b' ($EEI_b$).
\footnote{The $EEI_b$ do not belong to the FDR calculation procedure.
    They are introduced only for the sake of comparison with \FDH.}
These integrals contain now $1/\epsilon$  poles of UV origin suitable to be
combined with the 'naive' \FDH\ expressions. For the regarded processes, their
contribution corresponds to the evanescent $a_e$ terms in \eqref{fhbare:oneloop}
and \eqref{sig_eev_four}, respectively, times the difference
of the renormalisation constants
$\delta_Z\!=\!(Z_{\alpha_e}\!-\!Z_{\alpha_s})$,
whose value in the $\MS$-scheme is given in \eqref{eq:delta},
\begin{subequations}
\label{eq:EEIb}
\begin{align}
    \Gamma^{(vv)}_{EEI_b} \ =\ &
     \Gamma^{(0)}_{\text{}}\times\delta_Z^{}\times
        C_F\,\Neps\Big[\frac{2}{\epsilon}+4\Big]\,,\\
    \sigma^{(vv)}_{EEI_b} \ =\ &
     \sigma^{(0)}_{\text{}}\times\delta_Z^{}\times
        C_F\,\Neps\Big[\frac{1}{\epsilon}+1\Big]\,.
\end{align}
\end{subequations}
More precisely, \eqref{fhbare:oneloop} and \eqref{sig_eev_four} are reproduced
if the contribution of \eqref{eq:EEIb} is added to the 'naive' FDH results. 
Note that, since  $\delta_Z\!=\!\mathcal{O}(\epsilon^{\!-\!1})$ and
$\Neps\!=\!2\epsilon$, the $EEI_b$ are of $\mathcal{O}(\epsilon^{\!-\!1})$ and
that \eqref{eq:EEIb} refers to the full contributions, not only the $N_F$ parts.
Finally, it should be mentioned that the contributions in \eqref{eq:EEIb} have
been extracted from off-shell diagrams, while the unitarity restoring \FDR\
procedure to be used on-shell is slightly different~\cite{Page:2018ljf},
although differing at most by finite terms.
We leave a deeper study of this for future work.

\subsection{Discussion}
We have presented new results for $H\!\to\! b\bar{b}$ in \DRED\ and \FDH,
and have compared, up to NNLO, the \FDH\ and \FDR\ calculations of the
$N_F$ part of $H\!\to\! b\bar{b}$ and $\gamma\!\to\! q \bar q$.

The situation at NLO is very satisfactory. There is a universal
%(i.\,e.\ the same for real and virtual corrections)
transition rule for each individual part between
\FDR\ and \FDH. In principle, this allows one to perform the virtual
computation in one scheme, the real in another, and consistently
combine them to obtain the correct physical result. 

At NNLO, we have identified the prefactor which transforms from
\FDH\ to \FDR\ the double-virtual and the {\em sum} of real-virtual
and double-real components, i.e. \eqref{eq:phi2}--\eqref{eq:bijres2}.
However, the real-virtual and double-real contributions transform differently,
so that only their sum can be translated from one scheme to the  other.
Here we have focused on the transformation properties of the contribution
proportional to $N_F$, but it is conceivable that an analogous treatment
also exist for the $C_F$ and $C_A$ parts. At the moment, this cannot be
confirmed due to the lack of the \FDR\ NNLO calculations of the $C_F$ and $C_A$
components.

Finally, we have started a preliminary comparison between the unitarity
restoration mechanisms of \FDH\ and \FDR. 

Many open questions remain that could be interesting subject for further investigation.

\graphicspath{{valencia/}}
%%%%%%%%%%%%%%%%%%%%%%%%%%%%%%%%%%%%%%%%%%%%%
\section{FDU: Four-dimensional unsubtraction} 
\label{sec:fdu}
%%%%%%%%%%%%%%%%%%%%%%%%%%%%%%%%%%%%%%%%%%%%%

Even if subtraction methods have been widely used for the computation of higher-order corrections in perturbative QFT, their applicability to multi-particle multi-loop processes is being challenged by the intrinsic computational complexity. One of the main limitations is related to the treatment of the non-local cancellation of IR/UV~singularities which forces the introduction of counterterms in the real and virtual components. Besides, the non-local issue is enhanced by the fact that most of the computations require some kind of additional regularisation, such as \DREG. 

With the aim of by-passing these difficulties, the four-dimensional unsubtraction (\FDU)~\cite{Hernandez-Pinto:2015ysa, Sborlini:2016gbr, Sborlini:2016hat, Rodrigo:2016hqc, Driencourt-Mangin:2016dpf} approach constitutes a radically-new alternative to the traditional subtraction technique. It is based on the loop-tree duality (\LTD) theorem \cite{Catani:2008xa,Rodrigo:2008fp,Bierenbaum:2010cy,Bierenbaum:2012th}, which establishes a connection among loop and dual integrals. The main advantage of the dual representation is that integrals are defined in 
the Euclidean space, closely related to the usual real-emission phase-space. In this way, the method provides a natural way to implement an integrand-level combination of real and virtual contributions, thus leading to a fully local cancellation of IR\ singularities. Besides that, the \LTD~theorem leads to dual representations of local UV~counterterms~\cite{Driencourt-Mangin:2017gop,Driencourt-Mangin:2019aix,Driencourt-Mangin:2019yhu}, which allows to reproduce the proper results in standard renormalisation schemes by performing a purely four-dimensional numerical computation.

In the following, we describe general properties of the \LTD~theorem, 
focusing on the innovative multi-loop dual representation \cite{Verdugo:2020kzh,Aguilera-Verdugo:2020kzc,Ramirez-Uribe:2020hes,Aguilera-Verdugo:2020nrp}\footnote{Alternative representations have been presented by other authors~\cite{Runkel:2019yrs,Runkel:2019zbm,Capatti:2019ypt,Capatti:2019edf,Capatti:2020ytd}.}. Then, we discuss on the structure of the kinematical mappings that allow to combine the real and virtual corrections in a single integral. Also, general comments about the local renormalisation procedure are presented, making emphasis mainly on 
the two-loop extension of the formalism. Finally, we depict the application of the 
\FDU~framework to obtain the NNLO QED corrections to the $N_f$ terms associated to $\gamma^{*}\!\to\! q \bar q (g)$.

We would like to highlight that the \FDU/\LTD~framework has been extended and improved since the last WorkStop/ThinkStart meeting in 2016~\cite{Gnendiger:2017pys}. 
Therefore, we briefly review the new features and properties that 
have been recently improved. 

%%%%%%%%%%%%%%%%%%%%%%%%%%%%%%%%%%%%%%%%%%%%%%%%%%%%%%%%%%%
\subsection{Dual representations from the LTD theorem}
\label{ssec:FDUIntroduction}
%%%%%%%%%%%%%%%%%%%%%%%%%%%%%%%%%%%%%%%%%%%%%%%%%%%%%%%%%%%
The \LTD~theorem is based on a clever application of Cauchy's residue theorem (CRT) to the loop integrals. The original version was developed in Ref.~\cite{Catani:2008xa,Bierenbaum:2010cy}, 
where CRT was used to integrate out the energy component of each loop momenta. This procedure reduces loop amplitudes to collections of tree-level-like diagrams with a modification of the customary Feynman prescription. Thus, given a loop line associated to the momentum $q_i$, we should use the rule
\beq
G_F(q_i)=\frac{1}{q_i^2-m_i^2+\imath 0} \to G_D(q_i;q_j) = \frac{1}{q_i^2-m_i^2+\imath \, \eta\cdot(q_j-q_i)} \, ,    
\eeq
to replace the Feynman propagators when the momentum $q_j$ is set on-shell. In this expression, $\eta$ is a future-like vector that defines the explicit dependence of the prescription on the momentum flow, and the integration contours are closed on the lower half-plane such that only those poles with negative imaginary components are selected. It is important to notice that this dual representation is equivalent to the usual Feynman tree theorem (\FTT) \cite{Feynman:1963ax,Feynman:1972mt}, and the multiple cut information is encoded within the momentum-dependent dual prescription.

Recently, a new representation of dual integrals was achieved through the iterated calculation of residues, leading to what we call \emph{nested residues}~\cite{Verdugo:2020kzh,Aguilera-Verdugo:2020kzc,Ramirez-Uribe:2020hes,Aguilera-Verdugo:2020nrp}. This strategy turns out to be more efficient computationally, since it allows a straightforward algorithmic implementation. 
Hence, in order to elucidate  how  \LTD~formalism works, let us consider 
an $L$-loop scattering amplitude with $N$ external legs, $\{p_j\}_N$
and $n$ internal lines, 
in the Feynman representation,
\begin{align}
\mathcal{A}_N^{(L)}(1,\ldots, n) = \int_{\ell_1, \ldots, \ell_L} \mathcal{A}_F^{(L)}(1,\ldots, n) = \int_{\ell_1, \ldots, \ell_L} \mathcal{N}( \{ \ell_s\}_L,  \{ p_j\}_N) \, G_F(1,\ldots, n) \, .
\label{eq:amplitude}
\end{align}
This amplitude is naturally defined over the Minkowski space 
of the $L$ loop momenta, $\{\ell_s\}_L$. 
The singular structure is associated to the denominators introduced by the Feynman propagators, 
\begin{align}
G_F(1,\ldots, n) = \prod_{i\in 1\cup\ldots \cup n} \left( G_F(q_i) \right)^{a_i} \, ,
\end{align}
with 
\begin{align}
G_{F}\left(q_{i}\right) & =\frac{1}{q_{i}^{2}-m_{i}^{2}+\imath0}=\frac{1}{\left(q_{i,0}-q_{i,0}^{\left(+\right)}\right)\left(q_{i,0}+q_{i,0}^{\left(+\right)}\right)}\,,\label{eq:gf}
\end{align}
where the $q_{i},m_{i}$ and $+\imath0$ correspond to the loop momentum,
its mass, and the infinitesimal Feynman prescription. Furthermore,
we explicitly pull out the dependence on the energy component of
the loop momentum $q_{i,0}$ together with its on-shell energy, 
\begin{align}
q_{i,0}^{\left(+\right)} & =\sqrt{\boldsymbol{q}_{i}^{2}+m_{i}^{2}-\imath0}\,,
\end{align}
that is expressed in terms of the spatial components of the loop momentum. 

Besides, $a_i$  and $\{1,\ldots,n \}$ in Eq.~\eqref{eq:amplitude} correspond to 
the arbitrary positive integers, raising the powers of the propagators, and 
sets containing internal momenta of the form $q_{i_j}=k_{j}+\ell_i \in i$, respectively. 
In the following, the $a_j$ exponents will not be included in the notation because the treatment of the expressions is independent of their explicit values. 

As in the previous \LTD~representation, the dual contributions are obtained by integrating out one degree of freedom per loop through the Cauchy residue theorem. Iterating this procedure, we can write~\cite{Verdugo:2020kzh},
\begin{align}
\label{eq:nested}
&\mathcal{A}_D^{(L)}(1,\ldots, r; r+1,\dots, n)  
=-2\pi \imath %\\ & \quad \times 
\sum_{i_r \in r} {\rm Res} (\mathcal{A}_D^{(L)}(1, \ldots, r-1;r, \ldots, n), {\rm Im}(\eta\cdot q_{i_r})<0)\, , %\nn
\end{align}
starting from 
\begin{align}
&\mathcal{A}_D^{(L)}(1; 2, \ldots, n)  
=-2\pi \imath %\nn \\ & \qquad 
\sum_{i_1 \in 1} {\rm Res} (\mathcal{A}_F^{(L)}(1, \dots, n), {\rm Im}(\eta\cdot q_{i_1})<0)\, .
\end{align}
All the sets in~\Eq{eq:nested} before the semicolon contain one propagator that has been set on shell, while all the 
propagators belonging the sets that appear after the semicolon remain off shell. The sum over all possible on-shell configurations is implicit. Regarding the prescription, the contour choice is the same used in the original \LTD~formulation. It is worth mentioning that the \LTD\ representation is independent of the coordinate system, and that there are some non-trivial cancellations when iterating the residue loop-by-loop. The last result implies that only those poles whose loci is always on the lower complex half-plane will lead to non-vanishing contributions; the others, called \emph{displaced poles} will cancel at each iterative step~\cite{Aguilera-Verdugo:2020nrp}. Moreover, these contributions can be mapped onto the usual cut diagrams, thus allowing a graphical interpretation.

Finally, we would like to emphasise that the \LTD~representation corresponds to tree-level like objects integrated in the Euclidean space. In this way, the application of this novel representation of multi-loop scattering amplitudes allows to open loops into trees, 
aiming at finding a natural connection with the structures exhibited in the real emission contributions.

%%%%%%%%%%%%%%%%%%%%%%%%%%%%%%%%%%%%%%%%%%%%%%%%%%%%%%%%%%%
\subsection{Local cancellation of IR singularities through kinematic mappings}
\label{ssec:FDUMappings}
%%%%%%%%%%%%%%%%%%%%%%%%%%%%%%%%%%%%%%%%%%%%%%%%%%%%%%%%%%%
Once the \LTD\ theorem is applied to any multi-loop multi-leg scattering amplitude, we obtain a representation involving tree-level like objects and phase-space integrals. This leads to an important conceptual simplification to understand the origin of IR~and threshold singularities, at integrand level. By analysing the intersection of the integration hyperboloids associated to the dual contributions for different cuts~\cite{Buchta:2014dfa,Buchta:2015jea,Buchta:2015wna}, it is possible to detect those internal states that are simultaneously set on shell and lead to singular propagators. Moreover, it turns out that the intersection of positive (forward) and negative (backward) hyperboloids is responsible of the physical IR~singularities of multi-loop amplitudes, and it is localised within a compact region of the integration domain~\cite{Hernandez-Pinto:2015ysa,Sborlini:2016gbr,Sborlini:2016hat}. A recent re-interpretation and extension of this analysis \cite{Aguilera-Verdugo:2019kbz} was found useful to identify the origin of causal and anomalous thresholds, whose contributions are integrable but still introduce numerical instabilities.

The knowledge of the IR~and threshold singular structure of multi-loop scattering amplitudes is important for computing higher-order corrections to physical IR-safe observables. 
Due to the Kinoshita-Lee-Nauenberg (KLN) theorem~\cite{Kinoshita:1962ur,Lee:1964is}, 
summing over all the degenerated states contributing to a certain observable lead to a finite result. Thus, from the theoretical perspective, adding real-emission processes to the multi-loop amplitudes will produce a cross-cancellation of IR~singularities present in the different terms. 
Within \DREG, the divergent contributions manifest as $\epsilon$-poles after performing the $d$-dimensional integrals, which forces to use semi-numerical methods and reduces the efficiency of the cancellations. On the contrary, the \FDU\ approach aims at an early-stage cancellation, before the integration, by putting together the real and \emph{dualised} virtual contributions through proper momenta mappings.

The \FDU\ formalism has been successfully proven to deal with NLO corrections to physical observables, such as cross sections and decay rates \cite{Hernandez-Pinto:2015ysa,Sborlini:2016fcj,Sborlini:2016gbr,Sborlini:2016hat}. This involved the combination of one-loop scattering amplitudes with tree-level extra-radiation processes (i.e. processes with one additional particle), through the application of suitable momenta mapping which are very similar to the ones applied in the Catani-Seymour (CS)~\cite{Catani:1996jh,Catani:1996vz} or Frixione-Kunszt-Signer (FKS)~\cite{Frixione:1995ms} algorithms. These mappings relate the on-shell states in the virtual corrections with the momenta of the additional particles. 

For the sake of simplicity, let us consider a process involving only final-state radiation (FSR) singularities, such as an $n$-particle decay. The LO contribution is given by,
\beq
\sigma^{(0)} = \int {\rm dPS}^{1\to n} |{\cal M}^{(0)}_{n}|^2 \, {\cal S}_0(\{p_i\}) \, , 
\eeq
with $|{\cal M}^{(0)}_{n}|^2$ the Born squared amplitude and ${\cal S}_0$ the IR-safe measure function that defines the physical observable. On the one hand, the virtual one-loop contribution is
\beq 
\sigma_V^{(1)} = \int {\rm dPS}^{1\to n} \, \int_{\ell} 2 {\rm Re}({\cal M}^{(1)}_{n} {\cal M}^{(0),*}_{n}) \, {\cal S}_0(\{p_i\}) \, ,
\label{eq:VirtualNLO}
\eeq 
where ${\rm Re}({\cal M}^{(1)}_{n} {\cal M}^{(0),*}_{n})$ corresponds to the interference between the one-loop and the Born amplitude, including also factors that might be related to self-energy contributions. On the other hand, we need to take into account the real-emission contribution, 
\beq 
\sigma_R^{(1)} = \int {\rm dPS}^{1\to n+1} \, |{\cal M}^{(0)}_{n+1}|^2 \, {\cal S}_1(\{{p'}_i\}) \, ,
\label{eq:RealNLO}
\eeq 
which is characterised by the presence of an additional external particle. Notice that the measure function, ${\cal S}_1(\{{p'}_i\})$, is extended to include the extra radiation and it must fulfil the reduction property ${\cal S}_1 \to {\cal S}_0$ in the IR~limits. Moreover, the corresponding IR~singularities can be disentangled and isolated into disjoint regions of the real-emission phase-space. Following a slicing strategy, we introduce a partition ${\cal R}_i$ and define
\beq
\sigma_{R,i}^{(1)} = \int {\rm dPS}^{1\to n+1} \, d\sigma_R^{(1)} \, {\cal R}_i \, ,
\eeq
that fulfils $\sum_ i \sigma_{R,i}^{(1)} = {\sigma}_R^{(1)}$ and that only one IR~divergent configuration is allowed inside different ${\cal R}_i$.

Regarding the virtual contribution, the application of \LTD~to Eq. (\ref{eq:VirtualNLO}) will produce a sum of cuts leading to \emph{dual contributions}, namely
\beq 
\sigma_D^{(1)} = \int {\rm dPS}^{1\to n} \sum_{i=1}^{N}  \, \int_{\vec{\ell}} I_i(q_i) \, {\cal S}_0(\{p_i\}) \equiv \, \int {\rm dPS}^{1\to n} \sum_{i=1}^{N} \sigma_{D,i}^{(1)} \ ,
\label{eq:DualNLO}
\eeq 
with $N$ the number of different internal lines. Each line is characterised by a momenta $q_i$, which is set on shell in the different dual terms. 
The presence of this extra on-shell momenta allows to establish a connection with the real emission contributions, through a proper momentum mapping. 
Explicitly, for the NLO case, we have a bijective transformation,
\beq
{\cal T}_i(\{p_1,\ldots\,p_n,q_i\}) \to \{p'_1,\ldots\,p'_{n+1} \} \, 
\eeq
restricted to some partition ${\cal R}_i$. At this point, the dual contributions can be understood as \emph{local counterterms} for the real corrections, whilst the development of the transformations ${\cal T}_i$ is guided by the structure of the partition ${\cal R}_i$. This partition is based on the FKS or CS strategy, i.e. splitting the phase-space into disjoint regions containing at most one infrared singularity. Then, through a proper study of the cut structure, a connection among dual integrals and partitions is established, in such a way that a mapping ${\cal T}_i$ exists and leads to local cancellations of IR~singularities.

%%%%%%%%%%%%%%%%%%%%%%%%%%%%%%%%%%%%%%%%%%%%%%%%%%%%%%%%%%%\
\subsection{Multi-loop local UV~renormalisation}
\label{ssec:FDUrenormalisation}
%%%%%%%%%%%%%%%%%%%%%%%%%%%%%%%%%%%%%%%%%%%%%%%%%%%%%%%%%%%

The successful identification and cancellation of IR~singularities lead to cross sections with only 
UV~singularities. The standard procedure to remove those divergences requires renormalisation of the field wave-functions and couplings, this is what we aim to reach through the construction of local UV counterterms. Since all these elements have to cast only UV~divergences in the \LTD~framework, it is important to study carefully the analytic properties of the UV~integrands that will be added to the real radiation. At one-loop, it has been considered the regime of massless and massive particles propagating in the loop~\cite{Sborlini:2016gbr,Sborlini:2016hat,Hernandez-Pinto:2015ysa}. 
The transition between massless and massive renormalisation constants has found to be smooth and singularities are well understood in this framework. Let us point out a crucial difference between the standard renormalisation constant in \DREG~and \LTD. 

Wave-function renormalisation constants emerge from the computation of self-energy diagrams. In particular, massless bubble diagrams in \DREG~do not present any problem since the 
IR~and UV~divergences are considered as equal, therefore, the full integral vanishes. 
On the contrary, the same integral in the \FDU/\LTD~framework will contribute to the IR~and UV~regions because they are treated separately, therefore, 
the integral cannot be removed even if the full integral is actually zero. 
We stress that integrands in the \FDU/\LTD~are split into the IR~and UV~domains, and they have to be keep in this way in order to render the full integrands free of singularities in the \FDU~formalism. 

\par\bigskip Let us review the  basic ideas of one-loop renormalisation constant; these ideas are implemented and improved for the two-loop case and beyond. The massive wave-function renormalisation constant, in the Feynman gauge, at one-loop can be obtained from,
\begin{align}
\nn \Delta Z_2(p_1) = -\g^2 \, \CF \,
\int_{\ell} G_F(q_1) \, G_F(q_3) \,
\left[(d-2)\frac{q_1 \cdot p_2}{p_1 \cdot p_2} \right.
+4\, M^2 \left. \left(1- \frac{q_1 \cdot p_2}{p_1 \cdot p_2}\right)
  G_F(q_3)\right]\,, \nn \\
\label{eq:DeltaZ2expression}
\end{align}
which represent the unintegrated form. This integral is obtained by the standard Feynman rules procedure. It is important to remark that the limit of massless case is straightforward achieved from \Eq{eq:DeltaZ2expression}, so this expression is the most general of $\Delta Z_2(p_1; M)$. Since $\Delta Z_2(p_1; M)$ contains singularities associated to the UV domain, it is important to find the UV component of the \Eq{eq:DeltaZ2expression}, $\Delta Z_2^{\rm UV}$, and subtract it in order to find the UV-free wave-function renormalisation constant, $\Delta Z_2^{\rm IR}$. The UV part is extracted by performing an expansion of the integrand around the UV propagator $G_F(q_{\rm UV}) = (q^2_{\rm UV}-\mu_{\rm UV}^2 + \imath 0)^{-1}$. In particular, for \Eq{eq:DeltaZ2expression}, it is found
\begin{align}
\nn\Delta Z_2^{\text{UV}}(p_1) &= (2-d)\, \g^2 \, \CF \,
\int_{\ell} \big[G_F(q_\text{UV})\big]^2 \,
  \left(1+\frac{q_\text{UV} \cdot p_2}{p_1 \cdot p_2}\right) 
%\times
\Big[1\!-\!G_F(q_\text{UV})(2\, q_\text{UV} \cdot p_1 + \mu^2_\text{UV})\Big] \, . \\
\label{eq:ParteUV}
\end{align}
Finally, $\Delta Z_2^{\rm IR}$ is given by,
\begin{align}
\Delta Z_2^{\text{IR}} = \Delta Z_2 - \Delta Z_2^{\text{UV}}\, ,
\end{align}
which contains only IR~singularities and they are needed to cancel the remaining singularities at the cross section level. 
%There are some technicalities that are addressed in the references, but in here we take these ideas for the two-loop and multi-loop scenarios.

We focus now on subtraction of UV~singularities for two-loop amplitudes. In general, for the two-loop case, the integrand involved is cumbersome, therefore, we will use this document to emphasise the procedure to renormalise locally the UV behaviour of any two-loop amplitude. The procedure is valid if the integrand is free of infrared singularities. In this sense, it is mandatory to subtract first all IR divergences and then apply the following algorithm.
This algorithm has been extensively discussed in~\cite{Driencourt-Mangin:2019yhu,Driencourt-Mangin:2017gop,Driencourt-Mangin:2019aix}, 
with applications of one- and two-loop scattering amplitudes. 

Let us now consider a generic two-loop scattering amplitude free of IR~singularities,
\begin{align}
    \mathcal{A}^{(2)} = \int_{\ell_1}\int_{\ell_2} \mathcal{I}(\ell_1, \ell_2) \, ,
\end{align}
where the integrand is a function of the integration variables $\ell_1$ and $\ell_2$. UV divergences shall appear when the limit of $|\vec{\ell}_1|$ and $|\vec{\ell}_2|$ go to infinity. In the two-loop case, there are three UV limits to be considered. The limit when $|\vec{\ell}_1|$ goes to infinity and $|\vec{\ell}_2|$ remains fixed, the other way around, and when $|\vec{\ell}_1|$ and $|\vec{\ell}_2|$ go to infinity simultaneously. Based on the ideas developed at one-loop, the UV~divergences can be extracted from the integrand by making the replacement,\footnotew{
An interesting though not explored strategy is interplaying the Laurent expansion
in the UV region with the rewriting of Feynman propagators, discussed in 
Section~\ref{sec:ireg}.
This treatment of the UV might elucidate, for instance, the way 
how $|\vec{\ell}_1|$ and $|\vec{\ell}_2|$ will behave in
the UV limits. 
}
\begin{align}
    \mathcal{S}_{j,\rm UV} : \{ \ell_j^2\vert \ell_j \cdot k_i\} \to \{\lambda^2 q_{j, \rm UV}^2 + (1-\lambda^2)\mu_{\rm UV}^2 \vert \lambda \,  q_{j,\rm UV}\cdot k_i \} \, ,
\end{align}
for a given loop momentum $\ell_j$ and expanding the expression up to logarithmic order around the UV propagator. This construction is represented by the $L_{\lambda}$ operator. It is worth mentioning that the result shall generate a finite part after integration that has to be fixed to reproduce the correct value of the integral. Therefore, the first counterterms will be obtained by
\begin{align}
    \mathcal{A}_{j,\rm UV}^{(2)} = L_{\lambda} \left( \mathcal{A}^{(2)}\Big|_{\mathcal{S}_{j,\rm UV}}  \right) -d_{j,\rm UV} \, \mu_{\rm UV}^2 \int_{\ell_j} (G_F(q_{j,\rm UV}))^3 \, ,
\end{align}
where $d_{j,\rm UV}$ is the fixing parameter which makes the finite part of integral to be zero in the $\overline{\rm MS}$ scheme.

Now, after the complete subtraction of these counterterms, the remaining divergences shall occur when both $|\vec{\ell}_1|$ and $|\vec{\ell}_1|$ approach to infinity simultaneously. In this case, the following replacement is implemented,
\begin{align}
    \mathcal{S}_{\rm UV^2} : 
    &\{ \ell_j^2\vert \ell_j \cdot \ell_k\vert \ell_j \cdot k_i\} \to
    \nn \\
    &\{\lambda^2 q_{j, \rm UV}^2 + (1-\lambda^2)\mu_{\rm UV}^2
    \vert \lambda^2 q_{j,\rm UV} \cdot q_{k, \rm UV} +(1-\lambda^2) \, \mu_{\rm UV}^2/2
    \vert \lambda \,  q_{j,\rm UV}\cdot k_i \} \,
\end{align}
on the subtracted integrand. Then, the application of the $L_{\lambda}$ operation has to be made and the fixing parameter is again needed in order to build properly the counterterm, $\mathcal{A}^{(2)}_{\rm UV^2}$. Explicitly, 
\begin{align}
    \mathcal{A}_{\rm UV^2}^{(2)} = L_{\lambda} \left( \left(
    \mathcal{A}^{(2)} -\sum_{j=1,2}\mathcal{A}^{(2)}_{j,\rm UV} \right) \Bigg|_{\mathcal{S_{\rm UV^2}}}
    \right) -d_{\rm UV^2} \, \mu_{\rm UV}^4 \int_{\ell_1} \int_{\ell_2} (G_F(q_{1,\rm UV}))^3(G_F(q_{12,\rm UV}))^3 \, ,
\end{align}
where $d_{\rm UV^2}$ is the fixing parameter of the double limit.

Finally, the original amplitude can be renormalised by the subtraction of all UV counterterms, 
such that\footnotew{
We remark that~\eqref{sec:fduUVfinite}, differently from the approach of Section~\ref{sec:fdhdred}
is locally carried out. 
Namely, all singularities, IR and UV, are canceled out 
at integrand level. This allows for an evaluation of the integrals in four space-time dimensions
as carried out in~\cite{Driencourt-Mangin:2019aix,Driencourt-Mangin:2019yhu}.
}
\begin{align}
\mathcal{A}_{\rm R}^{(2)} = \mathcal{A}^{(2)} - \mathcal{A}_{1,\rm UV}^{(2)}- \mathcal{A}_{2,\rm UV}^{(2)} - \mathcal{A}_{\rm UV^2}^{(2)} \, ,
\label{sec:fduUVfinite}
\end{align}
is free of IR and UV singularities.

Before closing this discussion, let us deepen into the multi-loop case. In this scenario, multiple ultraviolet poles will appear, since all loops have the possibility to tend to infinity at different {\it speed}. However, if the amplitude is free of IR~singularities, the algorithm presented along this section is still valid. For an $L$-loop integral, there are $2^L-1$ UV~counterterms at most with the same number of fixing parameters. Therefore, after the proper knowledge of all UV~counterterms, the renormalisation of the original $L$-loop amplitude is achieved and the four-dimensional representation of the integrand can be obtained.

\subsection{$\gamma^{*}\to\,q\bar q$ at NNLO}
\label{ssec:FDUExample}

Let us first consider the kinematics of decay of $\gamma^{*}\to q\bar{q}$, in which,
to keep a simple structure at integrand level, we work with massless particles in the loop. 
We remark that this choice does not generate difficulties in the evaluation of 
integrals, within the \LTD\ framework. The latter pattern is because of the way how
integrands are expressed, which are inherit of the masses. 
Equivalently, their dependence is stored in the fixed energy components, 
$q_{i,0}^{\left(+\right)}$. 
Hence, here and in the following processes, $k=p_{1}+p_{2}+\hdots+p_{n}$,
with $p_{i}^{2}=0,k^{2}=s$ and $n\leq4$.
\begin{figure}[t]
\centering
\includegraphics[scale=0.75]{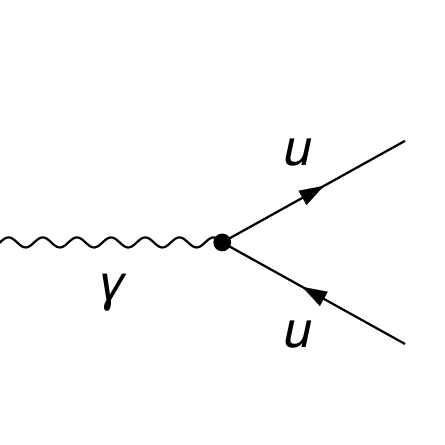}
\includegraphics[scale=0.75]{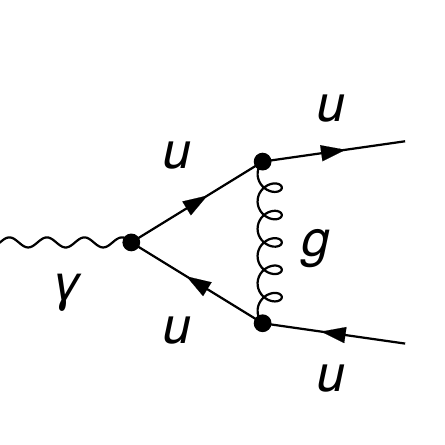}
\includegraphics[scale=0.75]{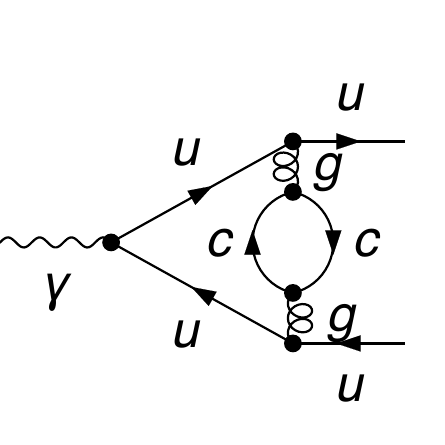}
\caption{Tree, one- and $N_F$ two-loop Feynman diagrams for $\gamma^*\to q\bar{q}$.}
\label{fig:fduqqbar}
\end{figure}
With this notation in mind, the squared tree-level amplitude can be
cast as, 
\begin{align}
\omega_{0}=\overline{\left|\mathcal{A}_{q\bar{q}}^{\left(0\right)}\right|^{2}} & =\left\langle\mathcal{A}_{q\bar{q}}^{\left(0\right)}\Big|\mathcal{A}_{q\bar{q}}^{\left(0\right)}\right\rangle=\frac{2}{3}e^{2}Q_{q}^{2}N_{c}\left(d-2\right)s\,,
\end{align}
with $Q_{q}=\frac{1}{3},-\frac{2}{3}$ and $N_{c}$ the electric charge
and the colour number of the quarks, respectively. 
%(Unlike in the
%former review, the flux factor $1/\left(2s\right)$ has not been included). 

\subsubsection{Virtual contributions}
The contribution at one-loop, after only performing Dirac algebra and 
expressing scalar product in terms of denominators, 
becomes, 
\begin{align}
\left\langle \mathcal{A}_{q\bar{q}}^{\left(0\right)}\Big|\mathcal{A}_{q\bar{q}}^{\left(1\right)}\right\rangle = & \imath\,C_{F}g_{s}^{2}\omega_{0}\Bigg[2sI_{111}^{\left(1\right)}-(d-8)I_{101}^{\left(1\right)}+\frac{2}{s}I_{1-11}^{\left(1\right)}\nonumber \\
 & +\frac{2}{s}\left(I_{010}^{\left(1\right)}-I_{100}^{\left(1\right)}-I_{001}^{\left(1\right)}\right)-2\left(I_{110}^{\left(1\right)}+I_{011}^{\left(1\right)}\right)\Bigg]\,,\label{eq:1LPV}
\end{align}
with, 
\begin{align}
I_{\alpha_{1}\alpha_{2}\alpha_{3}}^{\left(1\right)} & =\int_{\ell_{1}}\prod_{i=1}^{3}G_{F}\left(q_{i}^{\alpha_{i}}\right)\,,
 & q_{1}=\ell_{1}\,,\qquad q_{2}=\ell_{1}-p_{1}\,,\qquad q_{3}=\ell_{1}-p_{12}\,.
\end{align}
As mentioned in the former sections, 
we would like to emphasise that within \LTD~and, therefore, \FDU, scaleless 
integrals are not set directly to zero as conventionally carried out in \DREG. 
The main reason to do this is to achieve a complete cancellation
of singularities at integrand level by keeping as much as possible control on the 
local structure of the integrands. 
Thus, if we were in \DREG, one finds
that the second line in Eq.~(\ref{eq:1LPV}) vanishes and,
\begin{align}
I_{1-11}^{\left(1\right)} & =-\frac{s}{2}I_{101}\,.
\end{align}
These relations amount to 
%(Fix $\imath$ properly), 
\begin{align}
\left\langle \mathcal{A}_{q\bar{q}}^{\left(0\right)}\Big|\mathcal{A}_{q\bar{q}}^{\left(1\right)}\right\rangle = & \imath\,C_{F}g_{s}^{2}\omega_{0}\left[2sI_{111}^{\left(1\right)}-(d-7)I_{101}^{\left(1\right)}\right]\,,
\end{align}
where it is straightforward to identify the IR and UV singularities,
which come from $I_{111}^{\left(1\right)}$ and $I_{101}^{\left(1\right)}$,
respectively. This is indeed what is traditionally carried out
by means of the tensor reduction. 

In \LTD, we keep scaleless integrals to perform a local
UV renormalisation as well as IR subtraction from real corrections,
following the lines of the \FDU~scheme.
Thus, applying \LTD~to Eq.~(\ref{eq:1LPV}), 
\begin{align}
\left\langle \mathcal{A}_{q\bar{q}}^{\left(0\right)}\Big|\mathcal{A}_{q\bar{q}}^{\left(1\right)}\right\rangle =\imath\,C_{F}g_{S}^{2}\omega_{0}\int_{\boldsymbol{\ell}_{1}} & \bigg[-\frac{\mathcal{I}_{101}^{d}}{2s}\left(-4\left(q_{1,0}^{\left(+\right)}\right)^{2}+4\left(q_{2,0}^{\left(+\right)}\right)^{2}+(2d-15)s\right)\nonumber \\
 & +\frac{2}{s}\left(\mathcal{I}_{010}^{d}-\mathcal{I}_{100}^{d}\right)+2s\mathcal{I}_{111}^{d}-4\mathcal{I}_{110}^{d}\bigg]\,,\label{eq:AmpD1L}
\end{align}
with
\begin{subequations}
\begin{align}
&\mathcal{I}_{100}^{d}=\frac{1}{2q_{1,0}^{\left(+\right)}}\,,\qquad\mathcal{I}_{010}^{d}=\frac{1}{2q_{2,0}^{\left(+\right)}}\,,\\
&\mathcal{I}_{101}^{d}=-\frac{1}{4\left(q_{1,0}^{\left(+\right)}\right)^{2}}\left(\frac{1}{\lambda_{3}^{+}}+\frac{1}{\lambda_{3}^{-}}\right)\,,\\
&\mathcal{I}_{110}^{d}=-\frac{1}{4q_{1,0}^{\left(+\right)}q_{2,0}^{\left(+\right)}}\left(\frac{1}{\lambda_{1}^{+}}+\frac{1}{\lambda_{1}^{-}}\right)\,,\\
&\mathcal{I}_{111}^{d}=\frac{1}{4\left(q_{1,0}^{\left(+\right)}\right)^{2}q_{2,0}^{\left(+\right)}}\left(\frac{1}{\lambda_{1}^{+}\lambda_{1}^{-}}+\frac{1}{\lambda_{1}^{+}\lambda_{2}^{-}}+\frac{1}{\lambda_{1}^{-}\lambda_{2}^{+}}\right)\,,
\end{align}
\label{eq:DInt1L}
where, 
\begin{align}
 & \lambda_{1}^{\pm}=q_{1,0}^{\left(+\right)}+q_{2,0}^{\left(+\right)}\pm\frac{\sqrt{s}}{2}\,, &  & \lambda_{2}^{\pm}=2q_{1,0}^{\left(+\right)}\mp\sqrt{s}\,,
\end{align}
\end{subequations}
We remark that the integrands of Eq.~\eqref{eq:AmpD1L} are computed, without
the loss of generality, in the center-of-mass frame, 
allowing us to have $q_{3,0}^{\left(+\right)}=q_{1,0}^{\left(+\right)}$.
Additionally, the integrands~\eqref{eq:DInt1L} are expressed in the 
multi-loop \LTD~representation, displaying structure depending
only on physical singularities. 
A noteworthy comment on the structure of~\eqref{eq:DInt1L} is in order. 
The structure of these integrands can easily be related to one-, two- and three-point functions,
which have been explicitly computed, independently on the number of loops, 
in~\cite{Aguilera-Verdugo:2020kzc}. 
Although there a simple recipe for the calculation of dual integrals through \LTD, 
it as possible to profit of the explicit causal structure of multi-loop topologies.
This is indeed what is carried out in the two- and three-point integrands, 
where their structure correspond to the Maximal-Loop and Next-to-Maximal-Loop topologies, 
respectively. 
A detailed discussion of the structure and the features of these topologies 
is presented in~\cite{Aguilera-Verdugo:2019kbz,Aguilera-Verdugo:2020kzc,Aguilera-Verdugo:2020nrp}. 
Interestingly from Eq.~\eqref{eq:DInt1L}, 
the treatment of physical thresholds is straightforward because of the 
structure the latter hold. In this configuration, in fact, it is possible to obtained 
up to two ``entangled causal'' thresholds that are observed from 
the structure of $\lambda_{i}^{\pm}$. 

\bigskip Hence, by following the idea presented in the one-loop case,
we generate the two-loop contribution, in which we elaborate, for
the sake of simplicity, on the two-loop integral, 
\begin{align}
I_{\alpha_{1}\cdots\alpha_{7}}^{\left(2\right)} & =\int_{\ell_{1},\ell_{2}}\prod_{i=1}^{7}G_{F}\left(q_{i}^{\alpha_{i}}\right)\,,
\end{align}
with 
\begin{align}
 & q_{1}=\ell_{1}\,, &  & q_{5}=\ell_{1}-p_{12}\,,\nonumber \\
 & q_{2}=\ell_{2}\,, &  & q_{6}=\ell_{2}-p_{12}\,,\nonumber \\
 & q_{3}=\ell_{1}-p_{1}\,, &  & q_{7}=\ell_{2}-p_{1}\,.\nonumber \\
 & q_{4}=\ell_{1}+\ell_{2}-p_{1}\,,
\end{align}
Let us remark that the integrand obtained from the Feynman diagram
depicted in Figure~{\color{blue}\ref{fig:fduqqbar}c} one only has five propagators, the additional
ones, $q_{6}$ and $q_{7}$, are needed to express all scalar products
in terms of denominators. This is done to simplify the structure of
the integrand, but it is not mandatory and this step can be avoided.
In fact, the explicit dependence on the energy component of the loop
momenta can always pull out. 

Hence, with the above considerations, the two-loop contribution turns
out to be, 
\begin{align}
\left\langle \mathcal{A}_{q\bar{q}}^{\left(0\right)}\Big|\mathcal{A}_{q\bar{q}}^{\left(2\right)}\right\rangle  & =C_{F}N_{F}g_{S}^{4}\omega_{0}\left[\text{some two-loop integrals}\right]\,.
\end{align}
These two-loop integrals can be further reduced by means of the integration-by-parts
identities (IBPs). Moreover, in order not to alter the local structure
of the integrands, we do not make use of the zero sector symmetries
and loop momentum redefinition. This is done to carefully combine
and thus match virtual with real corrections.

\subsubsection{Real contributions}

In this section, we list the tree-level amplitudes
that are needed to perform a cancellation of the 
IR~singularities, $\gamma^{*}\to q\bar{q}g$ and $\gamma^{*}\to q\bar{q}q'\bar{q}'$.

\paragraph{$\boldsymbol{\gamma^{*}\to q\bar{q}g}$}

\begin{figure}[t]
\centering
\includegraphics[scale=0.75]{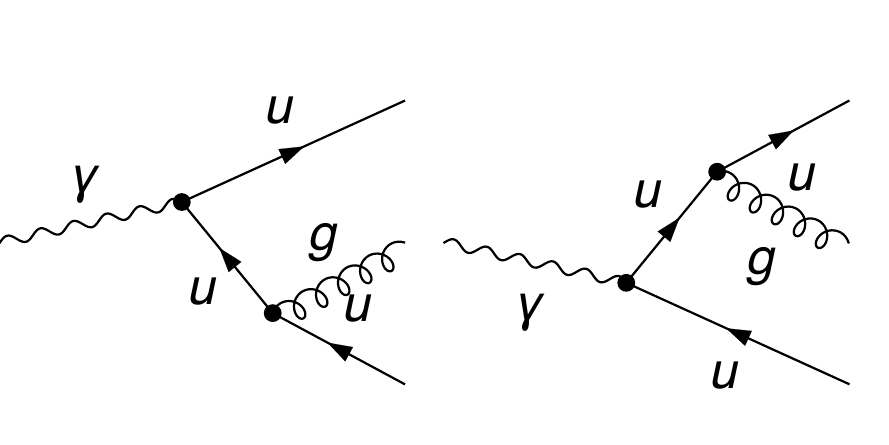}\\
\includegraphics[scale=0.75]{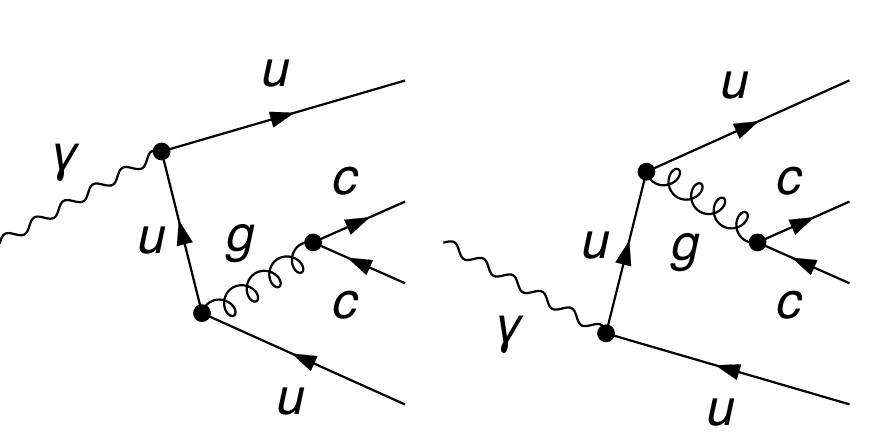}
\caption{Tree level Feynman diagrams for $\gamma^*\to q\bar{q}g$ 
and $\gamma^*\to q\bar{q}q'\bar{q}'$.}
\label{fig:fdureal}
\end{figure}

\begin{align}
\left\langle \mathcal{A}_{q\bar{q}g}^{\left(0\right)}\Big|\mathcal{A}_{q\bar{q}g}^{\left(0\right)}\right\rangle = & \frac{2(d-2)e^{2}N_{c}C_{F}Q_{f}^{2}g_{S}^{2}}{s_{12}s_{23}}\left((d-2)\left(s_{12}+s_{23}\right)^{2}+4\left(s_{13}s_{123}-s_{23}s_{12}\right)\right)\,,
\label{eq:real1L}
\end{align}
with $s_{123}=s_{12}+s_{23}+s_{13}$. 

\paragraph{$\boldsymbol{\gamma^{*}\to q\bar{q}q'\bar{q}'}$}

\begin{align}
\left\langle \mathcal{A}_{q\bar{q}q'\bar{q}'}^{\left(0\right)}\Big|\mathcal{A}_{q\bar{q}q'\bar{q}'}^{\left(0\right)}\right\rangle  & =\frac{4e^{2}N_{c}C_{F}Q_{f}^{2}g_{S}^{4}}{s_{23}^{2}s_{123}^{2}s_{234}^{2}}\Big[4(d-2)Q^{2}s_{23}s_{123}s_{234}\nonumber \\
 & -s_{1234}\Big((d-2)s_{123}^{2}\left((d-2)s_{23}^{2}+4s_{34}s_{23}+4\left(s_{34}-s_{234}\right){}^{2}\right)+2s_{234}s_{123}\nonumber \\
 & +\left(s_{23}\left(((d-10)d+20)s_{23}+2(d-2)s_{34}\right)+2(d-2)s_{12}\left(s_{23}+2s_{34}-2s_{234}\right)\right)\nonumber \\
 & +(d-2)s_{234}^{2}\left((d-2)s_{23}^{2}+4s_{12}^{2}+4s_{23}s_{12}\right)\Big)\nonumber \\
 & +s_{23}s_{123}s_{234}\Big(4(d-2)s_{12}^{2}+4s_{12}\left(-(d-2)s_{123}+(d-4)\left(s_{234}-2s_{34}\right)+2s_{23}\right)\nonumber \\
 & +4(d-2)s_{34}^{2}+(d-2)^{2}s_{123}^{2}+4s_{34}\left(2s_{23}-(d-2)s_{234}\right)\nonumber \\
 & +(d-2)\left((d-2)s_{234}^{2}+4s_{23}^{2}-4s_{234}s_{23}\right)\nonumber \\
 & +2s_{123}\left((d-4)\left((d-4)s_{234}+2s_{34}\right)-2(d-2)s_{23}\right)\Big)\Big]\,.
 \label{eq:real2L}
\end{align}
In this equation, we have not performed any additional collection of
terms. \\

\par It is remarkable the simplicity of the full squared amplitudes displayed
in ~\eqref{eq:real1L} and~\eqref{eq:real2L}. However, to keep track of
the divergencies, within the \LTD/\FDU~framework, one can consider individual
Feynman diagrams and perform the match between real and virtual corrections.
In fact, by keeping the ordering of the diagrams depicted in Figure~\ref{fig:fdureal}, 
one notices that, when squaring the amplitude, the interference
terms account for contributions coming from the virtual diagrams.
While the remaining diagrams account for the 
wave function corrections. 

\subsubsection{Mappings}
\label{sssec:Mappings}
As stated in the former sections, one of the distinctive features of the \FDU\ approach is the real-virtual integrand-level combination through kinematical mappings. At NNLO, these mappings should relate the one- and two-loop amplitudes with the double-real emission terms. A similar strategy to the phase-space slicing method must be applied, but now there will be more singular regions, which might also overlap. In this way, a generic NNLO mapping could be a complicated transformation with highly non-trivial dependencies. 

However, when computing the NNLO QCD corrections proportional to $N_f$ for the process $\gamma^* \to q \bar q$, a huge simplification takes place: only the double-virtual (i.e. two-loop) and the double-real emission contributes. Thus, we will need a transformation to generate a $1 \to 4$ physical configuration starting from double-cuts, which are described by a physical $1 \to 2$ process plus two additional on-shell momenta (i.e. the cut lines, $q_i$ and $q_j$). 
A preliminar proposal for such mapping is the following,
\beq
p'_k = q_i \ \  , \ \ p'_l = q_j \ \ , \ \ p'_1 = p_1 + \alpha_i q_i + \alpha_j q_j  \ \ ,  \ \ p'_1 = p_1 + (1-\alpha_i) q_i + (1-\alpha_j) q_j \, ,  
\label{eq:fdumaps}
\eeq
where momentum conservation is automatically fulfilled and the coefficients $\alpha_i$ and $\alpha_j$ must be adjusted to verify that $(p'_1)^2=p_1^2$ and $(p'_2)^2=p_2^2$. We can appreciate that the cut lines behave as real final state radiation, and we expect this transformation to link the IR singularities present in both contributions to achieve a fully local cancellation in four space-time dimensions\footnote{More details will be provided in a forthcoming article, in which we will carefully explain how to define the mappings in a more general case.}.

\subsection{Discussion}
\label{ssec:FDUOutlook}

The ultimate goal of the \LTD/\FDU\ framework consists in achieving a fully local regularisation of both IR and UV singularities, thus leading to a four-dimensional representation of physical observables. The key observations are:
\begin{itemize}
    \item IR singularities cancel among virtual and real contributions, which is supported by the well-known KLN theorem;
    \item and IR singularities inside the \emph{dualised virtual} terms can be isolated into a compact region of the integration domain.
\end{itemize}
In this way, the singular regions in both contributions can be mapped to the same points, leading to a complete cancellation and skipping the need of introducing additional regularisation techniques (such as \DREG). Alternatively, we can think that real emission is being used as an IR
\emph{local} counterterm for the dual contributions. Of course, renormalisation counterterms must be introduced, as well as potential initial-state radiation (ISR) subtraction terms.

Since the first proof-of-concept of the \FDU\ framework, we have developed several new strategies to tackle the problem of obtaining integrable representations of IR-safe observables in four space-time dimensions. In particular, during last year, we have progressed a lot in understanding the location of IR and threshold singularities in the virtual amplitudes, as well as elucidating a novel dual representation through the application of \emph{nested residues}. The path looks very promising to address some of the current limitations of our approach, namely a fully automated multi-loop local renormalisation and the cancellation of ISR singularities in a universal way. Other strategies, such as $q_T$-subtraction/resummation have shown to be perfectly adapted to attack these problems, although they still lack of locality. Thus, we believe that a conceptual combination of other methodologies might shed light to solve the current limitations to extend the \LTD/\FDU\ framework.

\graphicspath{{IReg/}}
\section{IREG: Implicit regularisation}
\label{sec:ireg}
%\subsection{Brief introduction of Ireg}
%\label{sec:rules}
Envisaging beyond NLO calculations, let us generalise the procedure discussed in \cite{Gnendiger:2017pys} within the non-dimensional \IReg\ framework. 
We summarise the rules applicable to a general $n$-loop Feynman amplitude ${\cal A}_N^{(L)}$ 
with $N$ external legs. Let $k_l$ be the internal (loop) momenta ($l= 1,\hdots, L$) and $p_i$ be the external momenta. 
After performing the usual Diracology and spacetime algebra in the physical dimension and internal symmetry contractions, the UV content of ${\cal A}_N^{(L)}$ can be cast in terms of well-defined basic divergent integrals (BDI's), which are independent on the  physical momenta. In order to define a massless renormalisation scheme, the explicit mass dependence in the BDI's can be removed via regularisation independent identities which gives rise to a renormalisation scale. 
It was shown in \cite{Cherchiglia:2010yd} that BDI, as defined in \IReg, comply with the Bogoliubov-Parasiuk-Hepp-Zimmerman (BPHZ) program \cite{Zimmermann:1969jj, Bogoliubov:1957gp, Hepp:1966eg, Piguet:1995er, Epstein:1973gw}, which  is a consistent renormalisation program applicable to arbitrary loop order in perturbation theory. Based on the topology of a Feynman graph, the subtraction of UV divergences is organised by the Zimmermann’s forest formula in a regularisation independent way. The forest formula can be cast into a counterterm language by means of Bogoliubov’s recursion formula, respecting locality, Lorentz invariance, unitarity and causality. Thus, in \IReg, after subtracting, using the counterterms of lower $(n-1)^{th}$, the $n^{th}$-order counterterms can themselves be cast as BDI, without explicit evaluation.

Clearly care must be exercised as the symmetry content of the underlying model increases because finite regularisation dependent terms can lead to spurious symmetry breakings. In order to evaluate finite Green's functions in a symmetry preserving fashion, the BPHZ program allied to quantum action principles can be used for an all-order proof of renormalisability of gauge field theories. By adopting a gauge invariant scheme a general proof can be constructed in a minimal subtraction scheme and then generalised to arbitrary gauge invariant schemes \cite{tHooft:1972tcz, Breitenlohner:1977hr}. A proof
for all order abelian gauge invariance of \IReg\ can be found in \cite{Ferreira:2011cv}.

\subsection{IREG procedure}
The steps that accomplish the above mentioned issues are as follow.

\begin{enumerate}
	\item Perform the internal symmetry group and the usual Dirac algebra in the physical dimension avoiding symmetric integration in divergent amplitudes as such an operation is ambiguous \cite{PerezVictoria:2001ej}.  The anticommutation $\{\gamma_5,\gamma_\mu\}=0$ inside divergent amplitudes must not be used {\it even} in the physical dimension as they lead to spurious terms as well  \cite{Bruque:2018bmy,Viglioni:2016nqc,Porto:2017asd}. 
	
	\item Starting at one loop remove external momenta dependence from the divergent part of the  amplitude by  applying the identity\footnotew{
We would like to remark that the method of pulling out the UV behaviour of the amplitude can be traced back to the original papers regarding the BPHZ theorem~\cite{Zimmermann:1969jj, Bogoliubov:1957gp, Hepp:1966eg, Piguet:1995er, Epstein:1973gw}, used in~\cite{Giavarini:1992xz, Misiak:1994zw,Chetyrkin:1997fm} and has recently been reconsidered in~\cite{Lang:2020nnl}.
}
	\begin{align}
	\frac{1}{(k_{l}-p_{i})^2-\mu^2}=\sum_{j=0}^{n_{i}^{(k_{l})}-1}\frac{(-1)^{j}(p_{i}^2-2p_{i} \cdot k_{l})^{j}}{(k_{l}^2-\mu^2)^{j+1}}
	+\frac{(-1)^{n_{i}^{(k_{l})}}(p_{i}^2-2p_{i} \cdot k_{l})^{n_{i}^{(k_{l})}}}{(k_{l}^2-\mu^2)^{n_{i}^{(k_{l})}}
		\left[(k_{l}-p_{i})^2-\mu^2\right]},
	\label{ident}
	\end{align}
	in the propagators, where $n_i$ is chosen so that the internal momentum $k_l$ of the l-th loop renders the integral power counting ultraviolet finite. Logarithmically BDI's appear as\footnotew{
We point out that this procedure of extracting the UV behaviour of multi-loop scattering amplitudes,
directly from the Feynman propagators, can be compared with the procedure described in 
Section~\ref{ssec:FDUrenormalisation}.
In fact, in \IReg, all propagators are rewritten without performing any Laurent expansion in the UV region as opposed to \FDU~\cite{Cherchiglia:2010yd}.
}	
	\be
	I_{log}(\mu^2)\equiv \int_{k} \frac{1}{(k^2-\mu^2)^{2}},\quad \quad
    I_{log}^{\nu_{1} \cdots \nu_{2r}}(\mu^2)\equiv \int_k \frac{k^{\nu_1}\cdots
		k^{\nu_{2r}}}{(k^2-\mu^2)^{r+2}}, 
	\ee
	with the definition $\int_{k} = \int \frac{ d^{4} k}{(2 \pi)^{4}} $. 
	The UV finite part in the limit where $\mu$ approaches zero from above $\mu \to 0$ has logarithmical dependence in the physical momenta which is the characteristic behaviour of the finite part of massless amplitudes.

\item BDI's with Lorentz indices $\nu_{1} \cdots \nu_{2r}$ may be written as  linear combinations of BDI's without Lorentz indices  plus well defined surface terms (ST's), e.g. 
\be
\Upsilon_{0}^{(1)\mu\nu}=\int_k\frac{\partial}{\partial k_{\mu}}\frac{k^{\nu}}{(k^{2}-\mu^{2})^{2}}=4\Bigg[\frac{g_{\mu\nu}}{4}I_{log}(\mu^2)-I_{log}^{\mu\nu}(\mu^2)\Bigg],\label{ST1L}
\ee
 ST's vanish if and only if momentum routing invariance (MRI) holds in the loops of Feynman diagrams. Moreover these requirements automatically deliver gauge invariant amplitudes \cite{Battistel:1998sz,BaetaScarpelli:2000zs,Dias:2008iz,Vieira:2015fra,Cherchiglia:2015vaa} which has been demonstrated for abelian gauge theories to arbitrary loop order \cite{Ferreira:2011cv,Viglioni:2016nqc} and verified for  non-abelian gauge models \cite{Sampaio:2005pc,Fargnoli:2010mf,Cherchiglia:2020iug}. Rephrasing it, unless MRI is verified, a  symmetric integration  leads to a finite definite value for the arbitrary surface term which potentially breaks  (gauge) symmetry. 
 By performing a general routing calculation it can be shown that setting ST's=0 cancels routing dependent terms (which they systematically multiply), see e.g. \cite{Ferreira:2011cv}.
This may explain why dimensional regularisation, where surface terms vanish in $d$ dimensions, ensures MRI.\footnotew{
By promoting the space-time dimensions from four to $d$ and 
taking into account \DREG, $\Upsilon_{0}^{(1)\mu\nu} = 0$, which 
can be understood from integration-by-parts identities~\cite{Chetyrkin:1981qh,Laporta:2001dd}. 
}

\item An arbitrary positive (renormalisation group) mass scale $\lambda$
appears via  regularisation independent identities,
\be
I_{log}(\mu^2) = I_{log}(\lambda^2) + \frac{i}{(4 \pi)^2} \ln \frac{\lambda^2}{\mu^2},
\label{SR1}
\ee
which enables us to write a BDI as a function of $\lambda^2$ plus logarithmic functions of $\mu^2/\lambda^2$.  The BDI can be absorbed in the renormalisation constants  \cite{Brito:2008zn} and renormalisation functions can be computed using the regularisation independent identity:
\be
\lambda^2\frac{\partial I_{log}(\lambda^2)}{\partial \lambda^2}= -\frac{i}{(4 \pi)^2}.
\ee

\end{enumerate}

At two-loop order a similar program can be  devised, which allows to express the UV divergent content in terms of BDI in one loop momentum only. As an example, consider an UV divergent two-loop massless scalar integral
%\begin{align}
%\mathcal{A}=\int_{k_{1},k_{2}}\frac{1}{k_{1}^{2}(k{1}-k_{2})^{2}k_{2}^{2}(k_{1}-p)^{2}(k_{2}-p)^{2}}
%\end{align}
\begin{align}
\mathcal{A}=\int_{k_{1},k_{2}}G(p_{1},\ldots,p_{L},k_{1},k_{2})H_{1}(p_{1},\ldots,p_{L},k_{1})H_{2}(p_{1},\ldots,p_{L},k_{2}).
\end{align}

Following the algorithm proposed in \cite{Cherchiglia:2010yd}, one identifies the different regimes in which the internal momenta can go to infinity ($k_{1}\rightarrow \infty$, $k_{2}$ fixed; $k_{2}\rightarrow \infty$, $k_{1}$ fixed; $k_{1}\rightarrow \infty$, $k_{2}\rightarrow \infty$); for each case, uses identity \ref{ident} in the internal momenta that goes to infinity regarding all other momenta as external. This procedure allows to automatically identify the UV-counterterms required by Bogoliubov's recursion formula in terms of BDI's. Explicitly,
\begin{align}
\mathcal{A}_{k_{1}\rightarrow\infty}=\int_{k_{2}}\bar{H}_{2}(p_{1},\ldots,p_{L},k_{2})I_{log}(\lambda^2),\nonumber\\
\mathcal{A}_{k_{2}\rightarrow\infty}=\int_{k_{1}}\bar{H}_{1}(p_{1},\ldots,p_{L},k_{1})I_{log}(\lambda^2),\nonumber\\
\mathcal{A}_{k_{1}\rightarrow\infty,k_{2}\rightarrow\infty}=\mathcal{F}(p_{1},\ldots,p_{L})I_{log}(\lambda^2),
\end{align}
where the function $\bar{H}_{1}$ contains terms generated by integrating in $k_{2}$, and similarly to $\bar{H}_{2}$. In this example, the first two terms are going to be canceled by 1-loop counterterms while the last one will contribute to the 2-loop counterterm. Further contributions to the 2-loop counterterm are also automatically identified, which will be of the form 
\begin{align}
\mathcal{\bar{A}}_{k_{1}\rightarrow\infty}=\int_{k_{2}}\bar{H}_{2}(p_{1},\ldots,p_{L},k_{2})\ln\!\left(\!-\frac{k_{1}^{2}-\mu^{2}}{\lambda^{2}}\right),\nonumber\\
\mathcal{\bar{A}}_{k_{2}\rightarrow\infty}=\int_{k_{1}}\bar{H}_{1}(p_{1},\ldots,p_{L},k_{1})\ln\!\left(\!-\frac{k_{2}^{2}-\mu^{2}}{\lambda^{2}}\right),
\end{align}
or integrals in $k_{1}$ ($k_{2}$) with no dependence on the scale $\lambda$. The above integrals give rise to BDI's of two-loop order defined by
\begin{align}
I_{log}^{(2)}(\mu^2)&\equiv \int\limits_{k} \frac{1}{(k^2-\mu^2)^{2}}
\ln{\left(-\frac{k^2-\mu^2}{\lambda^2}\right)},
\end{align}

This approach can be extended to tensorial and/or arbitrary loop order integrals, as sketched by the steps below

\begin{enumerate}

\item At higher loop order the divergent content can be expressed in terms of  BDI in one loop momentum after performing $n-1$ integrations. The order of such integrations is chosen systematically to display the counterterms to be subtracted in compliance with the Bogoliubov's recursion formula \cite{Zimmermann:1969jj, Bogoliubov:1957gp, Hepp:1966eg, Piguet:1995er, Epstein:1973gw,Cherchiglia:2010yd}. The general form of the terms of a Feynman amplitude after $l$ integrations is
\begin{align}
&I^{\nu_{1}\ldots \nu_{m}}\!=\!\!\int\limits_{k_{l}}\!\frac{A^{\nu_{1}\ldots \nu_{m}}(k_{l},q_{i})}{\prod_{i}[(k_{l}-q_{i})^{2}-\mu^{2}]}\ln^{l-1}\!\left(\!-\frac{k_{l}^{2}-\mu^{2}}{\lambda^{2}}\right)\!,
\label{I}
\end{align}
\noindent
where $l=1, \cdots , n$ and $q_{i}$ is an element (or combination of elements) of the set $\{p_{1},\ldots,p_{L},k_{l+1},\ldots,k_{n}\}$. $A^{\nu_{1}\ldots \nu_{m}}(k_{l},q_{i})$ represents all possible combinations of $k_{l}$ and $q_{i}$ compatible with the Lorentz structure.
\item Apply relation (\ref{ident}) in (\ref{I}) by choosing $n_{i}^{(k_{l})}$ such that all divergent integrals are free of $q_{i}$. Therefore, the divergent integrals are cast as a combination of
\begin{align}
I_{log}^{(l)}(\mu^2)&\equiv \int\limits_{k_{l}} \frac{1}{(k_{l}^2-\mu^2)^{2}}
\ln^{l-1}{\left(-\frac{k_{l}^2-\mu^2}{\lambda^2}\right)},\quad
\label{Ilogilog}\\
I_{log}^{(l)\nu_{1} \cdots \nu_{2r}}(\mu^2)&\equiv \int\limits_{k_{l}} \frac{k_{l}^{\nu_1}\cdots
	k_{l}^{\nu_{2r}}}{(k_{l}^2-\mu^2)^{r+1}}
\ln^{l-1}{\left(-\frac{k_{l}^2-\mu^2}{\lambda^2}\right)},
\label{IlogLorentz}
\end{align}
The surface terms derived from higher loop BDI's are obtained through the identity
\begin{align}
\Upsilon_{2i}^{(l)\nu_{1}\cdots\nu_{2j}}\equiv\int_k\frac{\partial}{\partial k_{\nu_{1}}}\frac{k^{\nu_{2}}\cdots k^{\nu_{2j}}}{(k^{2}-\mu^{2})^{1+j-i}}\ln^{l-1}\Bigg[-\frac{(k^{2}-\mu^{2})}{\lambda^{2}}\Bigg].
\label{tsdef}
\end{align}
For instance,
\begin{align}
&I_{log}^{(l)\,\mu
	\nu}(\mu^2)=\sum_{j=1}^{l}\left(\frac{1}{2}\right)^j\!\frac{(l-1)!}{(l-j)!}\!\left\{\frac{g^{\mu \nu}}{2}I_{log}^{(l-j+1)}(\mu^2)-\frac{1}{2}\Upsilon_{0}^{(l)\,\mu\nu}\right\}.
\label{identsurface1}
\end{align}
\item A renormalisation group scale is encoded in BDI's. At n$^{th}$-loop order a relation analogous to (\ref{SR1}) is obtained via the regularisation independent identity
\begin{align}
I_{log}^{(l)}(\mu^2)&=I_{log}^{(l)}(\lambda^2)-\frac{b}{l}\ln^{l}\left(\frac{\mu^2}{\lambda^2}\right)- b \sum_{j=1}^{l-1}\frac{(l-1)!}{(l-j)!}\ln^{l-j}\left(\frac{\mu^2}{\lambda^2}\right),
\label{scale}
%\\
%\mbox{where}\quad\lambda^{2}&\neq0,\;\;  b \equiv\frac{i}{(4\pi)^2}.
%\label{bd}
\end{align}
where $\lambda^{2}\neq0,\;\;  b \equiv\frac{i}{(4\pi)^2}$.
\item BDI's can be absorbed in renormalisation constants. A minimal, mass-independent scheme amounts to absorb only $I_{log}^{(l)}(\lambda^2)$. To evaluate RG constants, BDI's need not be explicitly evaluated as their derivatives with respect to the renormalisation scale $\lambda^2$ are also BDI's. For example \cite{Ferreira:2011cv},
\begin{align}
\lambda^2\frac{\partial I_{log}^{(n)}(\lambda^2)}{\partial \lambda^{2}}&=-(n-1)\, I_{log}^{(n-1)}(\lambda^2)- b \,\, \alpha^{(n)}\, ,\nonumber \\
\lambda^2\frac{\partial I_{log}^{(n)\,\mu\nu}(\lambda^2)}{\partial \lambda^{2}}&=-(n-1)I_{log}^{(n-1)\,\mu\nu}(\lambda^2)-\frac{g_{\mu\nu}}{2} \, b \, \beta^{(n)}.
\label{gerder}
\end{align}
where $n \geq 2$, $\alpha^{(n)} = (n-1) !$ and $\beta^{(n)}$ may be obtained from $\alpha^{(n)}$ via relation (\ref{identsurface1}).

\end{enumerate}

\subsection{Disentangling UV and IR divergences: a two-loop example}

\begin{figure}[t]
\centering
\includegraphics[width=5.0 cm]{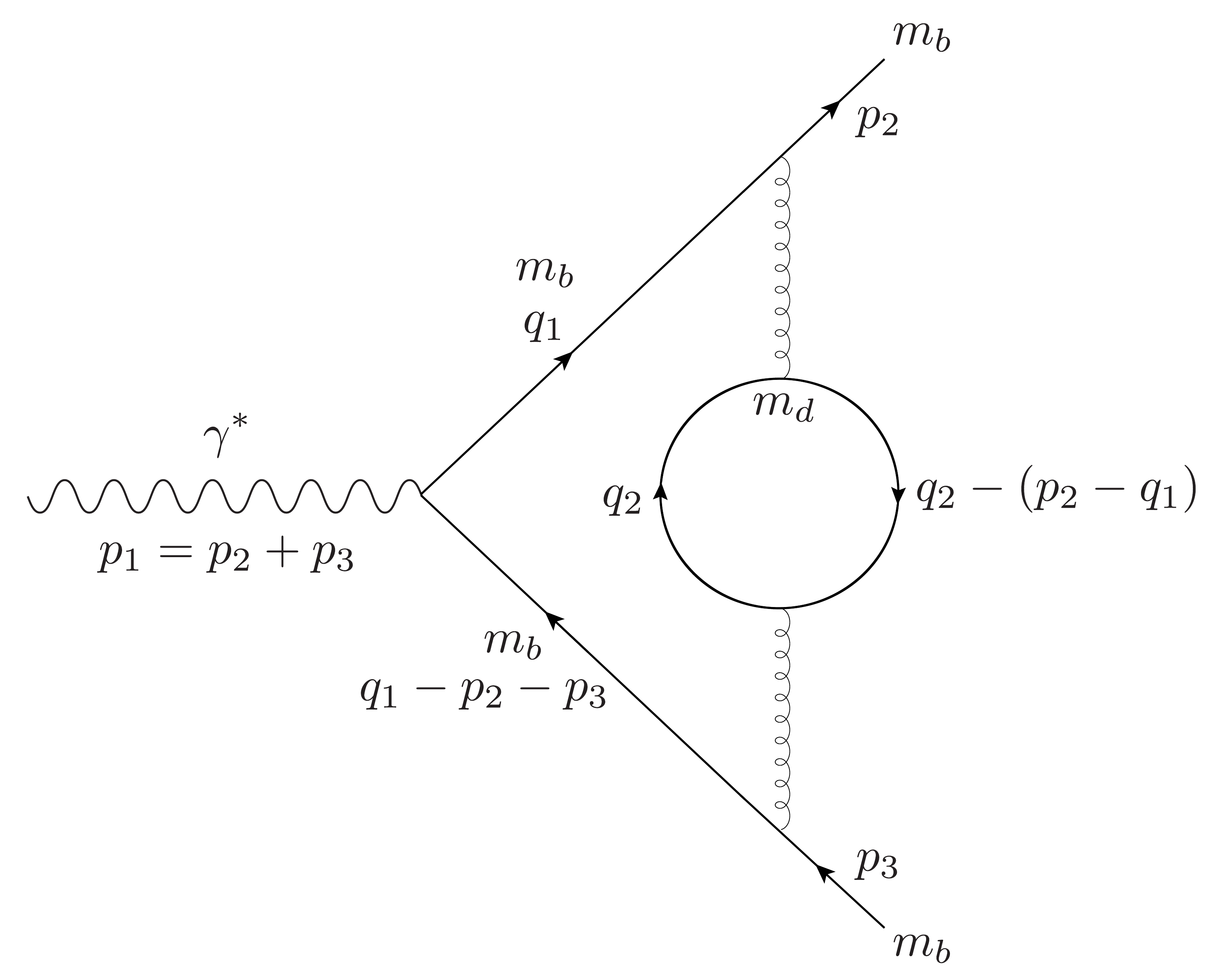}
\caption{$N_F$ two-loop Feynman diagram for $\gamma^*\to b\bar{b}$.}
\label{fig:diagram}
\end{figure} 

We consider the process $\gamma^{*}\rightarrow q\bar{q}$ at two-loop level. In order to illustrate the method, we evaluate the contribution that contains a quark loop of flavour $d$ and external quarks of flavour $b$ as depicted in Figure~\ref{fig:diagram}. The amplitude can be written as
\be
 {\cal{M}}^\mu_{ab}= \int_{q_1,q_2}\frac{C_F \, Q \, e \, g_s^4 \, \delta_{ab}\,\,\times \,\, \bar{u} (p_2,m_b) N^\mu(m_d,m_b,p_2,p_3,q_1,q_2) v (p_3,m_b) }{((q_1+p_2)^2-m_b^2)(q_1^2-\mu^2)^2[(q_1-p_3)^2-m_b^2](q_2^2-m_d^2)[(q_2+q_1)^2-m_d^2]}
\label{amplitude}
\ee
where $\int_q = \int d^4 q/(2 \pi)^4$ and we have shifted the momentum variables such that $q_1 \rightarrow q_1'=-(q_1 -p_2)$ and relabelled $q_1'$ back to $q_1$. $N$ in the numerator stands for 
\bq
N^\mu &=& 4 m_d^2 [-2 \slashed{q_1}(p_2-p_3+q_1)^\mu +  \gamma^\mu (-2p_2\cdot p_3 + 2p_2 \cdot q_1 - 2p_3 \cdot q_1 + q_1^2 ) + m_b (\slashed{q_1} \gamma^\mu + \gamma^\mu \slashed{q_1})]+ \nonumber \\
&+& 4 \gamma^\mu \slashed{q_1} \slashed{q_2} (p_2 \cdot q_1 + 2 p_2 \cdot q_2) - 8 \gamma^\mu (p_2 \cdot q_1)(p_3 \cdot q_2) - 4 \slashed{q_2} \slashed{q_1} \gamma^\mu (p_3 \cdot q_1 + 2 p_3 \cdot q_2) - \nonumber \\
&-& 4 \gamma^\mu (p_2 \cdot q_2) (2 p_3 \cdot q_1 + 4 p_3 \cdot q_2 - q_1^2)-2q_1^2 (\slashed{q_1} \gamma^\mu \slashed{q_2} + \slashed{q_2} \gamma^\mu \slashed{q_1} + 2 \gamma^\mu p_3 \cdot q_2) + 8 \slashed{q_1} (p_2-p_3)^\mu q_1 \cdot q_2 +\nonumber \\
&+& 8 \slashed{q_1} q_1^\mu q_1\cdot q_2 - 8 \gamma^\mu (p_2\cdot q_1) (q_1 \cdot q_2) + 8 \gamma^\mu (p_2 \cdot p_3 + p_3 \cdot q_1) (q_1 \cdot q_2) - 4 \gamma^\mu q_1^2 q_1 \cdot q_2 + 8 \slashed{q_2} \times \nonumber \\
&\times& ((q_1 - q_2)^\mu q_1^2 + 2 q_1^\mu q_1 \cdot q_2) + 8 \slashed{q_1}q_2^2 (p_2 - p_3)^\mu - 8 \gamma^\mu q_2^2 (p_2 \cdot q_1 + p_2 \cdot p_3 + p_3 \cdot q_1)-\nonumber \\
&-& 4 m_b (\slashed{q_1} \gamma^\mu + \gamma^\mu \slashed{q_1}) (q_2^2 + q_1 \cdot q_2), 
\eq
 $Q=-1/3$, $(a, b)$ are  colour indices of the external quarks, $C_{F}$ is the quadratic Casimir for the fundamental representation and $\mu$ is a fictitious mass for the gluon propagator.
 
 The integration over $q_2$ is performed according to the \IReg\ rules, by first applying Eq.~\eqref{ident} to separate its divergent content, which is expressed as an internal $I_{log}(\lambda^2)$, from the finite contribution  encoded as $Z_0(p^2, m_1^2, m_2^2)$ plus possible local terms, both multiplying the $q_1$ integrand, as follows
 \be
 {\cal{M}}^\mu_{ab} = - \frac{2}{9} Q \, e\, C_F\, g_s^4\, \delta_{ab} \times \bar{u} (p_2,m_b)\Bigg[ \underbrace{\int_{q_1} {\cal{I}}^\mu (p_2,p_3,m_b,m_d,q_1,\mu,\lambda)}_{\Sigma^\mu} \Bigg] v (p_3,m_b)
 \label{eq:mu}
 \ee
where
\bq
 &&  {\cal{I}}^\mu = \Big( [ q_1^2 (-4 p_2 \cdot p_3 - 2 p_3 \cdot q_1 + q_1^2) + 2 p_2 \cdot q_1 (2 p_3 \cdot q_1 + q_1^2)] \gamma^\mu + 4 q_1^2[ m_b q_1^\mu - (p_2 - p_3 + q_1)^\mu \slashed{q_1}] \Big)\times \nonumber \\ 
 &&  \Bigg[ \frac{-3\,  I_{log} (\lambda^2) + 3\, b \, Z_0 (q_1^2,m_d^2,\lambda^2)+b}{[(q_1+p_2)^2-m_b^2](q_1^2-\mu^2)^2[(q_1-p_3)^2-m_b^2]}
 + \frac{6\, b\, m_d^2 \, Z_0 (q_1^2, m_d^2, m_d^2)}{[(q_1+p_2)^2-m_b^2](q_1^2-\mu^2)^3[(q_1-p_3)^2-m_b^2]} \Bigg].\notag\\
 \label{eqn:q2}
\eq
Here
\be
Z_0 (p^2, m_1^2, m_2^2) \equiv \int_0^1 \, dx \,\, \ln \Big( \frac{p^2 x (x-1) + m_1^2}{m_2^2}\Big),
\label{eqn:Z0}
\ee
and we have used relation (\ref{SR1}). Notice that $Z_{0}$ is scale $\lambda$  dependent in the first term in the square brackets of \ref{eqn:q2}, and independent on it in the second term in the brackets. 

According to the features of BDI mentioned in the beginning with respect to the Bogoliubov's recursion formula, the term proportional to $I_{log} (\lambda^2)$ in \ref{eqn:q2} corresponds to a subdivergence and is exactly cancelled by the counterterm graph corresponding to the one mass-independent renormalisation of the down-quark loop.

\begin{figure}[h]
\centering
\includegraphics[width=8.0 cm]{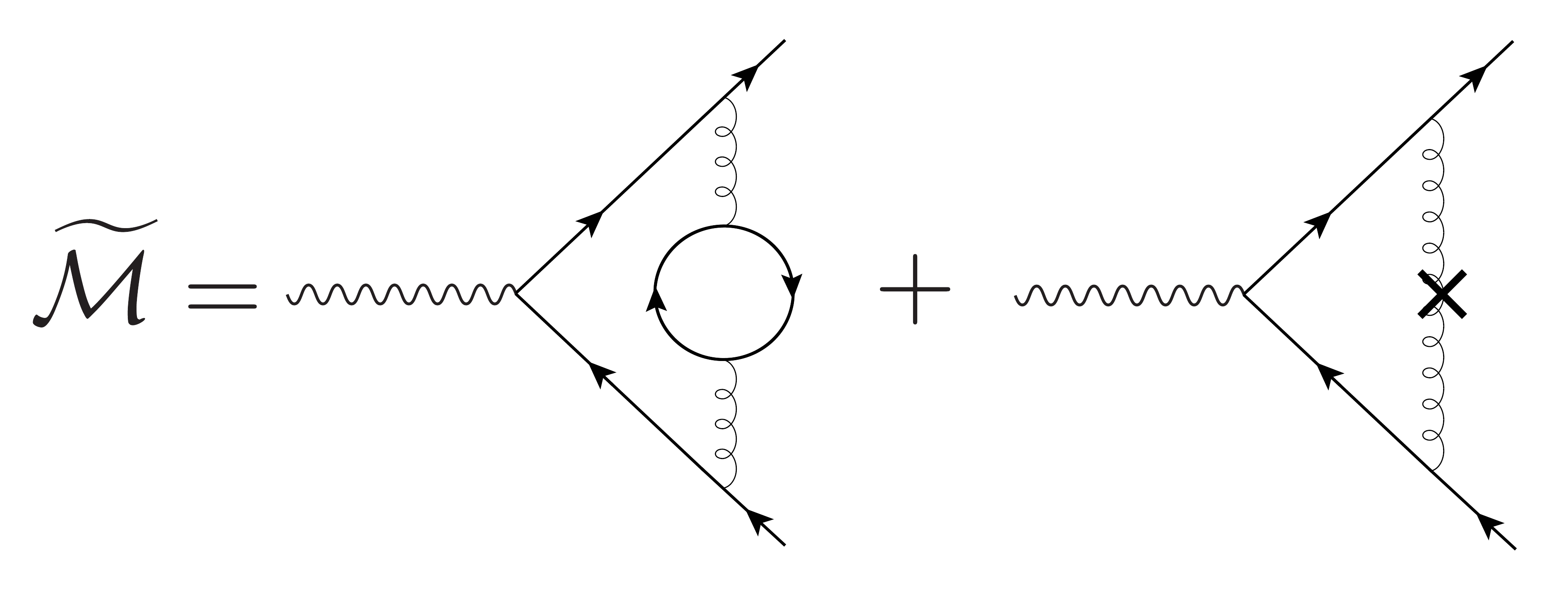}
\caption{Diagrams entering in the UV renormalisation of scattering process $\gamma^*\to b\bar b$.}
\label{fig:diagramCT}
\end{figure}

 Thus $ \widetilde{\cal{M}}^\mu_{ab}$, the amplitude which defines the two-loop order UV divergence after subtracting the subdivergences, is obtained by substituting ${\cal{I}}^\mu$ by
\bq
 && \widetilde{{\cal{I}}}^\mu = \Big( [ q_1^2 (-4 p_2 \cdot p_3 - 2 p_3 \cdot q_1 + q_1^2) + 2 p_2 \cdot q_1 (2 p_3 \cdot q_1 + q_1^2)] \gamma^\mu + 4 q_1^2[ m_b q_1^\mu - (p_2 - p_3 + q_1)^\mu \slashed{q_1}] \Big)\times \nonumber \\ 
 &&  \Bigg[ \frac{ 3\, b \, Z_0 (q_1^2,m_d^2,\lambda^2)+b}{[(q_1+p_2)^2-m_b^2](q_1^2-\mu^2)^2[(q_1-p_3)^2-m_b^2]}
 + \frac{6\, b\, m_d^2 \, Z_0 (q_1^2, m_d^2, m_d^2)}{[(q_1+p_2)^2-m_b^2](q_1^2-\mu^2)^3[(q_1-p_3)^2-m_b^2]} \Bigg],\notag
 \\
 \label{eqn:Nq2}
\eq
to define $\widetilde{{\cal{M}}}^\mu_{ab}$ as depicted in Figure~\ref{fig:diagramCT}.

In the next section we address the two-loop order  specific UV divergences and their removal.

\subsubsection{Virtual Contribution: UV part}

In this particular example, after the removal of the one-loop subdivergence, the divergent part of the amplitude is either UV or IR divergent, there is no overlap of UV and IR divergent contributions. The two-loop UV renormalisation constant can be extracted as a BDI from the UV divergent part of $\widetilde{\Sigma}^\mu$  (remember definition in \ref{eq:mu}):
\bq
 \widetilde{\Sigma}^\mu\Big|_{UV} = \int_{q_1} \Big( q_1^4 \gamma^{\mu} - 4 q_1^2 q_1^\mu \slashed{q_1} \Big) \Bigg[ \frac{ 3\, b \, Z_0 (q_1^2,m_d^2,\lambda^2)+b}{[(q_1+p_2)^2-m_b^2](q_1^2-\mu^2)^2[(q_1-p_3)^2-m_b^2]}\Bigg],
\eq
Using the rules of IReg, we isolate a BDI as a UV bare bone object free of masses by taking the limit where $q_1 \gg m_d$ in $Z_0 (q_1^2,m_d^2,\lambda^2)$ to yield
\bq
 \widetilde{\Sigma}^\mu\Big|_{UV} = \int_{q_1} \Bigg[ \gamma^{\mu} - \frac{4 q_1^\mu \slashed{q_1}}{(q_1^2-\mu^2)}\Bigg]\Bigg[\frac{3\,b \, \ln{\left(-\frac{q_{1}^2-\mu^2}{\lambda^2}\right)}-5b}{[(q_1+p_2)^2-m_b^2][(q_1-p_3)^2-m_b^2]}\Bigg].
\eq
The integrals above are only UV-logarithmic divergent, therefore, the BDI's can be easily extracted as 
\bq
 \widetilde{\Sigma}^\mu\Big|_{UV} = b \gamma^{\mu}\left[3I_{log}^{(2)}(\lambda^2)-5I_{log}(\lambda^2)\right]
 -4b\gamma^{\mu}\left[\frac{3}{4}I_{log}^{(2)}(\lambda^2)+\frac{3}{8}I_{log}(\lambda^2)-\frac{5}{4}I_{log}(\lambda^2)\right],
\eq
where we made use of Eqs.~(\ref{identsurface1}) and (\ref{scale}) while  setting ST's to zero. 
Finally, the two-loop UV counterterm corresponding to this amplitude reads
\bq
{\cal{M}}^\mu_{ab}\Big|_{UV}^{\text{2\,loop\,ct}} =   Q \, e\, C_F\, g_s^4\, \delta_{ab}\,  \bar{u} (p_2,m_b)\gamma^{\mu}v(p_3,m_b)\, \times \frac{b}{3}\,I_{log} (\lambda^2) 
\eq

Upon subtracting it from $\widetilde{{\cal{M}}}^\mu_{ab}$,  defined by \ref{eq:mu} with \ref{eqn:Nq2},  one obtains the renormalised amplitude. The next task is to identify the infrared divergences in the renormalised amplitude.

\subsubsection{Virtual Contribution: IR part}
\begin{table} [h!]
\centering
\begin{tabular}{ | c || c  | }
\hline
Integrals  & IR divergences \\ [0.5 ex]
\hline \hline 
$\int_{q_1} \frac{p_2\cdot p_3}{(q_1^2-\mu^2)[(p_2+q_1)^2-m_b^2][(q_1-p_3)^2-m_b^2]}$ & $\frac{-(s-2 m_b^2)}{4 \sqrt{s S}} {\cal{F}} (s, S, \mu)$  \\ [1.5 ex] \hline
$\int_{q_1} \frac{(p_2\cdot q_1)(p_3 \cdot q_1)}{(q_1^2-\mu^2)^2[(p_2+q_1)^2-m_b^2][(q_1-p_3)^2-m_b^2]}$ & $-\frac{1}{4} \ln \Big( \frac{m_b^2}{\mu^2}\Big)$  \\ [1.5 ex] \hline
$\int_{q_1} \frac{p_2\cdot p_3 Z_0 (q_1^2,m_d^2,\lambda^2)}{(q_1^2-\mu^2)[(p_2+q_1)^2-m_b^2][(q_1-p_3)^2-m_b^2]}$ & $\frac{-(s-2 m_b^2)}{4 \sqrt{s S}} \ln \Big(\frac{m_d^2}{\lambda^2}\Big) {\cal{F}} (s, S, \mu)$  \\ [1.5 ex] \hline
$\int_{q_1} \frac{(p_2\cdot q_1)(p_3 \cdot q_1) Z_0 (q_1^2,m_d^2,\lambda^2)}{(q_1^2-\mu^2)^2[(p_2+q_1)^2-m_b^2][(q_1-p_3)^2-m_b^2]}$ & $-\frac{1}{4} \ln \Big( \frac{m_b^2}{\mu^2}\Big)  \ln \Big(\frac{m_d^2}{\lambda^2}\Big)$  \\ [1.5 ex] \hline
 $\int_{q_1} \frac{ q_1^2 Z_0 (q_1^2,m_d^2,m_d^2)}{(q_1^2-\mu^2)^2[(p_2+q_1)^2-m_b^2][(q_1-p_3)^2-m_b^2]}$& $\frac{-\ln ( m_d^2/m_b^2)}{2 \sqrt{s S}} {\cal{F}} (s, S, \mu)$ \\ [1.5 ex] \hline
$\int_{q_1} \frac{ (p_2 \cdot q_1) Z_0 (q_1^2,m_d^2,m_d^2)}{(q_1^2-\mu^2)^2[(p_2+q_1)^2-m_b^2][(q_1-p_3)^2-m_b^2]}$ & $\frac{\ln ( m_d^2/m_b^2)}{4 m_b^2 s S} \Big[ m_b^2 \sqrt{s S} \, {\cal{F}} (s, S, \mu) + s S \ln \Big( \frac{m_b^2}{\mu^2}\Big)\Big]$ \\ [1.5 ex] \hline
$\int_{q_1} \frac{ (p_3 \cdot q_1) Z_0 (q_1^2,m_d^2,m_d^2)}{(q_1^2-\mu^2)^2[(p_2+q_1)^2-m_b^2][(q_1-p_3)^2-m_b^2]}$ &  $-\frac{\ln ( m_d^2/m_b^2)}{4 m_b^2 s S} \Big[ m_b^2 \sqrt{sS }\, {\cal{F}} (s, S, \mu)+ s S \ln \Big( \frac{m_b^2}{\mu^2}\Big)\Big]$\\ [1.5 ex] \hline
 $\int_{q_1} q_1^\mu \slashed{q_1} Z_0 (q_1^2,m_d^2,m_d^2) \,\,\, \times $ & $\frac{-\ln ( m_d^2/m_b^2)}{4 m_b s S^2} \Big\{ m_b S \sqrt{s S }\,\,\, {\cal{F}} (s, S, \mu) \gamma^{\mu} + $ \\ $\times \frac{1}{(q_1^2-\mu^2)^2[(p_2+q_1)^2-m_b^2][(q_1-p_3)^2-m_b^2]}$  &
 $+ \,\, 2 (p_2^\mu -p_3^\mu) \Big[  m_b^2 \sqrt{s S} \,\, {\cal{F}} (s,S,\mu) + s S \ln \Big( \frac{m_b^2}{\mu^2}\Big)  \Big]\Big\} $
 \\ [1.5 ex] \hline
 $\int_{q_1} \frac{ q_1^\mu  Z_0 (q_1^2,m_d^2,m_d^2)}{(q_1^2-\mu^2)^2[(p_2+q_1)^2-m_b^2][(q_1-p_3)^2-m_b^2]}$& 
 $ \frac{(p_2 - p_3)^\mu   \ln(m_d^2/m_b^2)}{2 m_b^2 s S^2}   \Big[  m_b^2 \sqrt{s S} \,\,\, {\cal{F}} (s,S,\mu) + s S \ln \Big( \frac{m_b^2}{\mu^2}\Big)\Big]$\\ [1.5 ex] \hline
$\int_{q_1} \frac{ (q_1 \cdot p_2)(q_1 \cdot p_3)  Z_0 (q_1^2,m_d^2,m_d^2)}{(q_1^2-\mu^2)^3[(p_2+q_1)^2-m_b^2][(q_1-p_3)^2-m_b^2]}$ & $\frac{\ln (m_d^2/m_b^2) {\cal{F}} (s,S,\mu)}{8 \sqrt{s S}} + \Big( \frac{\ln (m_d^2/m_b^2)}{4 m_b^2} + \frac{1}{24 m_d^2}\Big)\ln \Big( \frac{m_b^2}{\mu^2}\Big)$ \\ [1.5 ex] \hline
\end{tabular}
\caption{
We have denoted $s \equiv (p_2 + p_3)^2 = 2 (p_2 \cdot p_3 + m_b^2)$, $S \equiv s - 4 m_b^2 $, ${\cal{F}} (s, S, \mu) \equiv \ln^2 \Big( \frac{2 \mu^2}{\sqrt{sS}-S}\Big) - \ln^2 \Big( \frac{-2 \mu^2}{\sqrt{sS} + S}\Big) $ and $Z_0 (p^2,m_1^2,m_2^2)$ is defined as in equation (\ref{eqn:Z0}).
}~
\label{table:1}
\end{table}

In Table~\ref{table:1}, we summarise the results which are relevant to isolate the IR divergences of different natures as $\ln^n\mu$ as $\mu \rightarrow 0$ in a non-dimensional regularisation scheme\footnote{Some simplification due to the presence of external spinors was already applied. Also, an identity matrix in Dirac space should be implicitly understood in the second term inside the brackets.}. The IR divergent part is parametrized as $\ln$ of the parameter $\mu^2$. Notice that this is a fictitious mass introduced in propagators to regularise the IR div. We have made use of \Feyncalc~\cite{Shtabovenko:2016sxi, Mertig:1990an,Shtabovenko:2020gxv} and \PackageX~\cite{Patel:2015tea} to compute the integrals. Moreover in order to express the function $Z_0 (k^2, m_1^2, \lambda^2)$ in a propagator like form we have used 
 $Z_0 = \int (\partial Z_0 / \partial m_1^2) d m_1^2$, with
 \be
\frac{\partial  Z_0 (k^2,\tilde{m}^2,\lambda^2)}{\partial \tilde{m}^2} =  \int_{0}^{1} dx \frac{-1}{{k}^2 x (1-x)-\tilde{m}^2}
\label{eq:dZ0}
\ee
Notice that since one is dealing with massive quarks, the integral in the Feynman parameter x of \ref{eqn:Z0} is more involved as opposed to the example of the electron self-energy we performed in section 4.1 of \cite{Gnendiger:2017pys}.

Putting all the results displayed in table \ref{table:1} in the amplitude we finally obtain for its infrared content
\be
{\cal{M}}^\mu_{ab}\Big|_{IR}^{\text{2\,loop}} =   Q \, e\, C_F\, g_s^4\, \delta_{ab}\,  \bar{u} (p_2,m_b) [\,\, (p_2-p_3)^\mu \,\,\, {\cal{A}}+ \gamma^\mu \,\,\, {\cal{B}}\,\,] v(p_3,m_b)\, ,
\label{eq:IR}
\ee
with
\bq
{\cal{A}} &=&  \frac{16 \,b\, m_d^2}{3 \,S} \ln \Big( \frac{m_d^2}{m_b^2}\Big)
\Bigg[ - \frac{m_b}{\sqrt{s\, S}}   {\cal{F}} (s, S, \mu) -
\frac{1}{m_b}   \ln \Big( \frac{m_b^2}{\mu^2}\Big) \Bigg]\nonumber \\
{\cal{B}} &=& \frac{2 \,b}{3} \Bigg[ \ln \Big( \frac{m_b^2}{\mu^2}\Big) \ln \Big( \frac{m_d^2}{\lambda^2}\Big) - \frac{4\, m_d^2}{m_b^2} \ln \Big( \frac{m_d^2}{m_b^2}\Big)\ln \Big( \frac{m_b^2}{\mu^2}\Big) \nonumber \\
&+& \frac{2\, m_b^2}{3 \sqrt{s\, S}} \Bigg( 3 \ln \Big( \frac{m_d^2}{\lambda^2}\Big) + 1 \Bigg) {\cal{F}} (s, S, \mu)
- \frac{1}{3} \sqrt{\frac{s}{S}} \Bigg( 3 \ln \Big( \frac{m_d^2}{\lambda^2}\Big) + 1 \Bigg) {\cal{F}} (s, S, \mu)\nonumber \\
&-& 4 \, \frac{m_d^2}{\sqrt{s\, S}} \ln \Big( \frac{m_d^2}{m_b^2}\Big) {\cal{F}} (s, S, \mu) \Bigg].
\eq

The expression in \ref{eq:IR} corresponds to the IR-divergent part of the amplitude that is going to be cancelled when adding the adequate real contributions to the process $\gamma^{*}\rightarrow q\bar{q}$, leading to scales and finite terms that must still be obtained.

\subsection{Discussion}
One of the main goals of the \IReg\ scheme is to represent the UV-divergent content of a given multi-loop Feynman integral in terms of well-defined Basic Divergent Integrals (BDI) which do not need to be explicitly evaluated. Such program is compatible with the BPHZ theorem, assuring locality, causality and Lorentz invariance. Further understanding of the relation between \IReg\ and dimensional schemes was achieved recently~\cite{Cherchiglia:2020iug}, with prospects to map BDI's into poles in $\epsilon^{-n}$ for a UV $n$-loop calculation in general. 

Regarding symmetries, the method complies with abelian gauge symmetry at arbitrary loop level, and non-abelian theories have also been tested up to two-loop level. A general proof based on the quantum action principle is still lacking, however, some of the main ingredients were proved to be fulfilled by the method in recent years~\cite{Bruque:2018bmy}. Besides, a more general picture of dimension-specific objects and their properties (such as the $\gamma_{5}$ matrix) in the context of four-dimensional schemes has emerged. In particular, it was shown, quite surprisingly, that consistent four-dimensional methods such as \IReg\ need to deal with $\gamma_{5}$ problems, in a way similar to dimensional schemes.

Finally, from the point of view of infrared divergences, some proof-of-principle computations are still lacking for a NNLO calculation. In particular, the knowledge of the double-real and virtual-real contributions for the process studied in this work is underway.

\section{Local Analytic Sector Subtraction: the Torino scheme}
This section is devoted to the {\it Local Analytic Sector subtraction} scheme that has been recently proposed \cite{Magnea:2018hab} for
  NLO and NNLO QCD calculations with coloured particles in the final state only.
In order to present the NLO implementation of Local Analytic Sector subtraction we start by introducing a generic differential cross section with respect to an IR-safe observable $X$ 
\beq
\label{subtra}
\frac{d\sigma^{\rm NLO}}{dX}=\lim_{d\rightarrow4}
\Bigg\{ \int d\Phi_n \; V_n \, \delta_n(X)
+\int d\Phi_{n+1} \, R_{n+1} \; \delta_{n+1}(X) \Bigg\} \, ,
\eeq
where $d$ is the space-time dimension set equal to $4-2\epsilon$, with $\epsilon<0$, and $n$ is the number of final state, coloured partons involved in the Born process.  The symbol $d\Phi_i$ identifies the $i$-body phase space, while $V_n$ is the UV-renormalised one loop correction and $R_n$ is the tree level squared amplitude for a single real radiation. Finally, $\delta_i(X) \equiv \delta (X-X_i)$ sets the observable $X$ to be computed in the $i$-body kinematics. 
In dimensional regularisation the virtual matrix element features up to a double pole in $\epsilon$, while the real correction, finite in $d=4$, is characterised by up to two singular limits of IR nature in the radiation phase space. By integrating $R$ in $d$ dimensions over the phase space, its implicit singularities become manifest as $1/\eps$ poles. \\
Although the sum on the r.h.s. of Eq.\eqref{subtra} is finite in $d=4$ thanks to the KLN theorem \cite{Kinoshita:1962ur, Lee:1964is}, it is unfeasible to estimate the differential distribution in this form. 
The complexity of a typical collider process requires to 
exploit numerical algorithms to treat the phase space integration.
Indeed the integration over the unresolved phase space has to be performed in $d=4$, therefore the IR singularities
  must be canceled prior to the integration: this goal can be achieved via a {\it subtraction} method.

The subtraction procedure consists in adding ad subtracting in Eq.\eqref{subtra} a {\it counterterm} that reproduces the singular behaviour of the real matrix element, and can be integrated analytically in the single-radiative phase space. Such a counterterm and its integrated counterpart can be defined in full generality as 
\beq
\frac{d\sigma^{\rm NLO}_{ct}}{dX}=\int d\Phi_{n+1} \, \overline{K}_{n+1} \, ,
\qquad \qquad 
I_n=\int d\Phi_{\rm rad} \, \overline{K}_{n+1} \, ,
\label{eq:counterterm}
\eeq
with $d\Phi_{\rm rad}= d\Phi_{n+1}/d\Phi_{n}$ being the factorised single-radiative phase space. The subtracted differential cross section then reads 
\beq
\frac{d\sigma^{\rm NLO}}{dX}=
\int 
d\Phi_n \; \Big[V_n +I_n \Big]\, \delta_n(X)
+\int d\Phi_{n+1} \, \Big[ R_{n+1} \; \delta_{n+1}-\overline{K}_{n+1}  \; \delta_n(X) \Big] \, ,
\eeq
where both terms in squared brackets are separately finite and integrable in $d=4$.\\
The specific implementation of the counterterm is not unique and requires various technical aspects, which characterise the different subtraction methods. 
Several solutions to the IR subtraction problem are indeed already available and well tested at NLO, 
such as the schemes by Frixione-Kunszt-Signer \cite{Frixione:1995ms,Frixione:1997np} (FKS) and Catani-Seymour (CS) \cite{Catani:1996vz,Catani:2002hc}. 
At NNLO the variety of subtraction procedures is much richer \cite{GehrmannDeRidder:2005cm, Daleo:2006xa, Somogyi:2006da, Cacciari:2015jma, Czakon:2010td, Boughezal:2011jf, Herzog:2018ily, Sborlini:2016hat, Frixione:2004is}, but despite this considerably wide range of sophisticated schemes, many of them rely on demanding numerical calculations or involved integration procedures.
In order to overcome such bottlenecks, the local analytic sector subtraction scheme provides an alternative NLO subtraction method that aims at combining the most advantageous aspects of the FKS and the CS schemes,
complementing a minimal subtraction structure with an efficient integration strategy. These key aspects are
suitable for a natural generalisation to NNLO.

We stress that all matrix elements entering the NLO and the NNLO corrections, are assumed to be treated with conventional dimensional regularisation, and renormalised within the $\overline{\rm MS}$ scheme. Accordingly, the IR kernels that enter the counterterms have been then computed with the same regularisation approach. The radiative phase space parameterisation and the consequent integration strategy have been conveniently chosen according to dimensional regularisation. Changing the regularisation scheme would mean re-thinking the counterterm definition (and integration), as well as the precise correspondence between different contributions to the NNLO computation.

\subsection{NLO subtraction}
\label{sec:NLO}
The key characteristic of the Local Analytic Sector subtraction at NLO is the {\it locality} of the counterterm $\overline K_{n+1}$ as well as the {\it analytic} procedure adopted to compute its integrated counterpart $I_n$, as introduced in Eq.\eqref{eq:counterterm}. The counterterm has indeed to mimic the infrared behaviour of the real matrix element locally in the phase space, and at the same time, it has to be simple enough to allow for analytic integration in the single-unresolved phase space. These two fundamental features are also crucial for the NNLO generalisation, as will be explained in Sec.\ref{sec:NNLO}.
To present the method in its core structure at NLO, we introduce the following notation: we set the squared centre-of-mass energy to be $s$, the centre-of-mass four momentum to be $q^\mu=\big(\sqrt{s}, \vec{0} \, \big)$, and  $k_i^\mu$ the $i$-th final state momentum. Moreover, the Lorentz invariants $s_{ab}=2 k_a \cdot k_b$, the energy fraction $e_{i}=s_{qi}/s$ and the angular variable $w_{ij}=s \, s_{ij} /{s_{qi} \, s_{qj}}$ are also mentioned below.
The singular soft and collinear behaviour of the real matrix element is extracted by the relevant projector operators,
\beq
  \bS{a}  &:&   \text{soft single limit}  \, (e_a \rightarrow 0) \nn \\ 
  \bC{ab} &:&  \text{collinear single limit} \,  (w_{ab} \rightarrow 0) \; .
  \label{eq:NLO_projectors}
\eeq
To appropriately define the desired counterterms, we partition the real-radiation phase space into sectors by means of sector functions $\mathcal{W}_{ij}$ ($i, j = 1, \dots, n ; \,  i\neq j$), inspired by the FKS scheme \cite{Frixione:1995ms}. In analogy to FKS, the sectors functions are designed to fulfil a set of fundamental properties: 
\begin{itemize}
\item[\it 1)] they have to be a unitary partition of the phase space:
   $\sum_{i, j\neq i} \W{ij}=1$,
\item[\it 2)] they have to select a minimum number of singularities in each sector: $\bS{i} \W{ib} \neq 0, \, \bS{i} \W{ab} =0 \, , \, \forall i \neq a \, ; \, \bC{ij} \W{ij} \neq 0, \, \bC{ij} \W{ji} \neq 0 \, ; \, \bC{ij} \W{ab} =0 \, , \,  \forall ab \notin \{i,j\}$,
\item[\it 3)] the sum over sectors sharing the same singular configurations has to be one: $\bS{i} \sum_{k\neq i} \W{ik} =1 \, , \bC{ij} \big( \W{ij} + \W{ji} \big)=1$.
\end{itemize}
These constraints must hold for any explicit definition of $\W{ij}$. An efficient realisation of such sector functions is given by the following Lorentz invariants ratios,
\beq
\W{ab} = \frac{\sigma_{ab}}{\sum_{a' b'} \sigma_{a'b'}} \; , \qquad \sigma_{ab}= \frac{1}{e_a \, w_{ab}} \; .
\label{eq:NLO_sec_explicit}
\eeq
Then, sector by sector, the singular regimes of the real matrix element are collected into a candidate counterterm $K_{n+1}$
\beq
K_{n+1}=\sum_{i, j \neq i} (K_{n+1})_{ij} 
\equiv 
\sum_{i, j \neq i}
 \Big( \bS{i}+\bC{ij}-\bS{i} \, \bC{ij} \Big) \, R \, \W{ij}
 \equiv 
 \sum_{i, j \neq i} {\bf L}_{ij}^{(1)} \, R \, \W{ij} \; ,
\label{eq:cand_count}
\eeq
where the last term in round brackets avoids the double subtraction of the mixed soft-collinear divergences. As it is evident from the expression above, the introduction of sectors enables a minimal definition for $K_{ij}$: a remarkable feature that can be generalised also at NNLO. \\
Given the factorised structure of the real matrix element under unresolved limits, each contribution appearing in Eq.\eqref{eq:cand_count} can be written as a universal singular kernel and an appropriate Born-like matrix element (for a review see for instance \cite{Catani:1999ss} and the references therein). Let us stress that the kinematics which the Born-level matrix elements depends on is not on-shell or momentum conserving away from the corresponding exact singular limit. The next key ingredient of the local analytic sector subtraction is the momentum mapping. It enables the factorisation of the $n+1$ phase space into a single radiative phase space $d\Phi_{\rm rad}$ and a remaining $n$-body resolved phase space, so that the counterterm can be integrated only over the former. Moreover, with a momentum mapping we force the Born kinematics to be on-shell and momentum conserving in the entire phase space. \\
There is ample freedom in the mapping procedure, that can be exploited to simplify the integration procedure. To this aim, we introduce a Catani-Seymour final state mapping \cite{Catani:1996vz}, and define a set of $n$ on-shell momenta $\{\bar k\}^{(abc)}$ by choosing three final-state momenta $k_a, k_b, k_c$ and combining them as
\beq
&& \hspace{40mm}
\{\bar k\}^{(abc)} \equiv \Big\{\{k\}_{\slashed a \slashed b \slashed c}, \,  \kk{b}{abc}, \,  \kk{c}{abc} \Big\} \; , 
\label{eq:mapping}
\\
&& \hspace{-12mm}
\kk{b}{abc} = k_a+k_b-\frac{s_{ab}}{s_{ac}+s_{bc}} k_c \; ,  \;  \;  \qquad
\kk{c}{abc} = \frac{s_{abc}}{s_{ac}+s_{bc}} k_c \; , \;  \; \qquad
\kk{i}{abc} = k_i \, , \; \;  \forall i \neq a,b,c \; .
\eeq
The notation $\{k\}_{\slashed a \slashed b \slashed c}$ states that we are eliminating the momenta $k_i$, with $i=a,b,c$, from the initial set of $n+1$ momenta.
By choosing $a,b,c$ according to the specific counterterm contribution, its integration can be carried out analytically with standard techniques. The phase space factorises consequently in terms of the Catani-Seymour parameters $y=s_{ab}/s_{abc} \, , \, z=s_{ac}/(s_{ac}+s_{bc})$, where $0\leq y \leq 1$ and $0\leq z \leq 1$, as
\beq
d\Phi_{n+1} = d\Phi_n^{(abc)} \, d\Phi_{\rm rad}^{(abc)} \, ,
\quad
d\Phi_{\rm rad}^{(abc)} \equiv d\Phi_{\rm rad} \Big(\bar s_{bc}^{(abc)};y,z,\phi \Big) \, ,
\label{eq:PS_facto}
\eeq
with $\phi$ being the azimuthal angle between $\vec k_a$ and an arbitrary three-momentum taken as reference direction. We stress that, referring to Eq.\eqref{eq:PS_facto}, the integration involves only the variables occurring after the semicolon, while the remaining variables indicate a functional dependence. Thus, the single radiative phase space can be written as
\beq
\int
d\Phi_{\rm rad}^{(abc)} 
=
\frac{(4\pi)^{\eps-2}}{\sqrt{\pi \Gamma(1/2-\eps)}}
\Big(\bar s_{bc}^{(abc)}\Big)^{1-\epsilon}
\int_0^\pi d\phi \sin^{-2\epsilon} \phi
\int_0^1 dy \, dz \; [y (1-y)^2 z(1-z)]^{-\epsilon}(1-y) \, .
\label{eq:NLO_PS}
\eeq
As already mentioned, we are free to choose partons $a,b,c$ differently for each contribution to the counterterm. In particular, for the soft limit we set $a=i$ and $b=l, c=m$, where $i$ identifies the soft parton, and $l,m$ the emitters. For the hard-collinear component, the natural choice is $a=i$, $b=j$ and $c=r$, with $i,j$ being the collinear partons, and $r$ an on-shell spectator different from $i$ and $j$. In the remapped kinematics (remapped quantities are identified with a bar) the contributions to Eq.\eqref{eq:cand_count} are then given by
\beq
&&
\hspace{-10mm}
\bbS{i} R\big(\{k\} \big) = - \Norm \sum_{l, m \neq i} \mc I_{lm}^{(i)} \; B_{lm} \Big(\{\bar k\}^{(ilm)} \Big) \; ,
\qquad \quad
\bbC{ij} R\big(\{k\} \big) =  \Norm \, \frac{P_{ij}^{\mu \nu}}{s_{ij}} \, B_{\mu \nu} \Big(\{\bar k\}^{(ijr)} \Big) \; , 
\nn \\
&& \hspace{30mm}
\bbS{i} \bbC{ij} R\big(\{k\} \big) =  2 \, \Norm C_{f_j} \, \mc I_{jr}^{(i)} \; B \Big(\{\bar k\}^{(ijr)} \Big)  \, .
\label{eq:barred_limits}
\eeq
Here $\mathcal{I}_{lm}^{(i)}= \delta_{f_i g} \, s_{lm}/(s_{il} \, s_{im})$ is the eikonal kernel relative to parton $i$, $B_{lm}$ is the colour-correlated Born matrix element, $P_{ij}^{\mu \nu}$ is the spin-dependent Altarelli-Parisi splitting function, $B_{\mu \nu}$ is the spin-correlated Born matrix element, $C_{f_j}$ is the quadratic Casimir relevant for the colour representation of parton $j$ and $\Norm = 8\pi \as (\mu^2 e^{\gamma_E}/(4\pi))^\eps$ is a normalisation factor.
It is important to notice that the remapped contributions in Eq.\eqref{eq:barred_limits} are not uniquely defined.
Any definition of the barred counterterm is indeed acceptable, provided it fulfils a set of consistency relations. Such relations ensure $\overline K_{n+1}$ to reproduce the correct behaviour of $R_{\npo}$ in all singular regions of the real phase space. These constraints reduce to the following set of relations
\beq
&& \bC{ij} \, \bbC{ij} \, R= \bC{ij} \, R \, , \qquad \qquad \bS{i} \, \bbS{i} \, R= \bS{i} \, R \, , 
\nn \\
&&  \bC{ij} \, \bbS{i}\bbC{ij} \, R= \bC{ij} \, \bbS{i}\, R \, , \qquad \;   \bS{i} \, \bbS{i}\bbC{ij} \, R= \bS{i} \, \bbC{ij}\, R \; ,
\label{eq:R_consistency}
\eeq
which are verified by the definitions in Eq.\eqref{eq:barred_limits}. \\
Before integrating over the unresolved phase space, the sum over sectors appering in Eq.\eqref{eq:cand_count} can be organised according to
\beq
\overline{K}_{n+1} 
&=& 
\sum_{i, \,  j \neq i} \overline{K}_{ij} 
=
\sum_i \! \Big[ \sum_{j\neq i} \bS{i} \W{ij}\Big] \bbS{i} R
+  \sum_{i, j>i}  \Big[\bC{ij} \big( \W{ij}+\W{ji}\big) \Big] \bbC{ij} R
-  \sum_{i, j \neq i}  \Big[\bS{i}\bC{ij} \W{ij}\Big]  \bbS{i} \bbC{ij} R
\nn\\
&=&
\sum_i \bbS{i} \, R
+ \sum_{i, j >i } \bbC{ij} (1-\bbS{i}-\bbS{j}) \, R \; ,
\label{eq:cand_count_sum_sect}
\eeq
where the combinations in square brackets have been reduced to one, thank to the $\W{ij}$ properties, preventing the sectors to affect the integration procedure. We are then left with the evaluation of the integrated counterterm $I_{n}$. 
To maximally facilitate this crucial step, we choose to parametrise the phase space according to the kinematic mapping adopted for each contribution, as done in the Catani-Seymour scheme. Considering for instance the soft contribution, the integration proceeds trivially,
\beq
\int \! d\Phi_{\rm rad}^{(ilm)} \, \bbS{i} R\big(\{k\} \big)
\propto
- \! \sum_{l, m \neq i} \! B_{lm} \Big(\{\bar k\}^{(ilm)} \Big)
\int \!  d\Phi_{\rm rad}^{(ilm)} \,\mc I_{lm}^{(i)} 
\propto
- \! \sum_{l, m \neq i} \! B_{lm} \; \frac{(4\pi)^{\eps-2}}{\bar{s}_{lm}^{(ilm)}} \; \frac{\Gamma(1-\eps) \, \Gamma(2-\eps)}{\eps^2 \, \Gamma(2-3\eps)} \; ,
\nn
 \eeq
where the factor $\Norm$ is omitted for brevity. Similar approach can be also applied for the hard-collinear component, choosing $a=i, b=j, c=r$ in Eq.\eqref{eq:PS_facto} \cite{Magnea:2018hab}. As a conclusive remark we notice that the counterterm integration is performed exactly at all orders in $\eps$. This is not significant {\it per se}, but denotes an optimised integration strategy. \\
This completes the discussion at NLO and points out two remarkable aspects of the method: differently with respect to the {\it dipole subtraction}, our counterterm is composed by different contributions, which reproduce separately the soft, the collinear, and the soft collinear singularities of $R_{\npo}$. Moreover, in contrast with {FKS}, we have identified the counterterm before choosing an appropriate phase space parametrisation and mapping. This way, we have exploited the full freedom in adapting the parametrisation to the specific counterterm contribution, simplifying as much as possible the integration procedure.

\subsection{NNLO subtraction}
\label{sec:NNLO}

To generalise the subtraction method to NNLO \cite{Magnea:2018hab}, we have exploited the two fundamental ingredients mentioned above, namely the sector partition of the phase space, and the optimised mapping of the counterterms.
This way, the NNLO extension preserves the advantages of the NLO version of the scheme, relying
on its physically transparent interpretation and the minimal counterterm structure. These characteristics could be in principle exploited to investigate higher orders in perturbation theory, given the intrinsic complexity of the problem \cite{Magnea:N3LO}. In the first stages of the method implementation \cite{Magnea:2018hab} some key elements were missing to provide an efficient subtraction method at NNLO: the treatment of the real-virtual singularities and the integration of the double unresolved counterterm. The lack of such ingredients obviously affects the possibility to test the method for arbitrary processes. 
Efforts are ongoing to tackle the above mentioned missing ingredients, towards a general validation of the scheme \cite{Magnea:2020trj}. 

At NNLO the subtraction pattern manifests a non-trivial degree of complexity, due to the increased number of contributions to a generic observable $X$,
\beq
\frac{d\sigma^{\rm NNLO}}{dX}=\lim_{d\rightarrow4}
\Bigg\{ \int d\Phi_n \; VV_{n} \, \delta_n(X)
+\int d\Phi_{n+1} \, RV_{n+1} \; \delta_{n+1}(X) 
+\int d\Phi_{n+2} \, RR_{n+2} \; \delta_{n+2}(X) \Bigg\} \, ,
\nn
\eeq
where $VV_{n}$ is the UV-renormalised double-virtual matrix element, $RV_{n+1}$ is the real-virtual correction and $RR_{n+2}$ is the double-real configuration.
As a consequence, more counterterms are needed to cancel all the singularities arising from the unresolved radiation, and delicate cancellations have also to occur amongst the counterterms themselves, to enable a minimal and transparent pattern. 
To account for the double-real singularities, we introduce $\overline{K}^{(1)}$ that encodes the single-unresolved configurations, and the combination $\overline{K}^{(2)}-\overline{K}^{(12)}$ which cures the double-unresolved limits. In particular, $\overline{K}^{(2)}$ collects the homogenous limits, {\it i.e.} those configurations where the two unresolved partons become soft/collinear at the same rate, while $\overline{K}^{(12)}$ mimics the ordered limits, where one (one pair of) parton is more unresolved than the others. Finally, the unresolved regions of the real-virtual phase space are caught by $\overline{K}^{(\rm RV)}$. Each counterterm has to be integrated over the corresponding
unresolved phase space, as prescribed by the definitions
\beq 
I^{(i)}= \int d\Phi_{{\rm rad},i} \,  \overline{K}^{(i)} \; ,
\quad  I^{(12)}=\int d\Phi_{\rm{rad},1} \,  \overline{K}^{(12)} \; ,
\quad 
I^{(\rm{RV})}=\int d\Phi_{\rm{rad}} \, \overline{K}^{(\rm{RV})}\; ,
\quad i=1,2 \, ,
\label{integrated_count}
\eeq 
where $d\Phi_{\rm rad,2}=d\Phi_{n+2}/d\Phi_n$, $d\Phi_{\rm rad,1}=d\Phi_{n+2}/d\Phi_{n+1}$ and $d\Phi_{\rm rad}=d\Phi_{n+1}/d\Phi_n$.  The subtraction pattern at NNLO then reads
\beq  \nn
\frac{d\sigma^{\rm NNLO}}{dX} \! \! \! &=&  \! \! \! \int d\Phi_n \; \bigg[VV_n +I^{(2)}+I^{(\rm RV)} \bigg]\, \delta_n (X)
\\ \nn
&&+\int d\Phi_{n+1} \, 
\bigg[ \big( RV_{n+1}+I^{(1)}\big)\; \delta_{n+1}(X) - 
\big(\overline{K}^{(\rm RV)}+I^{(12)}\big)\; \delta_n(X)\bigg]
\\ 
&& \qquad
+\int d\Phi_{n+2} \, \bigg[ 
RR_{n+2} \; \delta_{n+2}(X)
- \overline{K}^{(1)} \delta_{n+1}(X)
- \big(\overline{K}^{(2)}-\overline{K}^{(12)}\big) \delta_{n}(X)
\bigg] \, .
\qquad 
\label{subtraNNLO}
\eeq
As anticipated, the last line is finite in the whole phase space by construction, and therefore it can be evaluated in $d=4$ dimensions. 
In the second line, the combination $RV_{n+1}- \overline{K}^{(\rm RV)}$ is free of phase space divergences, but both terms manifest explicit pole in $\epsilon$, that do not cancel in the sum. 
Those poles are subtracted in a non-trivial way: $I^{(1)}$ exposes the same $1/\epsilon$ poles as $RV$, due to a straightforward consequence of the KLN theorem, while we can  properly define $\overline{K}^{(12)}$, such that its integrated counterpart reproduces the same explicit poles as $\overline{K}^{(\rm RV)}$.
We stress that, in order to have the second line in Eq.\eqref{subtraNNLO} finite in $d=4$, the integrated counterterm $I^{(12)}$ has to play a double role. First, it has to cancel the explicit poles of the real-virtual counterterm. Second, it has to feature the same phase space singularities as $I^{(1)}$ (up to a sign). 
This is in principle not guaranteed by the KLN and indeed requires a delicate interplay between the definition of $\overline{K}^{(12)}$ and $\overline{K}^{(\rm RV)}$. Finally, in the first line the combination $I^{(\rm RV)}+I^{(2)}$ returns the explicit singularities of the double virtual matrix element.
Provided that proper counterterms are defined to satisfy the cancellations just described, the three lines in Eq.\eqref{subtraNNLO} are finite in $d=4$ and can be computed separately with numerical algorithms. \\
To identify the singular configurations contributing to this perturbative order 
we introduce the relevant projector operators 
\beq
  \bS{ab} &:&  \text{uniform double soft limit} \, (e_a, e_b \rightarrow 0, e_a/e_b \rightarrow \text{constant}) \nn \\
  \bC{abc}&:& \text{uniform double collinear limit involving three partons} \nn \\
  &&    (w_{ab}, w_{ac}, w_{bc} \rightarrow 0, \quad  w_{ab}/w_{ac}, \;  w_{ab}/w_{bc}, \; w_{ac}/w_{bc} \rightarrow \text{constant}) \nn \\
 \bC{abcd}&:& \text{uniform double collinear limit involving two pairs of partons} \nn \\
  &&    (w_{ab}, w_{cd} \rightarrow 0, \quad  w_{ab}/w_{cd} \rightarrow \text{constant}) \nn \\
 \bSC{abc}&:& \text{uniform soft-collinear limit} \,  (e_a , w_{bc} \rightarrow 0 \; e_a/w_{bc} \rightarrow \text{constant}).
\label{eq:limits}
\eeq
Nested compositions of these limits with the one in Eq.\eqref{eq:NLO_projectors} also contribute to the divergent behaviour of the double real matrix element, and in particular the mixed action of NLO limits onto NNLO singular configurations gives rise to the strongly-ordered terms, collected by $\overline K^{(12)}$. \\

\subsection{Real-virtual contribution}
\label{sec:RV_generalities}
We start by analysing the real-virtual contribution to the NNLO computation.
  The (unintegrated) counterterm $\overline{K}^{(\rm RV)}$ must be defined in such a way that it embeds all of the phase space singularities of the real-virtual matrix-element $RV$. To do so, we partition the phase space into NLO sectors by means of sector functions $\W{ij}$, (the same functions used for NLO subtractions). In each sector $\W{ij}$ we then identify a finite quantity by subtracting from $RV$ all its singular limits
 \beq
(1-\bbS{i})\; (1-\bbC{ij}) \, RV \, \W{ij} = \text{finite}  \, .
\label{eq:RV_finite_sector}
\eeq
The contributing limits are understood to feature the kinematics mapping discussed in Sec.\ref{sec:NLO}. Note that Eq.\eqref{eq:RV_finite_sector} is a symbolic statement, which can be embedded in an efficient subtraction procedure only after providing an explicit expression for the barred projectors.
%Assuming to have found such definitions, we introduce the real-virtual counterterm
We then introduce the real-virtual counterterm:
\beq
\overline{K}^{(\rm RV)}=\sum_{i,j\neq i}  \overline{\bf L}_{ij}^{(1)} \, RV \, \W{ij}  \; .
\label{eq:K_RV}
\eeq
The subtraction of the real-virtual singularities proceeds sector by sector. Once the proper counterterm has been subtracted from $RV$, the combination $(RV-\overline{K}^{(\rm RV)}) \,  \W{ij}$ is free of phase-space singularities by construction. We then have to add the counterterm back in its integrated form, {\it i.e.} we need to compute $I^{(\rm{RV})}$. Before tackling the integration problem, we get rid of the sector functions as done at NLO (analogously to what we have presented in Eq.\eqref{eq:cand_count_sum_sect} upon replacing $R$ with $RV$), obtaining 
\beq
\overline{K}^{(\rm RV)}
%&=&
%\sum_{i} \Big[\sum_{j\neq i} \bbS{i} \, \W{ij} \Big] \bbS{i} \, RV
%+ \sum_{i, j >i } \Big[\bbC{ij} \, ( \W{ij} + \W{ji})\Big]\, \bbC{ij} \, RV 
%- \sum_{i, j \neq i} \Big[ \bbS{i} \bbC{ij} \, \W{ij} \Big] \bbS{i} \bbC{ij} \, RV 
%\nn \\
&=&
\sum_i \bbS{i} \, RV
+ \sum_{i, j >i } \bbC{ij} (1-\bbS{i}-\bbS{j}) \, RV\; .
\label{eq:K_RV_sum_sector}
\eeq
As discussed at NLO, the quantities $\bbC{ij} \, RV, \,  \bbS{i} \, RV, \, \bbS{i} \bbC{ij} \, RV$ are in general constrained by a set of consistency relations forcing the barred limits reproduce
the correct behaviour of $RV$. This requirement implies the relations given in Eq.\eqref{eq:R_consistency}, provided we substitute $R$ with $RV$. The implementation of such relations mostly
relies on the peculiar properties of the mapping that are applied to the singular kernels of the real-virtual matrix.  \\
The freedom in defining the barred projectors implies that $\bbC{ij} \, RV, \,  \bbS{i} \, RV, \, \bbS{i} \bbC{ij} \, RV$ can benefit from extra terms that are not present in the off-shell singular regimes, provided the consistency relations are still satisfied.
This feature can be exploited to implement further properties of $\overline{K}^{(\rm RV)}$, as
the cancellation of its explicit poles against the one stemming from $I^{(12)}$. Such a cancellation is not protected by the KLN theorem, and represents a specific trait of our method.

\subsubsection*{Momentum mappings and integration procedure for the real-virtual counterterm}
%\label{sec:IRV}
The core structure of barred operators contributing to $\overline K^{\rm (RV)}$ is designed to mimic the singular kernels known from the literature \cite{Bern:1999ry,Somogyi:2006db}. To provide an example, we focus on the collinear contribution. The singular behaviour of the real-virtual matrix element reads \cite{Bern:1999ry,Somogyi:2006db}
\beq
\bC{ij}\,
RV
\, = \,
\frac{\Norm}{s_{ij}}
\Big[
P^{\mu\nu}_{ij}
V_{\mu\nu}
-
\frac{\alpha_s}{4\pi}
\frac{\beta_0}{\eps}
P^{\mu\nu}_{ij}
B_{\mu\nu}
+
\frac{\Norm}{s_{ij}^{\eps}} \, 
\frac{\cos(\pi\eps)}{(4\pi)^{2-\eps}}
\frac{\Gamma(1+\eps)\Gamma^2(1-\eps)}
{\Gamma(1-2\eps)}
\,
P^{(1)\mu\nu}_{ij}
B_{\mu\nu}
\Big]\,,
\label{crvdefinition}
\eeq
where $\beta_0=(11\,C_A-4\,T_R\,N_f)/3$, $B_{\mu \nu}$ and $V_{\mu\nu}$ are respectively the Born and the virtual spin-correlated matrix element, while $P_{ij}^{\mu \nu}$ and $P_{ij}^{(1) \mu \nu}$ are the spin-dependent Altarelli-Parisi (AP) kernel at tree level and one-loop accuracy. \\
The one loop splitting function can be more easily treated by identifying  its spin-averaged and a spin-dependent component as
\beq
\label{decomprv}
P^{(1)\mu\nu}_{ij}
\Bn_{\mu\nu}
=
\Big(
M_{ij}\,
P_{ij}
+
N_{ij}
\Big)
\Bn
+
\Big(
M_{ij}\,
Q^{\mu\nu}_{ij}
+
O^{\mu\nu}_{ij}
\Big)
\Bn_{\mu\nu}
\, ,
\eeq
where for each $X_{ij} \in \{ M_{ij}\,P_{ij}, \, N_{ij}, \, M_{ij}\,Q^{\mu\nu}_{ij}, \, O^{\mu\nu}_{ij} \}$ one has
\beq
X_{ij}
&=&
\delta_{f_ig}\delta_{f_jg}
\,X_{gg}
+
\delta_{f_ig}\delta_{f_j \{q \bar q\}}
\,X_{gq}
+\,
\delta_{f_i \{q \bar q\}}\delta_{f_jg}
\,X_{qg}
+
\delta_{ \{f_i f_j\} \{q \bar q\} }
\,X_{qq}
\, ,
\eeq
with $\delta_{f_i \{q \bar q\}}=\delta_{f_i q}+\delta_{f_i \bar q}$ 
and 
$\delta_{ \{f_i f_j\} \{q \bar q\} }=\delta_{f_i q} \, \delta_{f_j\bar q}+\delta_{f_i\bar q} \, \delta_{f_j q}$.
The functions $P_{ij}$ and $Q_{ij}^{\mu\nu}$ are the spin components of the AP splitting functions at tree-level, written for instance in Eqs.~(2.28-2.29) of Ref.\cite{Magnea:2018hab}. For the $gq$ splitting, relevant for the process $e^+e^- \rightarrow jj$, we have
\beq
&& M_{gq}
(z)
=
\frac{1}{\eps^2} \bigg[
 (2C_F-C_A)\left(1- 
 {}_2F_1 \left(1,-\eps;1-\eps,\frac{-z}{1-z} \right) \right)
- C_A \, {}_2F_1 \left(1,-\eps;1-\eps,\frac{1-z}{-z} \right) \bigg]\;
\nn \\
&& N_{gq}
(z)
\, = \, 
C_F
\frac{C_A-C_F}
{1-2\eps}\,
(1-\eps z) \; , \qquad \qquad 
O_{gq}^{\mu\nu}
(z)
\, =\, 
0 \; ,
\label{eq:coll_kernels}
\eeq
where $z$ is the collinear energy fraction of the emitted gluon. The structure in Eq.\eqref{crvdefinition} provides the core structure of the corresponding barred limit, once the virtual and the Born spin-correlated matrices have been mapped. By choosing the $(ijr)$ mapping, the variable $z$ appearing in Eq.\eqref{eq:coll_kernels} coincides with the Catani-Seymour parameter $z$, introduced in Eq.\eqref{eq:PS_facto}. Adopting this parameterisation, the terms in Eq.\eqref{crvdefinition} that are proportional to virtual matrix-elements, as well as those coming from UV renormalisation (proportional to $\beta_0$), can be integrated with standard techniques. 
With regards to the $P^{(1)\mu\nu}_{ij}$ contribution,
the spin-dependent kernels $Q_{ij}^{\mu\nu}$ and $O_{ij}^{\mu\nu}$ vanish when integrated over the azimuth, while $N_{ij}$ can be trivially integrated. The most involved integrals are due to the $P_{ij}M_{ij}$ term, whose main structure is of the type
\beq
\int_0^\pi d\phi \sin^{-2\epsilon} \phi \int_0^1dy dz \,  (1-y)^{1-2\eps} \, y^{-1-2\eps}(1-z)^{m-\eps} z^{n-\eps} \, _2F_1\left(1,-\eps ;1-\eps ;-\frac{z}{1-z}\right)\,,
\eeq
where $n, m \in \{-1,0,1\}$. For these values, the integral over $z$ is well defined and gives
\beq
\frac{\Gamma (m-\eps +2) \Gamma (n-\eps +1)}{\Gamma(m+n-2\eps+3)} \, _3{F}_2(1,1,n-\eps +1;m+n-2 \eps +3,1-\eps ;1)\,,
\eeq
that can be expanded in $\eps$ powers, using for example the \texttt{HypExp} code \cite{Huber:2005yg,Huber:2007dx}. The integration over the remaining radiation phase space variables $\phi$ and $y$ is straightforward. All the other splitting configurations ($g \rightarrow gg$, $g \rightarrow q\bar q$) feature the same degree of complexity as $q \rightarrow gq$. Similar conclusions also hold for the core structure of the soft-collinear barred limit, that gives at most polynomials in the $z$ and $y$ variables. \\ 
Moving to the soft contribution, we can consider as the core structure the expression in Eq.(3.30) of Ref. \cite{Somogyi:2006db}.

The integration of its contributions can be performed with standard machinery, except for the tripole-colour-correlated component which is slightly more involved. It is worth noting that neither the $I^{(2)}$ counterterm, nor the double-virtual matrix element manifest such a peculiar colour structure. Thus, the cancellation of singularities proportional to tripole-colour-correlated matrix elements is a crucial step of the method, whose treatment is detailed in Ref.~\cite{Magnea:2020trj}.

\subsection{Double-real contribution}
\label{sec:I2}

The methods developed to treat the NLO phase space singularities of the real-virtual matrix element can be generalised at NNLO to subtract the divergences of the double-real correction.
At this perturbative order, sector functions and phase space parametrisation 
combine in a more involved way to enable the analytic integrations of the relevant singular kernels.\\
The partition of phase space requires new sectors functions $\W{abcd}$, that include as many different indices as the maximum number of partons that can become unresolved simultaneously.
The indices run over the $n+2$ legs of the double-real matrix element.
In order to account for all NNLO singular configurations, to select a minimal set of them in each sector, and to avoid double counting, the four indices are chosen such that $a\neq b$ and $c\neq d$. Furthermore, $c$ and $d$ are allowed to equal $b$ but not $a$ ($c,d\neq a$).
Three topologies arise from the possible choices of indices,
\beq
\W{ijjk} \, , \qquad \W{ijkj} \, , \qquad \W{ijkl} \,  , \qquad \qquad  i \neq j \neq k \neq l \, . 
\eeq
As we have done for $RV$, we require such sector functions to be a unitary partition of phase space and sum to one when considering sectors that share the same singular configurations. The former condition is satisfied by defining
\beq
\W{abcd}= \frac{\sigma_{abcd}}{\sigma} \, , \qquad 
\sigma = \sum_{a', b'\neq a'} \sum_{\substack{c'\neq a' \\ d'\neq a',c'}}\sigma_{a'b'c'd'}
\quad \Longrightarrow \quad
\sum_{a, b\neq a} \sum_{\substack{c\neq a \\ d\neq a,c}} \W{abcd}=1 \, .
\label{eq:sectors}
\eeq 
The latter requirement can be trivially verified once an explicit form for $\W{abcd}$ has been implemented. One possibility is choosing $\sigma_{abcd}$ to be a generalisation of the NLO $\sigma_{ab}$ in Eq.\eqref{eq:NLO_sec_explicit} as
\beq
\sigma_{abcd}= \frac{1}{(e_a \, w_{ab})^\alpha} \frac{1}{(e_c+\delta_{bc} \, e_a) \, w_{cd}} \, , \qquad \qquad \alpha >1 \, , 
\label{eq:sigma_abcs}
\eeq
We stress that the choice of sector functions is not unique. For example, given the structure of Eq.~\ref{eq:sigma_abcs}, the exponent $\alpha$ can be conveniently modulated. Also different structures could be envisaged, \emph{e.g.} the energy fraction and the angular separation relative to the first pair of indices ($a,b$ in Eq.~\ref{eq:sigma_abcs}) could feature two different exponents.
The sectors in Eq.\eqref{eq:sectors} together with the definition in Eq.\eqref{eq:sigma_abcs} can be easily checked to fulfil the relation 
\beq
\bS{ik} \Big( \sum_{b\neq i} \sum_{d \neq i,k} \W{ibkd}+ \sum_{b\neq k} \sum_{d \neq k,i} \W{kbid} \Big) =1 \, 
\eeq
which provides an example of the sum rules mentioned above.
Analogous relations hold for the remaining projector operators listed in Eq.\eqref{eq:limits} and for their nested combinations.

The collection of all singular configurations contributing to a given sector gives,
\beq
&& \W{ijjk} \, RR: \quad \bS{i} \, , \; \bC{ij} \, , \; \bS{ij} \, , \; \bC{ijk} \, , \; \bSC{ijk} \, , 
\label{eq:list_limit_sectors1} \\ 
&& \W{ijjk} \, RR: \quad \bS{i} \, , \; \bC{ij} \, , \; \bS{ik} \, , \; \bC{ijk} \, , \; \bSC{ijk} \, , \, \bSC{kij} 
\label{eq:list_limit_sectors2} \\
&& \W{ijkl} \, RR: \quad \bS{i} \, , \; \bC{ij} \, , \; \bS{ik} \, , \; \bC{ijkl} \, , \; \bSC{ikl} \, , \, \bSC{kij}
\label{eq:list_limit_sectors3}
\eeq
As already discussed, this set of limits is a direct consequence of our choice of sector functions. Modifications in the definition in Eq.\eqref{eq:sigma_abcs} lead to adjustments in the lists of contributing primary limits reported in Eqs.\eqref{eq:list_limit_sectors1}-\eqref{eq:list_limit_sectors3}.
Sector by sector, we subtract from the double-real matrix element all its singular configurations (avoiding double counting), obtaining a finite object that provides our candidate counterterm. In the $\W{ijkl}$ sector the subtracted double-real matrix element reads
\beq
(1-\bbS{i})\; (1-\bbC{ij}) \; (1-\bbS{ik}) \; (1-\bbC{ijkl}) \; (1- \bbSC{ikl}) \;  (1-\bbSC{kij})\, RR \, \W{ijkl} = \text{finite}  \, .
\label{eq:finite_sector}
\eeq
Similar relations hold for the other topologies. In Eq.\eqref{eq:finite_sector} we recognise the contribution of sector $\W{ijkl}$ to the integrand function in the last line of Eq.\eqref{subtraNNLO}. It is then necessary to disentangle the single-, the double- and the mixed-unresolved counterterms by reorganising the listed limits in Eqs.\eqref{eq:list_limit_sectors1}-\eqref{eq:list_limit_sectors3} according to their kinematics. In particular, the first two parentheses in Eq.\eqref{eq:finite_sector} contain single-unresolved operators that we label collectively $\overline{\bf L}_{ij}^{(1)}$, as already done for $RV$, while the remaining combinations feature pure double-unresolved limits, that are collected by $\overline{\bf L}_{ijkl}^{(2)}$. The relation in Eq.\eqref{eq:finite_sector} can be then rewritten in the more compact form as
\beq
(1-\overline{\bf L}_{ij}^{(1)}) (1-\overline{\bf L}_{ijkl}^{(2)}) \, \W{ijkl} \, RR 
=
(RR-\overline{\bf L}_{ij}^{(1)} \, RR - \overline{\bf L}_{ij}^{(1)} \, \overline{\bf L}_{ijkl}^{(2)} \, RR) \, \W{ijkl} 
=
\text{finite}  \, ,
\label{eq:NNLO_finite}
\eeq
with $\overline{\bf L}_{ij}^{(1)} \, \overline{\bf L}_{ijkl}^{(2)}$ giving rise to the strongly-ordered singularities. The explicit expression for the $\overline{ \bf L}$ operators in the $ijkl$ sector reads
\beq
1-\overline{\bf L}_{ij}^{(1)}= (1-\bbS{i})\; (1-\bbC{ij}) \, , \quad \; \; 
1- \overline{\bf L}_{ijkl}^{(2)} = (1-\bbS{ik}) \; (1-\bbC{ijkl}) \; (1- \bbSC{ikl}) \;  (1-\bbSC{kij}) \, .
\eeq
A fundamental requirement for the above described structure
is that ts must account for all and only the actual phase space singularities of $RR$.
%to correctly take into account all and only the actual phase space singularities of $RR$.
This statement implies that whatever $\overline{ \bf L}$ is, it must match the $RR$ behaviour under the singular limits listed in Eq.\eqref{eq:limits}. For $\overline{\bf L}_{ij}^{(1)}$ this means to impose the equivalent set of relations introduced in Eq.\eqref{eq:R_consistency} upon considering $RR$ instead of $RV$.
For $\overline{\bf L}_{ijkl}^{(2)}$ and $\overline{\bf L}_{ijkl}^{(12)}$ the number of consistency relations is much larger: as a general statement, a counterterm contribution obtained nesting $n$ primary projectors has to fulfil 
 $n$ consistency relations. For this reason, finding counterterm definitions that simultaneously satisfy  all the constraints  is highly non-trivial. 

Assuming the existence of consistent definitions for all the barred operators in Eq.\eqref{eq:NNLO_finite} the counterterms can then be defined as
\beq
\overline{K}^{(1)}=\sum_{i,j\neq i} \sum_{\substack{k\neq i \\ l \neq i,k}} \overline{\bf L}_{ij}^{(1)} \, RR \, \W{ijkl} \, ,
&&
\qquad\qquad \quad 
\overline{K}^{(2)}=\sum_{i,j\neq i} \sum_{\substack{k\neq i \\ l \neq i,k}} \overline{\bf L}_{ijkl}^{(2)} \, RR \, \W{ijkl} \, ,
\nn \\
\overline{K}^{(12)}\!\!\!&=&\!\!\!\sum_{i,j\neq i} \sum_{\substack{k\neq i \\ l \neq i,k}} \overline{\bf L}_{ij}^{(1)} \, \overline{\bf L}_{ijkl}^{(2)} \, RR \, \W{ijkl} \, .
\label{counterterms}
\eeq
Each term has to be integrated over its proper phase space, as defined in Eq.\eqref{integrated_count}, and features different characteristics, therefore we will discuss separately their properties and the corresponding integration procedure.

\subsubsection*{Double-real: single- and mixed double- unresolved counterterms}
The single unresolved $\overline K^{(1)}$ and the mixed double unresolved $\overline K^{(12)}$ have been already analysed in Ref.~\cite{Magnea:2018hab}, therefore we only summarise the main aspects of their treatment. \\
Once $\overline K^{(1)}$ and $\overline K^{(12)}$ have {\it locally} subtracted the singularities of $RR$ sector by sector in the double-unresolved phase space, both the counterterms have to be integrated over a single radiative phase space, as prescribed by Eq.\eqref{integrated_count}. Their integrated counterparts are then combined with the real virtual matrix element and with $\overline K^{(\rm RV)}$ (see the second line of Eq.\eqref{subtraNNLO}), which are split into NLO sectors.
For this purpose, the sector functions appearing in Eq.\eqref{counterterms}, as defined in Eqs.\eqref{eq:sectors}-\eqref{eq:sigma_abcs},
must factorise into NLO functions under single-unresolved limits.
%must satisfy an other important property: they factorise into NLO functions under single-unresolved limits.
The generic expression of this property reads
\beq
\bS{i} \, \W{abcd} = \W{cd} \, \bS{i} \, \W{ab}^{(\alpha)} \, , \quad
\bC{ij} \, \W{abcd} =  \W{cd} \, \bC{ij} \, \W{ab}^{(\alpha)} \, , \quad
\bS{i} \, \bC{ij} \,  \W{abcd} = \W{cd} \, \bS{i} \, \bC{ij}\, \W{ab}^{(\alpha)} \, ,
\eeq 
where 
\beq 
\W{ij}^{(\alpha)}=\frac{\sigma_{ij}^\alpha}{\sum_{a,b\neq a} \sigma_{ab}^\alpha}
\quad \Longrightarrow \quad  
\W{ij}^{(1)} =\W{ij} \; .
\eeq 
Considering as an example the pure-soft content of $\overline K^{(1)}$ and $\overline K^{(12)}$,
\beq
\overline{K}^{(1), \, \rm s} \equiv  
\sum_{i,j\neq i} \sum_{\substack{k\neq i \\ l\neq i,k }} \bbS{i} \, RR \, \W{ijkl}  \, , \qquad \qquad 
\overline{K}^{(12), \, \rm s} \equiv
\sum_{i,j\neq i} \sum_{\substack{k\neq i \\ l\neq i,k }} \bbS{i} \bbS{ij}\, RR \, \W{ijkl} \, ,
\eeq
the factorisation of NNLO sector functions into NLO sectors guarantees the following equalities 
\beq
\overline{K}^{(1), \, \rm s} & =& 
\sum_{i,j\neq i} \sum_{\substack{k\neq i \\ l\neq i,k }}
\Big(\bS{i} \W{ij}^{(\alpha)} \Big) \big( \bbS{i} \, RR \big) \; \bW{kl}
=
\sum_{i,k\neq i} \sum_{l\neq i,k}
\big( \bbS{i} \, RR \big) \; \bW{kl} \, ,
 \\ 
\overline{K}^{(12), \, \rm s} & =& 
\sum_{i,j\neq i} \sum_{\substack{k\neq i \\ l\neq i,k }}
\Big(\bS{i} \W{ij}^{(\alpha)} \Big) \big( \bbS{i} \, \bbS{ik} \, RR \big) \; \bbS{k} \bW{kl}
= 
\sum_{i, k\neq i} \sum_{l\neq i,k} \big( \bbS{i} \, \bbS{ik} \, RR \big) \; \bbS{k} \bW{kl} \, ,
\eeq
where we have exploited the sector function sum properties introduced at NLO, which hold also for $\W{ij}^{(\alpha)}$.  The kinematic mapping of sector functions, namely $\bW{kl}$,
enables to factorise
the structure of NLO sectors out of the radiative phase space, and integrate analytically only the singular kernels. By adopting the Catani-Seymour mappings already discussed in Eq.\eqref{eq:PS_facto}, and parametrising $d\Phi_{\rm rad, 1}^{(abc)}=d\Phi_{\rm rad}^{(abc)}$ with the $(iab)$ mapping 
we can easily compute
\beq
I^{(1), \, \rm s} \!\!\!&\propto& \!\!\!
\sum_{i, k\neq i} \sum_{ l\neq i,k} \bW{kl} \int d\Phi_{\rm rad, 1}\,  \bbS{i} \, RR
\propto
\sum_{i, k\neq i} \sum_{ l\neq i,k} \sum_{\substack{a\neq i \\ b\neq i}} J^{s}
\big(\sk{ab}{iab}\big) \, R_{ab} \big( \{\overline{k}\}^{(iab)}\big) \;  \bW{kl}^{(iab)}  \, ,
\label{eq:I1_I12}
 \\ \nn 
I^{(12), \, \rm s} \!\!\! &\propto& \!\!\!
\sum_{i, k\neq i} \sum_{ l\neq i,k}\bbS{k} \bW{kl} \int d\Phi_{\rm rad, 1}\,  \bbS{i} \, \bbS{ik} \, RR
\propto
\sum_{i, k\neq i} \sum_{ l\neq i,k} \sum_{\substack{a\neq i \\ b\neq i}} J^{s}
\big(\sk{ab}{iab}\big) \,\bbS{k} \, 
\Big( R_{ab} \big( \{\overline{k}\}^{(iab)}\big) \, \bW{kl}^{(iab)} \Big) \; , 
\eeq
where the proportionality symbol understands constants and symmetry factors that are the same in the two lines above.
The soft function $J^{s}\big(\sk{ab}{iab}\big)$ is defined as the integral over the single phase space of the  Lorenz invariants occurring in the soft kernel and it can be easily computed to all orders in $\epsilon$,
\beq
J^{s}\big(\sk{ab}{iab}\big) \equiv \frac1{\sk{ab}{iab}} \int d\Phi_{\rm rad, 1}^{(iab)} \;  \frac{s_{ab}}{s_{ia} \, s_{ib}}
= 
\frac{(4\pi)^{\eps-2}}{\sk{ab}{iab}} \, \frac{\Gamma(1-\eps) \Gamma(2-\eps)}{\eps^2 \, \Gamma(2-3 \eps)} \, .
\eeq
From the explicit expressions of $I^{(1), \, \rm s}$ and $I^{(12), \, \rm s}$ on the r.h.s. of Eq.\eqref{eq:I1_I12} it is evident that the two counterterm share the same phase space singularities, which then cancel in the combination $I^{(1), \, \rm s}-I^{(12), \, \rm s}$. Analogous considerations apply to the hard-collinear component, so that $I^{(1)}-I^{(12)}$ is free of implicit poles. 

Applying a similar procedure also for the collinear component, the complete single-unresolved integrated counterterm reads
\beq
I^{(1)}&=& \frac{\as}{2\pi} \Big(\frac{\mu^2}{s} \Big)^\eps
\sum_{h,q\neq h} \bW{hq} \; 
\bigg\{ R\big(\{\bar{k}\}\big) 
\sum_a \Big(\frac{C_{f_a}}{\eps^2}+\frac{\gamma_a}{\eps} \Big) 
+\sum_{a, b\neq a} R_{ab} \big(\{\bar k\}\big) \frac1\eps \, \log \bar \eta_{ab}
\nn \\
&&
+  R\big(\{\bar{k}\}\big) \sum_a \Big[\delta_{f_a g} \, \frac{C_A+4T_R N_F}6 \Big(  \log \bar \eta_{ar}-\frac83 \Big)
+ \delta_{f_a g} C_A \Big(6-\frac72 \zeta_2 \Big) \Big]
\nn \\
&&
+ \sum_{a, b\neq a} R_{ab} (\{\bar k\}) \log \bar \eta_{ab} \Big( 2-\frac12 \log \bar \eta_{ab} \Big) \bigg\}
\label{eq:I1}
\eeq
Notice that after the integration, the barred variables can be relabelled to the same real kinematics $\{\bar k\}$, and that the sum over $h,q$ runs over the NLO partons, and barred momenta and invariants refer to the NLO kinematics.

\subsubsection*{Double-real: pure double-unresolved counterterm}
%\label{sec:I2}
In order to integrate the NNLO kernels in the double-unresolved phase space we need to implement NNLO mapping that can simplify the integration procedure. To this purpose, we introduce double Catani-Seymour mappings \cite{Magnea:2018hab}, designed as a generalisation of the NLO mapping in Eq.\eqref{eq:mapping}, and able to reduce the initial set of $n+2$ momenta to $n$ on-shell momenta without breaking total momentum conservation. The double mapping is defined as 
\beq
\{\, \bar k \, \}^{(abcd)}= \Big\{ \{k\}_{\slashed a \slashed b \slashed c \slashed e \slashed f}\, , \;  \bar k_e^{(abcd)} \, , \;  \bar k_f^{(abcd)}\Big\} \, , 
\qquad\qquad
\kk{n}{abcd} 
\, = \, 
k_n \, , \quad \, n \neq a,b,c,d \, , 
\nn \\
\kk{c}{abcd} 
\, = \, 
\kk{b}{abc} + \kk{c}{abc} 
- \frac{\sk{bc}{abc}}{\sk{bd}{abc}+\sk{cd}{abc}} \, \kk{d}{abc}
 \, ,  \qquad \qquad 
\kk{d}{abcd} 
\, = \, 
\frac{\sk{bcd}{abc}}{\sk{bd}{abc}+\sk{cd}{abc}} \, \kk{d}{abc} \, .
\label{eq: repar2 abcd} 
\eeq
We then introduce the Catani-Seymour parameters 
\beq
y' = \frac{s_{ab}}{s_{abc}} \, , 
\qquad
z' = \frac{s_{ac}}{s_{ac} + s_{bc}} \, ,  
\qquad 
y = 
\frac{\sk{bc}{abc}}{\sk{bcd}{abc}} \, , 
\qquad
z =  
\frac{\sk{bd}{abc}}{\sk{bd}{abc} + \sk{cd}{abc}}  
\, , 
\label{eq:CSparam NNLO}
\eeq
to factorise the $(\npt)$-body phase space as 
\beq
d \Phi_{\npt} 
& = & 
d \Phi_n^{\, (abcd)} \, d \Phi_{\rm rad, 2}^{\,(abcd)}  \; ,
\qquad
d \Phi_{\rm rad, 2}^{\,(abcd)}
\, = \,
d \Phi_{\rm rad}^{\,(abc)} \, d \Phi_{\rm rad}^{\,(abcd)}
\, .
\label{factophsp2}
\eeq
The double unresolved phase space can be written as an integral over the variables $(\phi, y, z, x', y', z')$ as
\beq
\int d \Phi_{\rm rad, 2}^{\,(abcd)} 
&=&
\frac{(4\pi^2)^{\eps - 4}}{\pi \, \Gamma^2 (1/2 - \eps)}
\, {(s_{abcd})}^{2 - 2 \eps}\, \!\!
\int_0^1 \!\!\! d x'  \, 
\big[ x'(1-x')\big]^{-1/2-\eps} 
\int_0^1 \!\!\! d y' \!
\int_0^1 \!\!\! d z' \!
\int_0^\pi \!\!\! d \phi \, \sin ^{- 2 \eps}\phi \; \times
\nn \\
&&  \times
\int_0^1 \!\!\! d y \!
\int_0^1 \!\!\! d z \, 
\Big[ y'(1-y')^2\,z'(1-z')\,y^2(1-y)^2\,z(1-z) \Big]^{- \eps} 
(1-y') \, y(1-y) \, ,
\label{phspproco}
\eeq
where $y'$ and $z'$ are the variables relative to the 
secondary-radiation phase space, and $x'$ parametrises the azimuth between 
subsequent emissions. \\
As an example, we consider the contribution to the double soft barred limit $\bbS{ij} \, RR$ stemming from the $q \bar q $ configuration. The core structure of such a limit embeds the NNLO soft current in Eq.(96) of Ref.\cite{Catani:1999ss}, which reads
\beq
\mc I_{cd}^{(ij)}= \frac{s_{ic} \, s_{jd}+s_{id} \, s_{jc}-s_{ij} \, s_{cd}}{s_{ij}^2 \, (s_{ic}+s_{jc}) \, (s_{id}+s_{jd})} \; ,
\label{eq:qq_soft}
\eeq
with $i,j$ referring to the unresolved partons, and $c,d$ to the emitting particles. To integrate this current we choose to parametrise the double-unresolved phase space according to the $(ijcd)$ mapping.
In this parametrisation, the denominators appearing in Eq.\eqref{eq:qq_soft} read
\beq
s_{ij} = y' y \, \sk{cd}{ijcd} , \quad 
s_{ic}+s_{jc} = (1-y') y\, \sk{cd}{ijcd}  \, , \quad  
s_{id}+s_{jd} = (y'+z-y' z) (1-y)\, \sk{cd}{ijcd}  ,
\label{invaproco_text}
\eeq
while in the numerators only polynomials in the Catani-Seymour parameters appear. Note that also the dependence on the azimuth is completely factorised and occurs only in the numerator through $s_{jd}$ and $s_{id}$, since
\beq
&& s_{jd} = (1-y) \left[y'z'(1-z) + (1-z')z + 2(1-2x')\sqrt{y'z'(1-z')z(1-z)} \right]  \sk{cd}{ijcd}  \, ,
\label{azimuth_dep} 
\eeq
and $s_{id}=(y'+z-y' z) (1-y)\, \sk{cd}{ijcd} -s_{jd} $.
The phase space integral assumes the following form
\beq
\int d \Phi_{\rm rad, 2}^{\,(abcd)} \, \mc I_{cd}^{(ij)}
&\propto&
 {(\sk{cd}{ijcd} )}^{ - 2 \eps}\, \!\!
\int_0^1 \!\!\! d x'  \, 
\big[ x'(1-x')\big]^{-1/2-\eps} 
\int_0^1 \!\!\! d y' \!
\int_0^1 \!\!\! d z' \!
\int_0^\pi \!\!\! d \phi \, \sin ^{- 2 \eps}\phi \; \times
 \\ \nn
&&
\int_0^1 \!\!\! d y \!
\int_0^1 \!\!\! d z \, 
\Big[ y'(1-y')^2\,z'(1-z')\,y^2(1-y)^2\,z(1-z) \Big]^{- \eps} 
\frac{\mc N}{(y')^{2} \;  y^{2} \; (y'+z-y'z)} \; ,
\label{phspproco}
\eeq
where the numerator reads in full generality
\beq
\mc N= z^{\ell_1} (1-z)^{\ell_2}  \; y^{m_1} (1-y)^{m_2} \; (z')^{n_1} (1-z')^{n_2} \; (y' )^{r_1} (1-y')^{r_2}
(1-2 x')^k  \; .
\eeq
Now we sketch the integration procedure by considering one variable at a time: the integration over $\phi$ is trivial, the one over $y$ returns a simple Beta function $B(m_1-1-2\eps, m_2+1-2\eps)$, with $m_1,m_2 \in \mathbb{Z}$, and the azimuth contribution is $B(1/2-\eps,1/2-\eps) \, \delta_{k \, 0}$. The trivial dependence on $z'$ in the numerator enables a straightforward integration that returns $B(n_1+1-\eps, n_2+1-\eps)$, with $n_1,n_2$ being integers or semi-integers. The $z$ variable features instead a less-trivial structure, which however can be integrated according to
\beq
\int_0^1 dz \; \frac{z^{\ell_1-\eps} \, (1-z)^{\ell_2-\eps}}{y'+z-y' \, z} = B(\ell_1+1-\eps,\ell_2+1-\eps) \, 
{}_2 F_1 (1,\ell_2+1-\eps,\ell_1+\ell_2+2-2\eps, 1-y') \; .
\eeq
The remaining integration over $y'$ is tackled by applying recursively the hypergeometric function properties until we obtain ${}_2 F_1 (-\eps,-2\eps,1-2\eps, 1-y')$.  The series expansion of such class of hypergeometric functions is known at all orders in $\eps$ in terms of Spence functions.
At this point the poles in $\epsilon$ can be extracted using the {\it plus} prescription and the remaining integration over $y$ can be carried out with standard techniques. \\
We stress that the $q\bar q$ case is particularly simple, since no denominators containing the azimuth appear in the current structure after the parametrisation. 
For the gg case (and for the collinear contributions) the integration is much more involved. 
However, in our approach it can be carried out with standard techniques~\cite{Magnea:2020trj}. 
Similar integrals have been computed in the context of other NNLO schemes,
 for instance by means of integration by parts identities 
and differential equations machinery~\cite{Caola:2018pxp,Delto:2019asp}.

\subsection{Double virtual contribution}
Given a general strategy to define double-unresolved and real-virtual counterterms (see Sec~\ref{sec:I2}) we have to identify the IR singularities of the double-virtual matrix element, that we assume to be already UV renormalised. From the studies carried on in the context of IR factorisation \cite{Becher:2009cu, Becher:2009qa, Gardi:2009zv}, the infrared poles of gauge theory scattering amplitudes are known to organise according to
\beq
\mc A \left(\frac{p_i}{\mu}, \alpha_s(\mu), \eps \right)
=
\mathbf{Z}\left(\frac{p_i}{\mu}, \alpha_s(\mu), \eps \right)
\mc H\left(\frac{p_i}{\mu}, \alpha_s(\mu), \eps \right) \, ,
\label{facto}
\eeq
where $\mc A$ is a generic $n$-parton amplitude, $\mc H$ is finite for $\eps \rightarrow 0$ and $\mathbf{Z}$ is a color operator with a universal form. In Eq.\eqref{facto} all the color indices are understood, to simplify the notation. The operator $\mathbf{Z}$ obeys a renormalisation group equation that can be solved in terms of the anomalous dimension $\mathbf \Gamma$ as described by the following expression
\beq
\mathbf{Z}\left(\frac{p_i}{\mu}, \alpha_s(\mu), \eps \right)= \mc P \exp \left\{ \int_0^\mu \frac{d\lambda}{\lambda} \; \mathbf{\Gamma} \left(\frac{p_i}{\lambda}, \alpha_s(\lambda), \eps \right)\right\} \; .
\eeq 
The operator $\mathbf \Gamma$, in turn, manifests a universal behaviour regulated by the {\it dipole formula} \cite{Gardi:2009qi,Gardi:2009zv,Becher:2009cu,
Becher:2009qa}
\beq
 \mathbf{\Gamma} \left(\frac{p_i}{\lambda}, \alpha_s(\lambda), \eps \right)
 = \frac12  \, 
 \widehat{\gamma}_k (\alpha_s(\lambda, \eps))
 \sum_{\substack{i, j>i=1}}^n
 \ln \left(\frac{2 \, p_i \cdot p_j \, \ e^{i \pi \sigma_{ij}}}{\lambda^2} \right)
 \mathbf{T}_i \cdot \mathbf{T}_j  
 -\sum_{i=1}^n \gamma_i(\alpha_s(\lambda, \eps)) \; ,
\eeq
where the $\sigma_{ij}$ is a phase factors that equals 1 if $i,j$ are both in the initial or in the final state, and vanishes otherwise. The function $\widehat{\gamma}_k$ is a universal quantity related to the cusp anomalous dimension, and the jet anomalous dimensions $\gamma_i$ are related to the anomalous dimensions of quark and gluon fields. Finally, $\mathbf{T}_a$ are color operators \cite{Bassetto:1984ik,Catani:1996vz}. By expanding $\mathbf{\Gamma}$ at two-loop order, and then deriving the expression of $\mathbf{Z}$, it is straightforward to obtain the singular part of the squared amplitude up to $\alpha^2_s$ by means of Eq.\eqref{facto}.
As a result, the double virtual matrix element features infrared poles that obey the following general structure \cite{Becher:2009qa}\cite{Aybat:2006wq}:
\beq
VV_{\rm poles }&=&
\Big( \frac{\alpha_s}{\pi}\Big)^2
\bigg[
-\frac{1}{8 \eps^4} \Big(\sum_i C_{f_i} \Big)^2 B
+ \frac{1}{4 \eps^3} \Big(\sum_i C_{f_i} \Big) \Big(\frac38 b_0 + 2\sum_j \gamma_j^{(1)} \Big)B
+ \frac{1}{4\, \eps^2} \bigg[
\Big(-\frac{b_0}2 \sum_i \gamma_i^{(1)}
\nn \\
&-& 
\frac{\widehat{\gamma}_k^{(2)}}{4} \sum_i C_{f_i}-2\Big(\sum_i \gamma_i^{(1)} \Big)^2\Big) B
+ \frac{b_0}4 \sum_{i,j\neq i} \ln \frac{s_{ij}}{\mu^2} \, B_{ij}
+ \frac14 \sum_{\substack{i,j\neq i \\ k,l \neq k}}
\ln \frac{s_{ij}}{\mu^2} \, \ln\frac{s_{kl}}{\mu^2} \, B_{ijkl}
 \bigg]
\nn \\
&+&  
 \frac{1}{2 \eps} 
 \bigg[\sum_i \gamma_i^{(2)} B
 - \frac{\widehat{\gamma}_k^{(2)}}{4} \sum_{i, j\neq i}\ln \frac{s_{ij}}{\mu^2} \, B_{ij}
 -\sum_{i, j\neq i} \ln \frac{s_{ij}}{\mu^2} H_{ij}
 \bigg]
\bigg]
 -
\frac{\alpha_s}{\pi} \bigg[
\frac{1}{2\eps^2} \sum_i C_{f_i} 
- \frac{1}{\eps} \sum_i \gamma_i^{(1)} \bigg] V
\nn\\
\label{eq:VV_poles}
\eeq
with $C_{f_i}$ being the Casimir eigenvalue for the leg $i$ and $b_0=(11C_A-4T_RN_f)/3$ being the one loop $\beta$-function coefficient. The 
quantity $H_{ij}$ is a process-dependent finite contribution that derives from the virtual matrix element, while $B_{ab}$ and $B_{abcd}$ are respectively the single and double colour-correlated Born matrix elements:
\beq
B_{ab}\equiv \langle \mc A_B | \mathbf{T}_a \cdot  \mathbf{T}_b |\mc A_B  \rangle \, , \qquad 
B_{abcd}\equiv \langle \mc A_B | \{\mathbf{T}_a \cdot  \mathbf{T}_b \, , \, \mathbf{T}_c \cdot  \mathbf{T}_d\} |\mc A_B  \rangle \, .
\eeq
From Eq.\eqref{eq:VV_poles} it is evident that such a structure can be implemented in the subtraction procedure only given the knowledge of the necessary anomalous dimensions, which however can be found in the literature.
Moreover, also the process-dependent quantity $H_{ij}$ has to be considered as an external input of the scheme.

\subsection{Application: $T_R C_F$ contribution to $e^+e^- \rightarrow j j$ at NNLO}

The validation of the subtraction scheme has been performed for the two jet production in $e^+e^-$ annihilation, considering for the moment only the $T_R C_F$ contribution \cite{Magnea:2018hab}.
The virtual $e^+e^- \rightarrow q_1 \bar q_2 $, real-virtual $e^+e^- \rightarrow q_1 \bar q_2 g_{[34]}$, and double real $e^+e^- \rightarrow q_1 \bar q_2 q'_3 \bar q'_4$ contributions to the inclusive cross-section are known analytically from Ref.\cite{GehrmannDeRidder:2004tv, Hamberg:1990np, Ellis:1980wv}
\beq
&&\!\!\!\!\!\!\!\! 
VV= B \Big(\frac{\as}{2\pi} \Big)^2
T_R C_F \bigg\{ \Big(\frac{\mu^2}{s} \Big)^{2\eps}
\Big[
\frac1{3\eps^3}
+\frac{14}{9\eps^2}
+\frac1\eps \Big(\frac{353}{54}-\frac{11}{18}\pi^2 \Big)
+\frac{7541}{324}-\frac{77}{27}\pi^2-\frac{26}{9}\zeta_3
\Big]
 \\
&&\hspace{35mm}
+
\Big(\frac{\mu^2}{s} \Big)^{\eps}
\Big[
-\frac4{3\eps^3}
-\frac2{\eps^2}
+\frac1\eps \Big(-\frac{16}{3} + \frac79\pi^2\Big)
-\frac{32}3
+\frac76\pi^2
+\frac{28}9\zeta_3
\Big]
\bigg\}
\nn \\
&& \!\!\!\!\!\!\!\! \int 
d\Phi_{\rm rad} \, RV =
%\int d\Phi_{\rm rad} \;  \frac{\as}{2\pi} \frac{2}{3\eps} T_R \, R
%\nn \\
%&& \hspace{16.5mm}=
 B \Big(\frac{\as}{2\pi} \Big)^2
T_R C_F \Big(\frac{\mu^2}{s} \Big)^{\eps}
\Big[
\frac4{3\eps^3}
+ \frac2{\eps^2}
+\frac1\eps \Big(\frac{19}3 -\frac79 \pi^2\Big) 
+\frac{109}6
-\frac76\pi^2
-\frac{100}9 \zeta_3
\Big]
\nn \\
&& \!\!\!\!\!\!\!\! \int 
d\Phi_{\rm rad, 2} \, RR =
-
 B \Big(\frac{\as}{2\pi} \Big)^2
T_R C_F \Big(\frac{\mu^2}{s} \Big)^{2\eps}
\Big[
\frac1{3\eps^3}
+ \frac{14}{9\eps^2}
+\frac1\eps \Big( \frac{407}{54}-\frac{11}{18}\pi^2 \Big) 
+\frac{11753}{324}
-\frac{77}{27}\pi^2
-\frac{134}9 \zeta_3
\Big] \; .
\nn
\eeq
We can now compute the local counterterms and their integrated counterparts, showing the cancellation of the singularities presented above.
The double real matrix element presents single phase space singularities corresponding to the single collinear limit only.
The double-unresolved singularities arise from the configurations where both the emitted quarks are soft, or they are collinear to one of the hard Born-level fermion. The relevant limits in the unbarred kinematics 
are $\bbC{ij} RR, \, \bbS{ij} RR, \, \bbC{ijk} RR , \, \bbS{ij} \bbC{ijk} RR$, on top of the NLO limits, relevant for the real-virtual counterterm. 
Here $\{i,j\}=\{3,4\}$, and $\{ijk\} = \{134, 234\}$, and $r=\{1,2,3,4\}, r\neq i,j,k$. The resulting complete set of counterterms is then given by
\beq
\overline{K}^{(1)} &=& \bbC{34} RR \; , 
\label{eq:example_K1} 
 \\
\overline{K}^{(2)} &=& \Big( \bbS{34}+  \bbC{123} (1-\bbS{34})+\bbC{234}(1-\bbS{34})
\Big) RR \; , 
\label{eq:example_K2} 
 \\
\overline{K}^{(12)} &=& \bbC{34} \Big( \bbS{34}+  \bbC{123} (1-\bbS{34})+\bbC{234}(1-\bbS{34})
\Big) RR  \; ,
\label{eq:example_K12}
 \\
\overline{K}^{(RV)} &=&
\frac{\as}{2\pi} \frac{2}{3\eps} T_R 
\Big[\bbS{[34]}+  \bbC{1[34]} \big(1-\bbS{[34]}\big)+\bbC{2[34]}\big(1-\bbS{[34]}\big) \Big] \, R  \; .
\label{eq:example_KRV}
\eeq
The explicit definitions of the contributing limits in the remapped kinematic are reported in Ref.\cite{Magnea:2018hab}.
In the evaluation of the corresponding integral we need to introduce an appropriate mapping, and then apply the integration strategy sketched in the previous sections. In particular
\beq
&& \!\!\!\!\!\!\!\!\!\!\!\!
\int \! d\Phi_{\rm rad, 2} \, \bbS{ij} \, RR
= \Norm^{\, 2} \, T_R C_F 
\sum_{c,d =1}^2
B_{cd} \Big(\{\bar k\}^{(ijcd)} \Big)\int d\Phi_{\rm rad, 2}^{(ijcd)} \; \times
\nn \\
&& \hspace{30mm}
\times \Big[ \frac{s_{ic} \, s_{jd}+s_{id} \, s_{jc}-s_{ij} \, s_{cd}}{s_{ij}^2 \, (s_{ic}+s_{jc}) \, (s_{id}+s_{jd})} 
- \frac{s_{ic} \, s_{jc}+s_{ic} \, s_{jc}}{s_{ij}^2 \, (s_{ic}+s_{jc})^2} 
- \frac{s_{id} \, s_{jd}+s_{id} \, s_{jd}}{s_{ij}^2 \, (s_{id}+s_{jd})^2} 
 \Big]
\nn\\
&&\hspace{22mm}=
- B \Big(\frac{\as}{2\pi} \Big)^2
T_R C_F \Big(\frac{\mu^2}{s} \Big)^{2\eps}
\Big[
\frac1{3\eps^3}
+ \frac{17}{9\eps^2}
+\frac1\eps \Big( \frac{232}{27}-\frac{7}{18}\pi^2 \Big) 
+\frac{2948}{81}
-\frac{131}{54}\pi^2
-\frac{38}9 \zeta_3
\Big],
\nn \\
&& \!\!\!\!\!\!\!\!\!\!\!\!
\int \! d\Phi_{\rm rad, 2} \, \bbC{ijk} \, RR
= \frac{\Norm^{\, 2}}2 \,
B_{\mu \nu} \Big(\{\bar k\}^{(ijkr)} \Big)\int d\Phi_{\rm rad, 2}^{(ijkr)} \; 
\frac{P_{ijk}^{\mu \nu}}{s_{ijk}}
\nn\\
&& \hspace{23.5mm}=
- B \Big(\frac{\as}{2\pi} \Big)^2
T_R C_F \Big(\frac{\mu^2}{s} \Big)^{2\eps}
\Big[
\frac1{3\eps^3}
+ \frac{31}{18\eps^2}
+\frac1\eps \Big( \frac{889}{108}-\frac12\pi^2 \Big) 
+\frac{23941}{648}
-\frac{31}{12}\pi^2
-\frac{80}9 \zeta_3
\Big].
\nn
\eeq
Let us stress that the spin-dependent component of the double-collinear Altarelli-Parisi splitting function vanishes upon integration. Finally, the composite limit $\bS{ij} \bC{ijk} RR$ coincides with the double soft contribution $\bS{ij} RR$, given the fact that $k$ and $r$ have to be different from $i,j$, and in this specific process they can only coincide with 1 and 2. Summing all the contributions, as prescribed by Eq.\eqref{eq:example_K2}, we easily obtain the double-unresolved integrated counterterm
\beq
I^{(2)}= B  \Big(\frac{\as}{2\pi} \Big)^2 T_R C_F  \Big(\frac{\mu^2}{s} \Big)^{2\eps}
\Big[ -\frac1{3\eps^3}
-\frac{14}{9\eps^2}
+\frac1\eps \Big(\frac{11}{18} \pi^2-\frac{425}{54} \Big)
+\frac{12149}{324}
+\frac{74}{27} \pi^2
+\frac{122}9 \zeta_3\Big] \; .
\label{eq:I2}
\eeq
The next contribution is due to the single unresolved configurations, which are entirely reproduced by the collinear limit $\bbC{34}$. The expression of $T_R C_F$ contribution to $I^{(1)}$ can be directly read from Eq.\eqref{eq:I1} returning 
\beq
I_{hq}^{(1)} = -\frac{\as}{2\pi} \Big(\frac{\mu^2}{s} \Big)^\eps
\frac23 T_R \Big(\frac1\eps
- \log \bar{\eta}_{[34] r}
+\frac83 \Big)
R  \, \bW{hq}
\eeq
here $\{h,q\}= \{1, 2, [34]\}$.
The mixed-double unresolved counterterm is given by
\beq
I^{(12)}_{hq} =- \frac{\as}{2\pi} \frac23 T_R \Big(\frac{\mu^2}{s} \Big)^\eps
 \Big( \frac1\eps -\log \bar \eta_{[34]r}+\frac83\Big)
\Big[\bbS{h}+\bbC{hq}(1-\bbS{h}) \Big]
R  \, \bW{hq}
\eeq
Finally, the real-virtual counterterm reads
\beq
\overline{K}_{hq}^{(\rm RV)} =\frac{\as}{2\pi} \,  \frac23 T_R \frac1\eps \Big[ \bbS{h}+\bbC{hq}(1-\bbS{h})\Big] R \, \bW{hq} \; ,
\eeq
and its integrated counterpart that results
\beq
I^{(RV)}
&=&
 \frac{\as}{2\pi} \frac23 \frac1\eps T_R \int d\Phi_{\rm rad} 
\Big[ \bbS{[34]}+\bbC{1[34]} (1-\bbS{[34]})+\bbC{2[34]}(1-\bbS{[34]})\Big] R
\nn \\
&=&
B  \Big(\frac{\as}{2\pi} \Big)^2 T_R C_F  \Big(\frac{\mu^2}{s} \Big)^\eps
\Big[ \frac4{3\eps^3}
+\frac2{\eps^2}
+\frac1\eps \Big(\frac{20}3-\frac79 \pi^2 \Big)
+20
-\frac76 \pi^2
+\frac{100}9 \zeta_3\Big]
\label{eq:Irv}
\eeq
We can check verify that all the expected cancellations take place.
The subtracted double real matrix element is finite by construction, the difference $\overline{K}^{(RV)} + I^{(12)}$ has to be finite sector-by-sector, and indeed we have
\beq
\overline{K}_{hq}^{(RV)} +
I^{(12)}_{hq} 
=
-\frac{\as}{2\pi} \,  \frac23 T_R 
\Big(\log \frac{\mu^2}{s_{[34]r}}+\frac83\Big)
\Big[\bbS{h}+\bbC{hq}(1-\bbS{h}) \Big]
R  \, \bW{hq} \; ,
\label{eq:KRV_I12}
\eeq
which is clearly free of explicit poles. The real-virtual matrix element has to be finite for $\eps \rightarrow 0$ when combined with $I^{(1)}$, thanks to the KLN theorem. Indeed we have
\beq
RV \,  \bW{hq} + I^{(1)}_{hq}
=
-\frac{\as}{2\pi} \,  \frac23 T_R 
\Big(\log \frac{\mu^2}{s_{[34]r}}+\frac83\Big) \, 
R  \, \bW{hq}  \; .
\label{eq:RV+I1}
\eeq
The comparison between Eq.\eqref{eq:KRV_I12} and Eq.\eqref{eq:RV+I1} makes evident that the phase space singularities of the two objects cancel in the combination $RV \,  \bW{hq} + I_{hq}-(\overline{K}_{hq}^{(\rm RV)} + I^{(12)}_{hq})$. Finally, the double-virtual poles are cancelled by the sum $I^{(2)}+I^{(\rm RV)}$, as one can easily deduce by looking at the expressions in Eqs.\eqref{eq:I2}-\eqref{eq:Irv}.

\subsection{Discussion}
In this report we have reviewed the main aspects of the
local analytic sector subtraction scheme.\\
We have recently computed all the integrated counterterms that are necessary to have a fully general subtraction method (for massless, final-state QCD).

It is evident from the discussion above that the scheme benefits from an optimised partition and parametrisation of the phase space, which allows for the analytic integration of the singular kernels arising both from the double real and the real-virtual matrix element. 

Currently, the subtraction scheme is fully validated at NLO for a generic process with final state partons. 
With regards to NNLO, we have computed the $C_F T_R$ contribution to the inclusive $e^+e^- \rightarrow jj$  cross-section.
Much work is in progress to further check the scheme in less trivial processes, towards its the generalisation to any final state QCD process. 

Beyond this, the next steps will concern the extension of the scheme to the treatment of initial state radiation. Very recently, a preliminary successful study has been carried out at NLO. Extending it to NNLO
is expected to be time-consuming, but not to present conceptual novelties. However we foresee to be able to propose a similar structure as the one for final-state radiation. The generalisation to the massive case is expected to be more involved, especially concerning the integration procedure.

\section{The $\qt$-subtraction method}
In the recent years a huge effort was made by the experimental community to
increase the accuracy of high-energy physics measurements. On the theory
side, then, it has become mandatory to aim at a deeper understanding of the
perturbative behaviour of the Standard Model (SM), which translates into a
need for better control on higher-order  calculations, as long as on the
issues they yield.

One of them addresses the handling of IR divergences appearing in
the intermediate stages of higher-order QCD computations. Once their cancellation is
under control at numerical level, order-by-order in the relevant coupling
constant, a step forward is made towards an attempt of a tentative SM
falsification, within the comparison with the experimental result.

Monte Carlo generators regularise the IR divergences appearing in real and
virtual contributions to scattering amplitudes with subtraction
prescriptions. Such subtraction methods are not only capable of producing
total cross sections as well as differential distributions, but they also
allow the implementation of the same selection cuts imposed by the
experiment.

In order to expose the cancellation of the IR divergences between real and
virtual contributions, the behaviour of the scattering amplitudes at the
boundaries of the phase space is the key ingredient used by subtraction
methods, such as the well-established ones proposed in
Refs.~\cite{Catani:1996vz, Frixione:1995ms} for NLO computations and those
developed for NNLO calculations. Examples of them are the
transverse-momentum~($\qt$) subtraction method~\cite{Catani:2007vq,
  Bozzi:2005wk, Catani:2013tia}, the $N$-jettiness
subtraction~\cite{Boughezal:2015eha, Gaunt:2015pea}, the
projection-to-Born~\cite{Cacciari:2015jma}, the residue
subtraction~\cite{Czakon:2011ve, Boughezal:2011jf} and the antenna
subtraction method~\cite{GehrmannDeRidder:2005cm, Daleo:2006xa,
  Currie:2013vh},\footnotew{
 An elaborated discussion of the antenna subtraction method shall be 
presented in Section~\ref{sec:antenna}.
}
which have all been successfully applied to LHC
phenomenology. The $\qt$-subtraction method was also applied for the first
time to differential cross sections (for hadron-hadron collisions) at N$^3$LO
in Ref.~\cite{Cieri:2018oms}. Other N$^3$LO differential computations can be
found in Refs.~\cite{Dreyer:2016oyx, Dreyer:2018qbw, Chen:2019lzz,
  Chen:2019fhs}. Also, N$^3$LO differential results to jet production in deep
inelastic scattering~(DIS) and charged current DIS were calculated using the
projection-to-Born method in Ref.~\cite{Currie:2018fgr}
and~\cite{Gehrmann:2018odt}, respectively.
Moreover, the $\qt$-subtraction method was also extended in order to deal
with massive partons in the final state~\cite{Bonciani:2015sha,
  Catani:2019iny}, with initial-state QED corrections~\cite{Cieri:2018sfk},
with final-state QED radiation~\cite{Buonocore:2019puv} and recently with
mixed QCD--QED corrections at full NNLO~\cite{Cieri:2020ikq}.

%The past recent years have witnessed a great activity in another kind of local subtraction methods: those regularisation schemes in four dimensions. In this report, the interested reader can find specific sections devoted to the presentation of the main features of the regularisation schemes in four dimensions: four dimensional helicity scheme (\FDH), dimensional reduction (\DRED), four-dimensional unsubtracted scheme (\FDU), four dimensional formulation regularisation/renormalisation (\FDR), and implicit regularisation (\IReg)

\subsection{The master formula of the $\qt$-subtraction method}
We consider the inclusive hard-scattering reaction 
\begin{equation}
h_1(p_1)+h_2(p_2)\to F(\{q_i\})+X\, ,
\label{class}
\end{equation}
where $h_1$ and $h_2$ are two hadrons colliding with momenta $p_1$ and $p_2$
and triggering the final-state system $F$ along with an arbitrary and
undetected final state $X$.
%% The colliding hadrons with centre--of--mass energy $\sqrt s$, are treated
%% as massless particles $(s= (p_1+p_2)^2 = 2p_1p_2)$.
The observed final state $F$ consists of a system of non-QCD partons composed
by one or more colour-singlet particles, e.g.~vector bosons, photons, Higgs
bosons, Drell--Yan lepton pairs, etc., with four-momenta $q_i$.  The total
four-momentum of the system $F$ is denoted by $q$, with
\begin{equation}
  q=\sum_i q_i \,,
\end{equation}
and can be expressed in terms of the total invariant mass $M$,
the transverse momentum ${\bf\qt}$ w.r.t. the direction of the
colliding hadrons, and the rapidity
\begin{equation}
  y = \frac{1}{2}\,\log\frac{p_2\cdot q}{p_1\cdot q}
\end{equation}
in the centre-of-mass system of the collision.  Since $F$ is colourless,
the LO partonic Born cross section can be either initiated by $qq^{\prime}$
annihilation, as in the case of the Drell--Yan process, or by gluon--gluon
fusion, as in the case of Higgs boson production.

In order to describe the structure of the subtraction formalism we first notice
that, at LO, the transverse momentum
\begin{equation}
  {\bf\qt} = \sum_i {\bf\qt}_i
\end{equation}
of the final state system $F$ is identically zero. Therefore, as long as $\qt
\neq 0$, the N$^{n}$LO QCD contributions, with $n=1,2,3$, are given by the
N$^{n-1}$LO QCD contributions to the triggered final state
$F$+jet(s).\footnote{The notation N$^{n}$LO stands for: N$^{0}$LO=LO,
  N$^{1}$LO=NLO, N$^{2}$LO=NNLO, and so forth.} Consequently, if $\qt \neq 0$
we have
\begin{equation}
\label{Eq:first}
\mathd\sigma^{\sss F}_{\sss\rm N^nLO}(\qt\neq0) \equiv \mathd\sigma^{\sss
  F+{\rm jets}}_{\rm N^{\sss n-1}LO}\,,\qquad n=1,2,3 \,.
\end{equation}
implying that, if $\qt\neq0$, the IR divergences
appearing in the computation of $\mathd\sigma^{\sss F}_{\sss\rm
  N^{n}LO}(\qt\neq0)$ are those already present in $\mathd\sigma^{\sss F+{\rm
    jets}}_{\sss\rm N^{n-1}LO}$.

The IR singularities involved in $\mathd\sigma^{\sss F}_{\sss\rm N^nLO}(\qt
\neq0)$ can be handled and cancelled by the available subtraction methods at
N$^{n-1}$LO. The only remaining singularities at N$^n$LO are associated with
the limit $\qt\to0$ and are treated with the $\qt$-subtraction method.\footnotew{
This point is a great advantage for $\qt$-subtraction, since the method profits from lower-order results. However, this also alters the specific IR behaviour of the contributions, preventing a fully local cancellation.
}
Since the small-$\qt$ behaviour of the transverse-momentum cross section is well
known through the resummation program of logarithmically-enhanced
contributions to transverse-momentum distributions, see
Refs.~\cite{Dokshitzer:1978yd, Dokshitzer:1978hw, Parisi:1979se,
  Curci:1979bg, Collins:1981uk, Kodaira:1981nh, Kodaira:1982az,
  Collins:1984kg, Catani:1988vd, deFlorian:2000pr}, we exploit this knowledge
to build the necessary counterterms in order to subtract the remaining
singularity, thus promoting the $\qt$-subtraction method proposed in
Refs.~\cite{Catani:2007vq,Bozzi:2005wk,Bozzi:2003jy} to N$^3$LO.

The sketchy form of the $\qt$-subtraction method for the N$^n$LO cross
section, see Ref.~\cite{Catani:2007vq}, is
\begin{equation}
 \label{masterEqqT}
 \mathd\sigma^{\sss F}_{\sss\rm N^nLO} = {\cal H}^{\sss F}_{\sss\rm N^nLO} \otimes
 \mathd\sigma^{\sss F}_{\rm LO} + \lq \mathd\sigma^{\sss F+{\rm jets}}_{\sss \rm N^{n-1}LO} -
   \mathd\sigma^{\sss F\;{\rm CT}}_{\sss\rm N^nLO} \rq , \qquad n=1,2,3\,,
 \end{equation}
where $\mathd\sigma^{\sss F\;{\rm CT}}_{\sss\rm N^nLO}$ is the contribution
of the counterterm to the N$^n$LO cross section which cancels the divergences
of $\mathd\sigma^{\sss F+{\rm jets}}_{\rm N^{n-1}LO}$ in the limit $\qt
\to0$. The $n$-order counterterm can be written as
\begin{equation}
  \mathd\sigma^{\sss F\;{\rm CT}}_{\sss\rm N^nLO} = \Sigma^{\sss F}_{\rm
    N^nLO} \lp\frac{\qt^2}{M^2}\rp \mathd^2{\bf\qt} \otimes
  \mathd\sigma^{\sss F}_{\sss\rm LO}\;,
\end{equation}
where the symbol $\otimes$ stands for convolutions over momentum fractions
and sum over flavour indices of the partons.  More precisely, the function
$\Sigma^{\sss F}_{\sss\rm N^nLO}(\qt^2/M^2)$ is the $n$-order truncation of
the perturbative series in $\as$
\begin{equation}
  \Sigma^{\sss F}_{c\bar{c}\leftarrow a_{1}a_{2}}
  \!\lp\frac{\qt^2}{M^2}\rp = \sum_{n=1}^\infty \asopi^n \Sigma^{\sss
    F;(n)}_{c\bar{c}\leftarrow a_{1}a_{2}} \!\lp\frac{\qt^2}{M^2}\rp,
\end{equation}
where the labels $a_{1}$ and $a_{2}$ stands for the partonic channels of the
N$^n$LO correction to the Born cross section ($\mathd\sigma^{\sss F}_{\rm LO}
\equiv \mathd\lq\sigma^{\sss F;(0)}_{c\bar{c}}\rq$). Notice that at LO the
only available configuration is $a_{1}=c$ and $a_{2}=\bar{c}$, where
$c\bar{c}$ is (are) the partonic channel(s) at which the LO cross section is
initiated.  The function $\Sigma^{\sss F}(\qt^2/M^2)$ embodies all the
logarithmic terms that are divergent in the limit $\qt\to0$, reproducing the
singular behaviour of $\mathd\sigma^{\sss F+{\rm jets}}$ in the small-$\qt$
limit.\footnotew{
These counterterms have a universal structure, and are local in the variable $q_T$.
}
The counterterm is defined free of terms proportional to
$\delta(\qt^2)$, which are all considered in the perturbative factor ${\cal
  H}^{\sss F}$. The hard coefficient function ${\cal H}^{\sss F}_{\sss\rm
  N^nLO}$, that encodes all the IR finite terms of the $n$-loop
contributions, is obtained by the N$^n$LO truncation of the perturbative
function
\begin{equation}
\label{Hstexpand}
{\cal H}^{\sss F}_{c\bar{c}\ito a_{1}a_{2}}(z;\as) =\delta_{c\,a_{1}}
\,\delta_{\bar{c}\,a_{2}}\;\delta(1-z)+
\sum_{n=1}^{\infty}\lp\asopi\rp^n {\cal
  H}^{\sss F;(n)}_{c\bar{c}\ito a_{1}a_{2}}(z) \;\; ,
\end{equation}
where $z=M^2/s$.\footnotew{
The definition of the hard coefficients requires the computation of 
the virtual matrix elements in $d$-dimensions, in order to explicitly remove the poles. Thus, the real-virtual cancellation of singularities is not fully local, as in \FDU.
}
According to the transverse momentum resummation formula,
see Eq.~(10) of Ref.~\cite{Bozzi:2005wk}, and using the Fourier
transformation between the conjugate variables ${\bf\qt}$ and $b$ (the impact
parameter), the perturbative hard function ${\cal H}^{\sss F}$ and the
counterterm are obtained by the fixed order truncation of the following
identity
\begin{eqnarray}
\label{reslean}
&&\!\!\!\!\!\!\!\lq \Sigma^{\sss F}_{c\bar{c}\leftarrow a_{1}a_{2}} \lp\frac{\qt^2}{M^2}\rp
+ {\cal H}^{\sss F}_{c\bar{c}\leftarrow a_{1}a_{2}} \lp\frac{M^2}{s};\,\as\rp \rq
\otimes \mathd\lq\hat{\sigma}^{\sss F;(0)}_{c\bar{c}} \rq = 
\nn\\
&=& \frac{M^2}{s}\, \int_0^\infty \mathd b \, \frac{b}{2}\, J_0(b\,{\qt}) \,
S_c(M,b) \nn
\nn\\
&& \times \int_{x_1}^1 \frac{\mathd z_1}{z_1} \, \int_{x_2}^1 \frac{\mathd
  z_2}{z_2} \,
\mathd\hat{\sigma}^{\sss F;(0)}_{c\bar{c}} \,
f_{a_1/h_1}\!\lp \frac{x_1}{z_1},\,\frac{b_0^2}{b^2}\rp
f_{a_2/h_2}\!\lp \frac{x_2}{z_2},\,\frac{b_0^2}{b^2}\rp
\otimes \lq H^{\sss F}\,C_1\,C_2 \rq_{c\bar{c};a_{1}a_{2}}\,,
\nn\\
\end{eqnarray}
where $J_0(b\,{\bf\qt})$ is the 0th-order Bessel function, $f_{c/h}$
corresponds to the distribution of a parton $c$ in a hadron $h$ and $b_0=2
e^{-\gamma_{\sss\rm E}}$ ($\gamma_{\sss\rm E}=0.5772...$ is the Euler number). The symbolic
factor $\mathd\hat{\sigma}^{\sss F;(0)}_{c\bar{c}}$ for the partonic Born
cross section $\hat{\sigma}^{\sss F;(0)}_{c\bar{c}}$ denotes
% $\mathd\big[\sigma^{F;(0)}_{c\bar{c}}\big]$ 
\begin{equation}
\mathd\hat{\sigma}^{\sss F;(0)}_{c\bar{c}}\equiv
\frac{\mathd\hat{\sigma}^{\sss F;(0)}_{c\bar{c}}}{\mathd\phi}\,,
\end{equation}
where $\phi$ represents the phase space of the final-state system $F$. 
In the l.h.s. of Eq.~(\ref{reslean}) the convolution (as well as the sum
over the flavour indices of the partons) between the resummation functions
$\Sigma^{\sss F}_{c\bar{c}}$ and ${\cal H}^{\sss F}_{c\bar{c}}$, the partonic Born
cross section and the parton distributions is symbolically denoted by
$\otimes \,\mathd\lq\hat{\sigma}^{\sss F;(0)}_{c\bar{c}}\rq$.

The large logarithmic corrections are exponentiated in the Sudakov form
factor $S_c(M,b)$ of the quark ($c=q, {\bar q}$) or of the gluon ($c=g$),
that has the following expression
\begin{equation}
\label{formfact}
S_c(M,b) = \exp \lg - \int_{b_0^2/b^2}^{M^2} \frac{\mathd q^2}{q^2} 
\lq A_c\lp\as\lp q^2\rp\rp \, \log\frac{M^2}{q^2} + B_c\lp\as\lp q^2\rp\rp \rq\rg. 
\end{equation}
The functions $A$ and $B$ in Eq.~(\ref{formfact}) are perturbative series in
$\as$:
\begin{eqnarray}
\label{aexp}
A_c(\as) &=& \sum_{n=1}^\infty \lp\asopi\rp^n A_c^{(n)} \,,
\, \\
\label{bexp}
B_c(\as) &= &\sum_{n=1}^\infty \lp\asopi\rp^n B_c^{(n)} \,.
\end{eqnarray}
The structure of the symbolic factor denoted by $\lq H^{\sss F=H} \,C_1\,C_2
\rq_{c\bar{c};a_{1} a_{2}}$, that strongly depends on the initial-state
channel of the Born subprocess, is explained with detail in
Refs.~\cite{Catani:2010pd, Catani:2013tia}.

\subsection{Higher-order power corrections at NLO}
There are subtraction methods which are independent of any regularising
parameter and proceed by building local counterterms and point-wisely
subtracting the IR divergences along the phase space -- thus they are
mentioned to be {\it local}, while other, such as the $\qt$-subtraction
method, introduce a regularising, or {\it slicing}, parameter, i.e.~a cutoff
scale, in order to separate different IR regions.\footnotew{
The main advantage of $\qt$-subtraction is its universality for achieving a cancellation of singularities, which allow to apply the method to several processes up to NNLO. The local structure of the required counterterms is much more complicated than the one obtained within this formalism.
}

Such separation of the phase space introduces instabilities in the numerical
evaluation of cross sections and differential
distributions~\cite{Catani:2011qz, Catani:2018krb, Grazzini:2017mhc,
  Boughezal:2016wmq}, and some care has to be taken in order to obtain stable
and reliable results. Furthermore, the knowledge of logarithmic and
power-correction terms in the cutoff plays a relevant role in the
identification of universal structures, in the development of regularisation
prescriptions and in resummation programs~\cite{Dokshitzer:1978yd,
  Dokshitzer:1978hw, Parisi:1979se, Curci:1979bg, Collins:1981uk,
  Kodaira:1981nh, Kodaira:1982az, Collins:1984kg, Catani:1988vd,
  deFlorian:2000pr, Alioli:2015toa}.

In Ref.~\cite{Cieri:2019tfv} a study was conducted about how power
corrections in the cutoff may affect the application of the $\qt$-subtraction
method to the production of a colourless system at next-to-leading order in
the strong coupling constant $\as$ -- in particular, to Drell--Yan vector (V)
 and Higgs (H) boson production in gluon fusion at NLO in QCD, in the
infinite top-mass limit.

In fact, the singular terms in the small-cutoff limit are universal and are
cancelled by the application of the $\qt$-subtraction (or other methods),
while finite and vanishing terms are, in general, process dependent and thus,
after the subtraction procedure, a residual dependence on the cutoff remains
as power corrections.  While these terms formally vanish in the null cutoff
limit, they give a non-zero numerical contribution for any finite choice of
the cutoff.

If one is able to take into account such terms, not only our understanding of
the perturbative behaviour of QCD cross sections increases from a theoretical
point of view, but also the numerical implementation of the subtraction
becomes more robust, since the power terms weaken the dependence of the final
result on the arbitrary cutoff. Notice that this becomes more relevant from a
numerical point of view, when applied to higher-order calculation, as pointed
out, for example, in the evaluation of NNLO cross sections in
Refs.~\cite{Grazzini:2017mhc, Boughezal:2016wmq}.

Power corrections at NLO have been extensively studied in refs.~\cite{Moult:2016fqy, Boughezal:2016zws, Boughezal:2018mvf, Moult:2017jsg, Ebert:2018lzn, Bhattacharya:2018vph, Campbell:2019gmd, Moult:2018jjd, Boughezal:2019ggi, Ebert:2019zkb} both for $N$-jettiness and transverse momentum distributions, in the context of the $N$-jettiness subtraction method, and in refs.~\cite{Bauer:2000ew, Bauer:2000yr, Bauer:2001ct, Bauer:2001yt, Bauer:2002aj, Moult:2019mog} within SCET-based subtraction methods. Power corrections at NLO for the transverse momentum of a colour singlet have been derived for the first time at differential level in ref.~\cite{Ebert:2018gsn} within the SCET framework. This study has been followed, with a different method, by ref.~\cite{Cieri:2019tfv}, which is considered in more detail in the following, and by ref.~\cite{Buonocore:2019puv}, which among other new results was able to confirm the former. A numerical extraction of power corrections in the context of NNLL'+NNLO calculations was done in $N$-jettiness~\cite{Alioli:2015toa}, and a general discussion in the context of the fixed-order implementation of the $N$-jettiness subtraction can be found in ref.~\cite{Gaunt:2015pea}.

\subsection{Power corrections for V and H production at NLO in QCD}
In Ref.~\cite{Cieri:2019tfv} it is considered the production of a colourless
system $F$ of squared invariant mass $Q^2$ plus a coloured system $X$ at a
hadron collider
\begin{equation}
\label{eq:proc}
h_1+h_2 \to F+X \,.
\end{equation}
The hadronic cross section can be written as
\begin{eqnarray}
  \sigma &=& \sum_{a,b} \int_\tau^1 \mathd x_1 \int_\frac{\tau}{x_1}^1 \mathd
  x_2\,
  f_a\!\lp x_1\rp f_b\!\lp x_2\rp \int  \mathd \qt^2\, \mathd z \,\dsigdqtz\,
  \delta\!\lp\! z- \frac{Q^2}{s}\rp,
  \nonumber\\
  %% &=& \sum_{a,b} \tau \int_\tau^1 \frac{dz}{z}
  %% \int_\frac{\tau}{z}^1 \frac{dx_1}{x_1}\, f_a\!\lp x_1\rp f_b
  %% \!\lp\frac{\tau}{z\,x_1}\rp \frac{1}{z} \int\! d\qt^2 \, \dsigdqtz
  %% \nonumber\\
  \label{eq:hadro}
  &=&  \sum_{a,b} \tau \int_\tau^1 \frac{\mathd z}{z}  \, {\cal L}_{ab}\!
  \lp\frac{\tau}{z}\rp \frac{1}{z} \int\! \mathd \qt^2 \,  \dsigdqtz \,,
\end{eqnarray}
where 
\begin{equation}
\label{eq:tau_def}
\tau = \frac{Q^2}{S}\,, \qquad z = \frac{Q^2}{s}\,,
\end{equation}
$f_{a/b}$ are the parton densities of the partons $a$ and $b$, in the hadron
$h_1$ and $h_2$ respectively, $S$ is the hadronic squared centre-of-mass
energy, $s$ is the partonic squared centre-of-mass energy, equal to
\begin{equation}
\label{eq:s_S}
s = S \, x_1\, x_2\,,
\end{equation}
$\mathd \hat\sigma_{ab}$ is the partonic cross section for the process $a+b \to F +
X$, $\qt$ is the transverse momentum of the system $F$ with respect to the
hadronic beams and the luminosity function is defined as
\begin{equation}
  {\cal L}_{ab}(y) \equiv \int_y^1 \frac{\mathd x}{x}\, f_a\!\lp x\rp f_b\!
  \lp\frac{y}{x}\rp .
\end{equation}
The dependence on the renormalisation and factorisation scales and on the
other kinematic invariants of the process are implicitly assumed. \\

In the small-$\qt$ region, i.e.~$\qt\ll Q$, the real contribution to the
perturbative partonic cross sections appearing in Eq.~(\ref{eq:hadro})
contains well-known logarithmically-enhanced terms that are singular in the
$\qt\to 0$ limit~\cite{Dokshitzer:1978yd, Dokshitzer:1978hw, Parisi:1979se,
  Curci:1979bg, Collins:1981uk, Kodaira:1981nh, Kodaira:1982az,
  Collins:1984kg, Catani:1988vd, deFlorian:2000pr}.
%
%% In the context of inclusive NLO fixed-order calculations, the logarithmic
%% terms are cancelled when using the subtraction prescriptions. For more
%% exclusive quantities, such as the transverse-momentum distribution of the
%% colourless system, the same logarithmic terms need to be resummed at all
%% orders in the strong coupling constant to produce reliable results.
The general structure of the power-correction terms, at variance with that of
the singular logarithms, is unknown. Thus, it is useful to inquire about it,
in order to find out whether it can as well be derived a universal structure,
or whether at least part of it follows a universal behaviour in connection
with its infrared limit.

On the other hand, in order to actually extract the power corrections, the
starting point is the real contribution at small $\qt$ to the processes,
first at parton level:
\begin{equation}
  \hat\sigma_{ab}^{\sss  <}(z) \equiv
  \int_0^{\lp\qtcut\rp^2}\!\! \mathd \qt^2 \,\frac{\mathd \hat\sigma_{ab}(\qt,z)}{\mathd \qt^2} \,.
\end{equation}
Since in this case the total cross section is analytically known, one may
refer to the above-$\qtcut$ region
\begin{equation}
  \label{eq:above}
  \hat\sigma_{ab}^{\sss >}(z) = \int_{\lp\qtcut\rp^2}^{\lp\qtmax\rp^2}\!\!
  \mathd \qt^2\,\frac{\mathd \hat\sigma_{ab}(\qt,z)}{\mathd \qt^2} \,,
\end{equation}
where $\qtmax$ is the maximum value for $\qt$ allowed by the kinematics, and
derive the the below-$\qtcut$ contribution as a difference.

At hadron level, when a cut on the transverse momentum is imposed, the
reality of the parton-level cross sections restricts the $z$-integration
\begin{eqnarray}
  \label{eq:Rabmin}
  \sigma_{ab}^{\sss <} &=& \tau \int_\tau^{1-f(a)} \frac{\mathd z}{z}  \, {\cal
    L}_{ab}\!\lp\frac{\tau}{z}\rp  
  \frac{1}{z} \,\hat\sigma_{ab}^{\sss <}(z)
  \nonumber\\
  &\equiv&
  \tau \int_\tau^1 \frac{\mathd z}{z}  \, {\cal L}_{ab}\!\lp\frac{\tau}{z}\rp 
  \hat{\sigma}^{\sss (0)} \, \hat{R}_{ab}(z) \,,
\end{eqnarray}
where
\begin{equation}
  a = \frac{\qt^2}{Q^2} 
\end{equation}
is the chosen basis for presenting the power corrections and
$\hat{\sigma}^{\sss (0)}$ is the partonic Born-level cross section for the
production of the colourless system $F$.

In the second line of Eq.~(\ref{eq:Rabmin}) the $z$-integration limit is
extended to one, aiming to make contact with the transverse-momentum
subtraction formulae, that describe the behaviour of the cross sections in
the soft and collinear limits, namely a Born-like kinematics is required. The
function $ \hat{R}_{ab}(z)$ is defined within this purpose and admits a
perturbative expansion in $\as$, whose coefficient functions $\hat
R_{ab}^{\sss(n)}(z)$ can be computed as power series in $a$. Here the
interest is driven to the NLO coefficient, whose form is well-known in
literature~\cite{Catani:2011kr} up to the vanishing power-correction terms
\begin{equation}
  \label{eq:Rab}
  \hat{R}_{ab}^{\sss(1)}(z) = \log^2(a) \, \hat R_{ab}^{\sss(1,2,0)}(z) +
  \log(a) \, \hat R_{ab}^{\sss(1,1,0)}(z) + \hat R_{ab}^{\sss(1,0,0)}(z) +
      {\cal O}\!\lp a^\frac{1}{2} \log{a}\rp .
\end{equation}
Notice that in Refs.~\cite{Catani:2011kr, Bozzi:2005wk}, the following
associations hold
\begin{eqnarray}
  R_{ab}^{(1,2,0)}(z) \quad &\leftrightarrow& \quad \Sigma_{c \bar{c} \leftarrow
    ab}^{F(1;2)}(z)
  \\
  R_{ab}^{(1,1,0)}(z) \quad &\leftrightarrow& \quad \Sigma_{c \bar{c} \leftarrow
    ab}^{F(1;1)}(z)
  \\
  {\cal H}_{c \bar{c} \leftarrow ab}^{F(1)}(z) \quad &\leftrightarrow& \quad
  R_{ab}^{(1,0,0)}(z) \,.
\end{eqnarray}
The aim of Ref.~\cite{Cieri:2019tfv} is the computation of the missing orders
in Eq.~(\ref{eq:Rab}). In order to achieve such result, in a way similar to
what is showed in Eq.~(\ref{eq:above}),\footnote{The same method was used in
  Refs.~\cite{Catani:2011kr, Catani:2012qa}, at leading power in $a$, to
  extract the soft constant of the $\qt$-subtraction hard function and the
  second-order collinear coefficient functions for the $\qt$-resummation.}
the function $\hat{G}_{ab}^{\sss}(z)$ is introduced via the definition
\begin{equation}
  \sigma_{ab}^{\sss >} =
 \tau \int_\tau^{1-f(a)} \frac{\mathd z}{z}  \, {\cal L}_{ab}\!\lp\frac{\tau}{z}\rp 
 \frac{1}{z} \,\hat\sigma_{ab}^{\sss >}(z)
 \equiv
  \tau \int_\tau^1 \frac{\mathd z}{z}  \, {\cal L}_{ab}\!\lp\frac{\tau}{z}\rp 
  \hat{\sigma}^{\sss (0)} \, \hat{G}_{ab}(z)\,.
\end{equation}
At first order in $\as$ it holds
\begin{equation}
 \label{eq:Gab1}
 \sigma_{ab}^{\sss > (1)} = \tau \int_\tau^{1-f(a)} \frac{\mathd z}{z}  \, {\cal
   L}_{ab}\!\lp\frac{\tau}{z}\rp  
 \frac{1}{z} \,\hat\sigma_{ab}^{\sss >(1)}(z)=
 \tau \int_\tau^1 \frac{\mathd z}{z}  \, {\cal L}_{ab}\!\lp\frac{\tau}{z}\rp 
  \hat{\sigma}^{\sss (0)} \, \hat{G}_{ab}^{\sss(1)}(z)  \,.
\end{equation}
and a process-independent formula is elaborated in the paper in order to
transform an integral of the form of the first one in Eq.~(\ref{eq:Gab1})
into the form of the second one, producing the series expansion of
$\hat{G}_{ab}^{\sss(1)}(z)$ in $a$. Also, the procedure enables to reach any order
in the transverse momentum cut-off.

The results are lengthy and the reader is referred to the original
paper. Here, it will suffice to remember that, for the calculation of these
functions, all the terms originating from the manipulation of the
contributions proportional to the Altarelli--Parisi splitting functions at
the level of the partonic cross sections constitute the so-called ``universal
part'' of the results.

The general form of the $\hat{G}_{ab}^{\sss(1)}(z)$ functions reads
\begin{eqnarray}
  \label{Ghat_one_expans}
\hat{G}_{ab}^{\sss(1)}(z) &=& {} \log^2(a) \, \hat G_{ab}^{\sss(1,2,0)}(z) +
\log(a)  \, \hat  G_{ab}^{\sss(1,1,0)}(z) + \hat G_{ab}^{\sss(1,0,0)}(z)
\nonumber\\[2mm]
&& {} +   a\log(a) \,\hat G_{ab}^{\sss(1,1,2)}(z)  +  a \, \hat G_{ab}^{\sss(1,0,2)}(z)
\nonumber\\[2mm]
&& {} + a^2 \log(a) \,\hat G_{ab}^{\sss(1,1,4)}(z)  + a^2 \, \hat G_{ab}^{\sss(1,0,4)}(z)  +
{\cal O}\!\lp a^\frac{5}{2}\log(a)\rp ,
\end{eqnarray}
all the other coefficients being zero. The terms in the first line of
Eq.~(\ref{Ghat_one_expans}) are referred to as as leading terms~(LT). These
terms are either logarithmically divergent or finite in the $a\to 0$ limit.
The terms in the sum in the second line of Eq.~(\ref{Ghat_one_expans}) are
referred to as next-to-leading terms~(NLT), the first two terms in the
third line as next-to-next-to-leading terms~(N$^2$LT), and so forth.

The results display some important features:
\begin{enumerate}[i)]
  \item no odd-power corrections of $\sqrt{a} = \qtcut/Q$ appear in the NLT
    and N$^2$LT terms;
  \item the NLT and N$^2$LT terms are at most linearly dependent on
    $\log(a)$;
  \item the non-universal contribution in fact appears to be highly dependent
    on the process at stake, thus making impossible a generalisation of the
    procedure to any process.
\end{enumerate}

\subsection{Discussion}

Although there is not a general proof, what is found for the inclusive cross
section expanded up to $(\qtcut)^4$, i.e.~the absence of odd-power
corrections in $\qtcut$, is thought to be valid even at higher orders. One is
not to expect this to be true, in general, for more exclusive quantities --
to this regard, see also Ref.~\cite{Ebert:2019zkb}.

Aside from this, it is useful to remark the importance of the knowledge of
power-corrections terms within the $\qt$-subtraction method.

In the original paper~\cite{Catani:2007vq}, the expansion in $\as$ of the
transverse-momentum resummation formula generates exactly the three terms in
Eq.~(\ref{eq:Rab}), plus extra power-correction terms. In the formula for
$\hat{R}_{ab}^{\sss (1)}(z)$ that one can build from the new expression of
$\hat{G}_{ab}^{\sss (1)}(z)$, by changing the overall sign and adding the
$\delta(1-z)$ contribution from the virtual correction, the power-correction
terms are exactly those produced by the expansion of the real amplitudes.
If one is interested in using the formula for $\hat{R}_{ab}^{\sss (1)}(z)$ to
reduce the dependence on the transverse-momentum cutoff, within the
${\qt}$-subtraction method, the aforementioned extra terms need then to be
subtracted from our expression of $\hat{R}_{ab}^{\sss (1)}(z)$.

On the other hand, the knowledge of power terms is also crucial for
understanding both the non-trivial behaviour of cross sections at the
boundaries of the phase space, and the resummation structure at subleading
orders. At the same time, within the $\qt$-subtraction method, the knowledge
of the power terms helps in reducing the cutoff dependence of the cross
sections.

While the application of the $\qt$-subtraction method in NLO calculations is
superseded by well-known local subtraction methods, at NNLO it still plays a
major role, also in view of the fact that, as shown in
Refs.~\cite{Grazzini:2017mhc, Cieri:2018oms}, the sensitivity to the
numerical value of the cutoff increases at higher orders. This also explains
why it is of interest the calculation of the power corrections to an NNLO
cross section. 
%Indeed, there is an on-going project that is devoted to a fully analytic computation of the subleading-power terms in a transverse-momentum cutoff for $Z$ production at NNLO~\cite{Cieri:202*}, with the parallel aim of finding universal patterns within them. Hence, the new terms could be used in order to build a local version of the $\qt$-subtraction method.

\graphicspath{{antenna/}}
\section{Antenna subtraction scheme}
\label{sec:antenna}
In this section, we present the antenna subtraction scheme for perturbative QCD calculations. This method has been derived 
in~\cite{GehrmannDeRidder:2005cm} and successfully applied to the calculation of the NNLO corrections to 3-jet production and related event shape observables in electron-positron annihilation 
in~\cite{GehrmannDeRidder:2007jk}. The extension of the scheme to the treatment of initial state radiation relevant for calculations of jet observables in hadronic collisions at LHC has been
established at NLO in~\cite{Daleo:2006xa} and at NNLO 
in~\cite{Daleo:2009yj,Glover:2010im,Boughezal:2010mc,Gehrmann:2011wi,GehrmannDeRidder:2011aa,GehrmannDeRidder:2012ja,Ridder:2012dg,Currie:2013vh}.
In a first subsection we review the NLO version of the scheme followed by the generalisation to NNLO. In the last subsection we provide our conclusions on the status of the
method and its current implementation for precision phenomenological studies at the LHC.

\subsection{Antenna subtraction at NLO}
\label{sec:nlosub}
To specify the notation, we define the LO contribution to an $m$-jet
cross section by,
\begin{equation}
{\rm d}\sigma_{LO}=\int_{{\rm d}\Phi_{m}}{\rm d}\sigma^{B}\;J_{m}^{(m)}(\{p_{m}\})
\end{equation}
where the partonic cross section ${\rm d}\sigma^{B}$ is related to the square of the tree level amplitude of the process, 
integrated over the appropriate $m$-particle phase space ${\rm d}\Phi_{m}$, subject to the kinematical constraint that precisely $m$-jets are observed. The latter constraint is imposed 
by the jet function $J_{m}^{(m)}(\{p_{m}\})$, that at this order selects $m$-jets from $m$-final state particles within the four-momentum set $\{p_{m}\}$ using an IR safe jet-algorithm.

At NLO, we consider the following $m$-jet cross section,
\begin{equation}
{\rm d}\sigma_{NLO}=\int_{{\rm d}\Phi_{m}}{\rm d}\sigma^{V}\;J_{m}^{(m)}(\{p_{m}\})+\int_{{\rm d}\Phi_{m+1}}{\rm d}\sigma^{R}\;J_{m}^{(m+1)}(\{p_{m+1}\})
\label{eq:nlo}
\end{equation}
where ${\rm d}\sigma^{V}$ is the UV-renormalised one loop virtual correction to the $m$-parton Born cross section ${\rm d}\sigma^{B}$, and ${\rm d}\sigma^{R}$ is the tree-level
squared amplitude for a single real radiation emission from the Born process. 

Although the sum in Eq.~\eqref{eq:nlo} is finite in $d=4$ dimensions, each of the two integrals is separately divergent if $d=4$. Using dimensional regularisation with
space-time dimension equal to $4-2\epsilon$, the divergences (arising from the integration over the loop-momentum in  ${\rm d}\sigma^{V}$) appear as \emph{explicit}
double $1/\epsilon^{2}$ and single $1/\epsilon$ poles. On the other hand, the real correction ${\rm d}\sigma^{R}$ being finite in $d=4$, has singularities when it is integrated over 
the phase space regions corresponding to soft and collinear emission which are allowed by the jet function $J_{m}^{(m+1)}(\{p_{m+1}\})$, which selects $m$-jets from an $(m+1)$ particle 
phase space. It is precisely the contribution of unresolved emission to the $m$-jet cross section from the real correction that generates the \emph{implicit} IR singularities in this contribution. 

Given that the individual contributions in Eq.~\eqref{eq:nlo} live in phase spaces of different dimensionality and in particular, both contribute to the evaluation of an arbitrary
observable, which often requires the imposition of arbitrary sets of experimental cuts on the phase space integration, it is necessary that the IR singularities must be cancelled 
prior to any numerical calculation.

The antenna subtraction method is a subtraction procedure which allows for the isolation of the infrared singularities present in intermediate steps of higher-oder perturbative QCD calculations. 
The procedure consists in
adding and subtracting a \emph{countertem} that reproduces the singular behaviour of the real correction, that is simple enough 
that it can be integrated analytically in the single-radiative phase space
and combined with the virtual contribution. The NLO cross section becomes,
\begin{eqnarray}
{\rm d}\sigma_{NLO}&=&\int_{{\rm d}\Phi_{m}}\bigg({\rm d}\sigma^{V}\;J_{m}^{(m)}(\{p_{m}\})+\int_{1}{\rm d}\sigma^{S}\;J_{m}^{(m)}(\{p_{m})\}\bigg)\nonumber\\
&+&\int_{{\rm d}\Phi_{m+1}}\bigg({\rm d}\sigma^{R}\;J_{m}^{(m+1)}(\{p_{m+1}\})-{\rm d}\sigma^{S}\;J_{m}^{(m)}(\{\tilde{p}_{m}\})\bigg)\,.
\label{eq:nlo2}
\end{eqnarray}
The contribution ${\rm d}\sigma^{S}$ in Eq.~\eqref{eq:nlo2} is a counterterm which reproduces the same singular divergent behaviour as the real emission matrix element 
${\rm d}\sigma^{R}$ in all appropriate limits. In particular, for an IR-safe observable in a singular soft or collinear phase space region, the following conditions are satisfied, 
\begin{eqnarray}
J_{m}^{(m+1)}(\{p_{m+1}\}) &\to& J_{m}^{(m)}(\{p_{m}\})\nonumber\\
J_{m}^{(m)}(\{\tilde{p}_{m}\}) &\to&  J_{m}^{(m)}(\{p_{m}\})\nonumber\\
{\rm d}\sigma^{S}&\to&{\rm d}\sigma^{R}\;,
\label{eq:nlo3}
\end{eqnarray}
such that the bottom line in Eq.~\eqref{eq:nlo2} can be integrated numerically in four dimensions. 

We note that the first condition in Eq.~\eqref{eq:nlo3} is satisfied automatically for all IR-safe observables,
while the remaining conditions are enforced by the subtraction scheme. In particular, any QCD amplitude with the emission of one unresolved parton in ${\rm d}\sigma^{R}$ can be written as a product 
of the Born amplitude times a soft and collinear factor which contains all the singular terms. As it will be shown below, the antenna subtraction counterterms will have the same 
factorised structure and employ a remapping of the real emission phase $\{p_{m+1}\}\to\{\tilde{p}_{m}\}$ space that preserves the on-shellness and momentum conservation in 
the underlying Born configuration in the counterterm contribution. This guarantees that the remaining conditions in~\eqref{eq:nlo3} are satisfied. Finally, the counterterm contribution ${\rm d}\sigma^{S}$ 
has to be integrated analytically over all singular regions of the 1-parton radiative subspace, leading to explicit $1/\epsilon$ poles that can be combined with the virtual contribution in Eq.~\eqref{eq:nlo2},
thus cancelling all the divergences and allowing the remaining numerical integration over the $m$-parton phase space in the first line of Eq.~\eqref{eq:nlo2} to be performed in $d$=4 dimensions.

A key characteristic in the antenna subtraction scheme is the subtraction of the infrared singularities following the singularity structure in colour-ordered amplitudes. In a given colour basis, QCD amplitudes
decompose into leading and subleading colour contributions with the singularities in colour-ordered amplitudes only occurring between colour adjacent partons. In this way, a NLO real emission squared tree-level 
colour ordered amplitude factorises as,
\begin{eqnarray}
|{\cal M}^0_{m+1}(1,\ldots,i,j,k,\ldots,m+1)|^2&\stackrel{j_g\to0}{\longrightarrow}&S_{ijk}\;|{\cal M}_m^{0}(1,\ldots,i,k,\ldots,m+1)|^2,
\label{eq:ssoft}
\end{eqnarray}
when gluon $j$ is soft between colour adjacent partons $i$ and $k$, with the singular eikonal factor given by,
\begin{eqnarray}
S_{ijk}=\frac{2s_{ik}}{s_{ij}s_{jk}}\quad\textrm{with}\quad s_{ij}=(p_{i}+p_{j})^2\label{eq:seik}\;.
\end{eqnarray}
Similarly in the limit where a quark and gluon pair become collinear, the colour-ordered amplitudes factorise. If quark $i$ and gluon $j$ become collinear and form quark $k$, then the colour adjacent $i,j$ pair gives 
a singular contribution,
\begin{eqnarray}
|{\cal M}^0_{m+1}(1,\ldots,i,j,\ldots,m+1)|^2&\stackrel{i//j}{\longrightarrow}&\frac{1}{s_{ij}}P_{qg\to q}(z)
|{\cal M}^0_m(1,\ldots,k,\ldots,m+1)|^2\label{eq:collimit}
\end{eqnarray}
while a separated quark/gluon pair does not,
\begin{eqnarray}
|{\cal M}^0_{m+1}(1,\ldots,i,\ldots,j,\ldots,m+1)|^2&\stackrel{i//j}{\longrightarrow}&\textrm{finite}.
\end{eqnarray}
In Eq.~(\ref{eq:collimit}), $z$ is the fraction of momentum carried by one of the collinear partons and the collinear splitting function $P_{qg\to q}$ is given by,
\begin{eqnarray}
P_{qg\to q}(z)=\left(\frac{1+(1-z)^2-\epsilon z^2}{z}\right).\label{eq:Pqg}
\end{eqnarray}

At NLO with one unresolved emission, the only kinematical configurations that generate IR singular contributions in the real emission tree-level squared amplitudes 
in ${\rm d}\sigma^{R}$ are the configurations corresponding to a single soft or single collinear emission. Looking at Eqs.~\eqref{eq:ssoft} and~\eqref{eq:collimit} we observe that in these limits,
the real emission amplitudes obey a factorisation formula in terms of universal singular factors multiplied by a born-like reduced matrix element. The basic idea of the antenna subtraction 
approach is to derive the subtraction terms with antenna functions which encapsulate all singular limits due to the emission of unresolved partons between two colour-connected hard partons. 
The full antenna subtraction term is then obtained by summing products of antenna functions with reduced matrix elements over all possible unresolved configurations. At NLO the subtraction term reads,
\begin{equation}
{\rm d}\sigma^{S}=\sum_{j}\,X_{ijk}^{0}\;|{\cal M}_m^{0}(1,\ldots,\tilde{I},\tilde{K},\ldots,m+1)|^2.
\label{eq:NLOsubt}
\end{equation}
In Eq.~\eqref{eq:NLOsubt}, $X_{ijk}^{0}$ is a tree-level three parton antenna, derived from a properly normalised physical matrix element that smoothly interpolates 
the single soft and single collinear configurations. In the subtraction term, the particles $\tilde{I}$ and $\tilde{K}$, form a colour connected hard antenna that radiated particle $j$. In doing so, 
the momenta of the radiators change to form particles $i$ and $k$. Depending on the flavour of the pair of hard radiators, the antennae can be quark-antiquark antennae, quark-gluon antennae, 
or gluon-gluon antennae. As an example, the quark-antiquark antennae can be derived from the decay of a virtual photon into a quark-antiquark pair $\gamma^*\to q\bar{q}$+(partons). For the
quark-gluon-antiquark final state the corresponding antenna is:
\begin{equation}
A_3^0(1_q,3_g,2_{\bar{q}})=\frac{1}{s_{123}}\left(\frac{s_{13}}{s_{23}}+\frac{s_{23}}{s_{13}}+\frac{2s_{12}s_{123}}{s_{13}s_{23}}\right)+{\cal O}(\epsilon)\;,
\end{equation}
which in the IR limits reproduces the universal soft and collinear singularities of tree-level QCD matrix elements,
\begin{eqnarray}
A_3^0(1,3,2)&\stackrel{3_g\to 0}{\longrightarrow}&S_{132}\;,\\
A_3^0(1,3,2)&\stackrel{1_q//3_g}{\longrightarrow}&\frac{1}{s_{13}}P_{qg\to q}(z)\;,\\
A_3^0(1,3,2)&\stackrel{2_q//3_g}{\longrightarrow}&\frac{1}{s_{23}}P_{qg\to q}(z)\;.
\end{eqnarray}
A key ingredient in the evaluation of the subtraction term in equation~\eqref{eq:NLOsubt} is the phase space mapping which relates the original momenta
$p_i,p_j,p_k$ describing the two hard radiator partons $i$ and $k$ and the emitted parton $j$ to a redefined on-shell set $p_{\tilde{I}},p_{\tilde{K}}$ which are linear combinations 
of $p_i,p_j,p_k$~\cite{Kosower:1997zr,Kosower:2003bh}\footnotew{
Similar mappings aiming to have a local cancellation of IR divergencies 
were studied for \FDU\ in Section~\ref{sec:fdu} for NLO. 
}
\begin{eqnarray}
p_I^{\mu} &=&x\,p_i^{\mu}+r\,p_j^{\mu}+z\,p_k^{\mu}\nonumber\\
p_K^{\mu}&=&(1-x)\,p_i^{\mu}+(1-r)\,p_j^{\mu}+(1-z)\,p_k^{\mu}\;
\label{3to2FFmap}
\end{eqnarray}
where, 
\begin{eqnarray}
x&=&\frac{1}{2(s_{ij}+s_{ik})}\Big[(1+\rho)\,s_{ijk} -2\,r\,s_{jk}     \Big],\nonumber\\
z&=&\frac{1}{2(s_{jk}+s_{ik})}\Big[(1-\rho)\,s_{ijk} -2\,r\,s_{ij}     \Big],\nonumber\\
&&\nonumber\\
\rho^2&=&1+\frac{4\,r(1-r)\,s_{ij}s_{jk}}{s_{ijk}s_{ik}}.\;
\end{eqnarray}
The parameter $r$ can be chosen conveniently~\cite{Kosower:1997zr,Kosower:2003bh} and we use $ r=s_{jk}/(s_{ij}+s_{jk}).$
The mapping \eqref{3to2FFmap} implements momentum conservation $p_{\tilde{I}}+p_{\tilde{K}}=p_i+p_j+p_k$ and satisfies the following properties:
\begin{eqnarray}
p_{\tilde{I}}^2=0,\qquad&&\qquad p_{\tilde{K}}^2=0,\nonumber\\
p_{\tilde{I}}\to p_i,\qquad&&\qquad p_{\tilde{K}}\to p_k \qquad\qquad\textrm{when \textit{j} is soft},\nonumber\\
p_{\tilde{I}}\to p_i+p_j,\qquad&&\qquad p_{\tilde{K}}\to p_k \qquad\qquad\textrm{when \textit{i} becomes collinear with \textit{j}},\nonumber\\
p_{\tilde{I}}\to p_i,\qquad&&\qquad p_{\tilde{K}}\to p_j+p_k \qquad\textrm{when \textit{j} becomes collinear with \textit{k}}.\nonumber
\end{eqnarray}
This guarantees the proper subtraction of infrared singularities. With this
mapping, the phase space factorises,
\begin{equation}
{\rm d\Phi}_{m+1}(p_{1},\hdots,p_{i},p_{j},p_{k},\hdots,p_{m+1})={\rm d\Phi}_{m}(p_{1},\hdots,p_{\tilde{I}},p_{\tilde{K}},\hdots,p_{m+1})\cdot{\rm d}\Phi_{X_{ijk}}(p_{i},p_{j},p_{k};p_{\tilde{I}}+p_{\tilde{K}})\\
\label{eq:psfact}
\end{equation}
such that the integration over the unresolved radiative degrees of freedom can be decoupled from the integration over the Born configurations. We then use~\eqref{eq:psfact} in~\eqref{eq:NLOsubt} to obtain 
the integrated counterpart of each of the subtraction terms, in a form that is suitable for the cancelation of the IR-singularities with the virtual contribution,
\begin{eqnarray}
\int_{1}{\rm d}\sigma^{S}\;J_{m}^{(m)}(\{p_{m}\})&=&|{\cal M}_{m}^{0}|^2\,J_{m}^{(m)}(\{p_{m}\})\;{\rm d}\Phi_{m}\int {\rm d}\Phi_{X_{ijk}} X_{ijk}^{0}\nonumber\\
&=&|{\cal M}_{m}^{0}|^2\,J_{m}^{(m)}(\{p_{m}\})\;{\rm d}\Phi_{m}\;{\cal X}_{ijk}^{0}\;.
\end{eqnarray} 
This integration is performed analytically in $d=4-2\epsilon$ dimensions to make the infrared singularities
explicit, yielding the integrated three-parton antenna function ${\cal X}_{ijk}^{0}$. For the quark-gluon-antiquark final state the corresponding integrated antenna is,
\begin{eqnarray}
{\cal A}_{3}^{0}(s_{123})=(s_{123})^{-\epsilon}\Bigg[\frac{1}{3\epsilon^2}+\frac{3}{2\epsilon}+\frac{19}{4}-\frac{7\pi^2}{12}+{\cal O}(\epsilon)\Bigg]
=-2I_{q\bar{q}}^{(1)}(\epsilon,s_{123})+\frac{19}{4}\;,
\end{eqnarray}
where in the last equality the infrared singularity structure of the integrated antenna is written using the ${\bold{I}}^{(1)}$-operator~\cite{Catani:1998bh} which
describes the singularity structure of virtual one-loop amplitudes. This results makes the cancellations between real and virtual corrections explicit using integrated antennae,
establishing the universality of the subtraction algorithm.

The extension of the scheme to the treatment of initial state radiation requires antennae with one or two radiators in the initial state (initial-final or initial-initial antennae). For those, the IR-singularity structure of
the integrated antennae contains the poles of the virtual one-loop contribution and simultaneously collinear poles originating from radiation off incoming partons~\cite{Daleo:2006xa}. The latter are cancelled by redefinition 
(mass factorisation) of the parton distributions yielding a finite contribution, free of any poles in $\epsilon$ that can be integrated numerically. The appropriate phase space factorisations and allowed
phase space mappings for these kinematical configurations are given in~\cite{Daleo:2006xa}.

\subsection{Antenna subtraction at NNLO}
\label{sec:AntNNLO}
At NNLO, there are three distinct 
contributions due to double real radiation ${\rm d}\sigma_{NNLO}^{RR}$, mixed real-virtual radiation ${\rm d}\sigma_{NNLO}^{RV}$
and double virtual radiation ${\rm d}\sigma_{NNLO}^{VV}$. The NNLO cross section becomes,
\begin{eqnarray}
{\rm d}\sigma_{NNLO}&=&\int_{{\rm d}\Phi_{m+2}}{\rm d}\sigma_{NNLO}^{RR}\;J_{m}^{(m+2)}(\{p_{m+2}\})\nonumber\\
&+&\int_{{\rm d}\Phi_{m+1}}{\rm d}\sigma_{NNLO}^{RV}\;J_{m}^{(m+1)}(\{p_{m+1}\})\nonumber\\
&+&\int_{{\rm d}\Phi_{m}}{\rm d}\sigma_{NNLO}^{VV}\;J_{m}^{(m)}(\{p_{m}\})\;.
\end{eqnarray}
For each matrix element the integration is over the appropriate phase space subject to the constraint that precisely $m$-jets are observed. As usual, the individual contributions in the $m$, $(m +
1)$ and $(m+2)$-parton final states are all separately infrared divergent. In the $(m+2)$-parton final state, two particles can become unresolved in several
possible configurations: double soft, soft/collinear, double single collinear, triple collinear. In each of these limits, the $(m+2)$-parton matrix element factorises into a reduced $m$-parton matrix element 
times a generalised double unresolved factor. A detailed discussion of the kinematical definition of double unresolved limits is 
available in~\cite{GehrmannDeRidder:1997gf,Campbell:1997hg,Catani:1998nv,Catani:1999ss}. In addition, in the $(m+1)$-parton final state, single unresolved soft and collinear
singularities arise in the real-virtual one-loop process.

As at NLO, one has to introduce subtraction terms for the $(m + 1)$- and $(m + 2)$-parton contributions. In this case we will explore
the factorised structure of QCD amplitudes at NNLO to derive the form of the antenna subtraction terms. Schematically, the NNLO $m$-jet cross section reads,
\begin{eqnarray}
&{\rm d}\sigma_{NNLO}&=\int_{{\rm d}\Phi_{m}}\Bigg({\rm d}\sigma_{NNLO}^{VV}\;J_{m}^{(m)}(\{p_{m}\})-{\rm d}\sigma_{NNLO}^{U}\;J_{m}^{(m)}(\{p_{m}\})\Bigg)\nonumber\\
&+&\int_{{\rm d}\Phi_{m+1}}\Bigg({\rm d}\sigma_{NNLO}^{RV}\;J_{m}^{(m+1)}(\{p_{m+1}\})
-{\rm d}\sigma_{NNLO}^{T,1}\;J_{m}^{(m+1)}(\{p_{m+1}\})
-{\rm d}\sigma_{NNLO}^{T,2}\;J_{m}^{(m)}(\{\tilde{p}_{m}\})\Bigg)\nonumber\\
&+&\int_{{\rm d}\Phi_{m+2}}\Bigg({\rm d}\sigma_{NNLO}^{RR}\;J_{m}^{(m+2)}(\{p_{m+2}\})
-{\rm d}\sigma_{NNLO}^{S,1}\;J_{m}^{(m+1)}(\{\tilde{p}_{m+1}\})
-{\rm d}\sigma_{NNLO}^{S,2}\;J_{m}^{(m)}(\{\tilde{p}_{m}\})\Bigg)\;.\nonumber\\
\label{eq:nnlosub}
\end{eqnarray}

By construction the last line is finite after the introduction of the subtraction terms for one-unresolved parton ${\rm d}\sigma_{NNLO}^{S,1}$ and two unresolved partons ${\rm d}\sigma_{NNLO}^{S,2}$ in 
the double-real ($m+2$)-contribution. For the mixed real-virtual ($m+1$)-contribution, the explicit poles in the one-loop real-virtual matrix element cancel against the
integrated single unresolved real-radiation counterms as guaranteed by the KLN theorem, which are collected in ${\rm d}\sigma_{NNLO}^{T,1}$. The remaining counterterm ${\rm d}\sigma_{NNLO}^{T,2}$
is by construction free of explicit $1/\epsilon$-poles and subtracts all phase space singularities of the physical real-virtual matrix-element and of the antenna subtracted ${\rm d}\sigma_{NNLO}^{T,1}$ countertem.
Its contribution encodes the exact factorisation formula of one-loop matrix elements in the soft and collinear limits. Finally, the contribution ${\rm d}\sigma_{NNLO}^{U}$ contains the integrated 
counterparts of the antenna subtraction terms introduced at the double-real and real-virtual level and returns the explicit singularities of the double virtual matrix element. In this way, the three lines
in~\eqref{eq:nnlosub} are finite in $d=4$ and can be safely evaluated with numerical methods.  

\subsubsection{Double-real contribution}
In this section we establish the factorised form of all antenna subtraction terms for the double-real contribution. We begin by deriving the subtraction term for a single unresolved parton in the 
double-real process. Since this configuration is NLO-type we can immediately use the result obtained in Section~\ref{sec:nlosub} and obtain,
\begin{equation}
{\rm d}\sigma^{S,1}={\rm d}\sigma^{S,a}=\sum_{j}\,X_{ijk}^{0}\;|{\cal M}_{m+1}^{0}(1,\ldots,\tilde{I},\tilde{K},\ldots,m+1)|^2.
\label{eq:RRa}
\end{equation}
With~\eqref{eq:RRa}, singly unresolved limits involving parton-$j$ in the antenna $X_{ijk}$ cancel directly against the double-real matrix element. 
However, contrary to the NLO case, the single-unresolved double-real subtraction term at NNLO factorises into an ($m+1$)-reduced matrix element and the jet function constrains 
that precisely $m$-jets are observed. For this reason, singly unresolved limits as well as genuine double unresolved limits 
involving the reduced matrix element in~\eqref{eq:RRa} are allowed and need to cancel with the genuine double-real doubly unresolved subtraction term ${\rm d}\sigma_{NNLO}^{S,2}$.

For the derivation of ${\rm d}\sigma_{NNLO}^{S,2}$ we must distinguish the following configurations according to the colour connection of the double-unresolved partons:
\begin{itemize}
\item[(b)] Two colour-connected unresolved partons (colour-connected).
\begin{figure}[h]
\begin{center} 
\includegraphics[height=1.2cm]{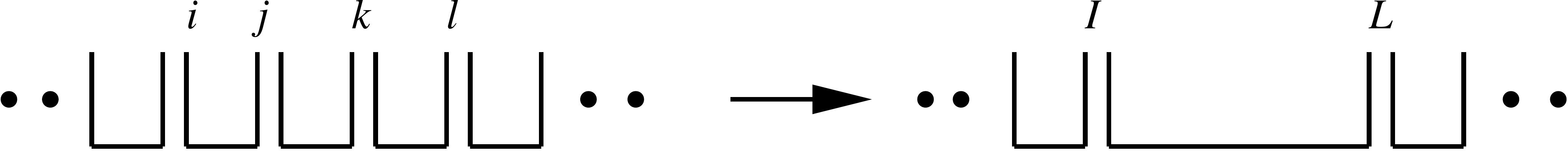}
\end{center}
\end{figure}
\end{itemize}
When two unresolved partons $j$ and $k$ are adjacent between radiators $i$ and $l$ the subtraction term is:
\begin{equation}
{{\rm d}\sigma^{S,b}}=\sum_{j}\,\Big(X_{ijkl}^{0}-X_{ijk}^{0}X_{IKl}^{0}-X_{jkl}^{0}X_{iJL}^{0}\Big)|{\cal M}_{m}^{0}(1,\ldots,\tilde{I},\tilde{L},\ldots,m)|^2\;,
\label{eq:sigB}
\end{equation}
where $X_{ijkl}^{0}$ is a tree-level four parton antenna that smoothly interpolates all colour connected double unresolved limits. As an example, the final state quark-gluon-gluon-antiquark 
antenna derived from $\gamma^{*}\to qgg\bar{q}$ obeys the following factorisation properties,
\begin{eqnarray}
A_4^0(q_1,g_3,g_4,\bar{q}_2)&\stackrel{3_g\to 0,4_g\to0}{\longrightarrow}&S_{1342}\;,\\
A_4^0(q_1,g_3,g_4,\bar{q}_2)&\stackrel{1_q//3_g//4_g}{\longrightarrow}&P_{qgg\to Q}(x,y,z)\;,\\
A_4^0(q_1,g_3,g_4,\bar{q}_2)&\stackrel{4_g\to0,1_q//3_g}{\longrightarrow}&S_{q;gg\bar{q}}P_{qg\to Q}(z)\;,\\
A_4^0(q_1,g_3,g_4,\bar{q}_2)&\stackrel{1_q//3,2_{\bar{q}}//4}{\longrightarrow}&P_{qg\to Q}(z)P_{\bar{q}g\to Q}(y)\;,
\end{eqnarray}
where the universal double soft, triple collinear, soft/collinear and double single collinear limits listed above have been extensively discussed in the 
literature~\cite{GehrmannDeRidder:1997gf,Campbell:1997hg,Catani:1998nv,Catani:1999ss}. The integrated counterpart of the four-parton antenna contribution exploits
the factorisation of the double-real radiation phase space,
\begin{eqnarray}
{\rm d}\Phi_{m+2}(p_1,\hdots,p_{m+2})={\rm d}\Phi_{m}(p_1,\hdots,p_{\tilde{I}},p_{\tilde{L}},\hdots,p_{m+2})\cdot {\rm d}\Phi_{X_{ijkl}}(p_i,p_j,p_k,p_l)\;,
\end{eqnarray}
obtained by redefining a set of four massless on-shell momenta (radiator, two unresolved partons, radiator) into two on-shell massless momenta. The
mapping is defined as:\footnotew{
A preliminar proposal for  mappings in \FDU\ was provided in~\eqref{eq:fdumaps}. 
A detailed comparison between both approaches should be considered. 
In fact, \FDU\ could profit from the way how the various IR regions are split
in the antenna subtraction method. 
}
\begin{eqnarray}
p_{I}^\mu \equiv \widetilde{p_{(i,j,k)}}&=&x\,p_{i}+r_1\,p_{j}+r_2\,p_{k}+z\,p_{l},\nonumber\\
p_{L}^\mu \equiv\widetilde{p_{(l,k,j)}}&=&(1-x)\,p_{i}+(1-r_1)\,p_{j}+(1-r_2)\,p_{k}+(1-z)\,p_{l}\;,
\end{eqnarray}
with $p_{I}^2=p_{L}^2=0$.
Defining $s_{kl}=(p_{i_k}+p_{i_l})^2$, 
the coefficients are given by \cite{Kosower:2002su}:
\begin{eqnarray}
r_1&=&\frac{s_{jk}+s_{jl}}{s_{ij}+s_{jk}+s_{jl}}\nonumber\\
r_2&=&\frac{s_{kl}}{s_{ik}+s_{jk}+s_{kl}}\nonumber\\
x&=&\frac{1}{2(s_{ij}+s_{ik}+s_{il})}\Big[(1+\rho)\,s_{ijkl}-r_1\,(s_{jk}+2\,s_{jl})   -r_2\,(s_{jk}+2\,s_{kl})  \nonumber\\
&&+(r_1-r_2)\frac{s_{ij}s_{kl}-s_{ik}s_{jl}}{s_{il}}     \Big]\nonumber\\
z&=&\frac{1}{2(s_{il}+s_{jl}+s_{kl})}\Big[(1-\rho)\,s_{ijkl} -r_1\,(s_{jk}+2\,s_{ij})   -r_2\,(s_{jk}+2\,s_{ik})  \nonumber\\
&&-(r_1-r_2)\frac{s_{ij}s_{kl}-s_{ik}s_{jl}}{s_{il}}     \Big]\nonumber\\
\rho&=&\Big[1+\frac{(r_1-r_2)^2}{s_{il}^2\,s_{ijkl}^2}\,\lambda(s_{ij}\,
s_{kl},s_{il}\,s_{jk},s_{ik}\,s_{jl})\nonumber\\
&&  +\frac{1}{s_{il}\,s_{ijkl}}\Big\{
2\,\big(r_1\,(1-r_2)+r_2(1-r_1)\big)\big( s_{ij}s_{kl}+s_{ik}s_{jl}-s_{jk}s_{il} \big)\nonumber\\
&&\qquad\qquad +\,4\,r_1\,(1-r_1)\,s_{ij} s_{jl}+4\,r_2\,(1-r_2)\,s_{ik}s_{kl}\Big\}\Big]^{\frac{1}{2}}\;,
\nonumber\\
\lambda(u,v,w)&=&u^2+v^2+w^2-2(uv+uw+vw)\;.\nonumber
\end{eqnarray}
This mapping smoothly interpolates all colour connected double unresolved singularities. It satisfies the following properties:
\begin{align} 
&\widetilde{p_{(ijk)}}\to p_{i}, 
&\widetilde{p_{(lkj)}}\to p_{l}\hspace{2.5cm} 
&\textrm{when $j,k\to0$,}\nonumber\\
&\widetilde{p_{(ijk)}}\to p_{i}+p_{j}+p_{k},
&\widetilde{p_{(lkj)}}\to p_{l}\hspace{2.5cm}  
&\textrm{when $i//j//k$,}\nonumber\\
&\widetilde{p_{(ijk)}}\to p_{i}, 
&\widetilde{p_{(lkj)}}\to p_{l}+p_{k}+p_{j}\hspace{0.6cm}   
&\textrm{when $j//k//l$,}\nonumber\\
&\widetilde{p_{(ijk)}}\to p_{i}, 
&\widetilde{p_{(lkj)}}\to p_{l}+p_{k}\hspace{1.5cm}  
&\textrm{when $j\to 0 +k//l$,}\nonumber\\
&\widetilde{p_{(ijk)}}\to p_{i}+p_{j}, 
&\widetilde{p_{(lkj)}}\to p_{j}\hspace{2.4cm} 
&\textrm{when $k\to 0 +i//j$,}\nonumber\\
&\widetilde{p_{(ijk)}}\to p_{i}+p_{j}, 
&\widetilde{p_{(lkj)}}\to p_{k}+p_{l} \hspace{1.5cm}
&\textrm{when $i//j+k//l$.}\nonumber\;
\end{align}
which guarantee that all double unresolved colour connected IR-singularities are properly subtracted. Moreover, in single unresolved limits, the momentum mapping above collapses into an 
NLO mapping~\eqref{3to2FFmap}, thereby allowing the products of three-parton antenna functions in~\eqref{eq:sigB} to subtract the single unresolved limits of the associated four parton antenna.
\begin{itemize}
\item[(c)] Two unresolved partons that are not colour connected but share a common
radiator (almost colour-unconnected).
\begin{figure}[h]
\begin{center} 
\includegraphics[height=1.2cm]{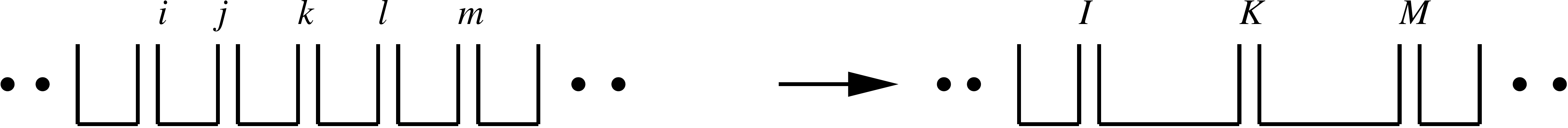}
\end{center}
\end{figure}
\end{itemize}
There are double unresolved configurations where the unresolved partons are separated by a hard radiator parton, for example, $i, j, k, l, m$ where $j$ and $l$ are unresolved. In this case
we take the strongly ordered approach where $i, j, k$ form an antenna with hard partons $I$ and $K$ yielding an ordered amplitude involving $I, K, l, m$. The case where $l$ is
unresolved is then treated using an antenna $K, l, m$ with hard partons $K'$ and $M'$. The other case where first $k, l, m$ form an antenna followed by $i, j, K$ is also included where the
momenta are obtained by iterative use of the NLO momentum mappings. The subtraction term is,
\begin{eqnarray}
{\rm d}\sigma_{NNLO}^{S,c}=&-&\sum_{j,l}X_{ijk}^{0}X_{Klm}^{0}|{\cal M}^0_{m}(p_{1},\hdots,p_{I},p_{K'},p_{M'},\hdots,p_{m})|^2\nonumber\\
&-&\sum_{j,l}X_{mlk}^{0}X_{ijK}^{0}|{\cal M}^0_{m}(p_{1},\hdots,p_{I'},p_{K'},p_{M},\hdots,p_{m})|^2\;.
\end{eqnarray}

\begin{itemize}
\item[(d)] Two unresolved partons that are well separated from each other 
in the colour 
chain (colour-unconnected).
\begin{figure}[h]
\begin{center} 
\includegraphics[height=1.2cm]{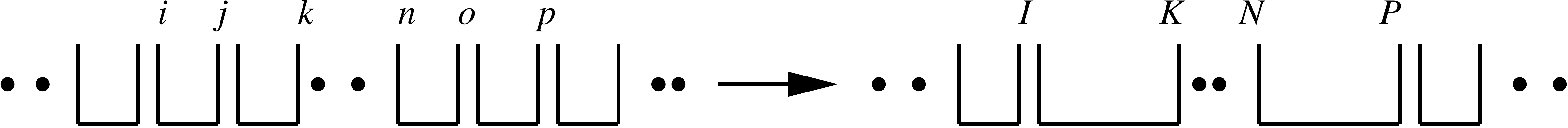}
\end{center}
\end{figure}
\end{itemize}

When two unresolved partons $j$ and $o$ are completely disconnected i.e. for colour ordered amplitudes of the type ${\cal M}(. . . , i, j, k, . . . , n, o, p, . . .)$, the double-real matrix element
factorises into the product of two uncorrelated single unresolved factors with hard partons $I, K$ and $N, P$ respectively. The subtraction term is,
\begin{eqnarray}
{\rm d}\sigma_{NNLO}^{S,d}&=&-\sum_{j,o}X_{ijk}^{0}X_{nop}^{0}|{\cal M}^0_{m}(p_{1},\hdots,p_{I},p_{K},\hdots,p_{N},p_{P},\hdots,p_{m})|^2\;.
\end{eqnarray}

\begin{itemize}
\item[(e)] Large angle soft emission.
\end{itemize}

By taking a strongly ordered subtraction of the unresolved limits, an uncanceled contribution involving the inner and outer antennae in the iterated subtracted structures defined above,
leads to an incomplete subtraction of large-angle soft gluon radiation. To account for the left over single soft-gluon emission contribution at large angles, an additional subtraction term is defined,
\begin{eqnarray}
{\rm d}\sigma_{NNLO}^{S,e}&=&\sum_{j}X_{IlK}^{0}\Big(S_{I'jK'}-S_{IjK}-S_{ajI'}+S_{ajI}-S_{K'jb}+S_{Kjb}\Big)\nonumber\\
&\times&|{\cal M}^0_{m}(p_{1},\hdots,p_a,p_{I'},p_{K'},p_b,\hdots,p_{m})|^2\;.
\end{eqnarray}
The large-angle soft subtraction term contains soft antenna functions of the form $S_{ajb}$ which are simply the eikonal factor for a soft gluon $j$ emitted between hard partons $a$ and $b$. The
soft factors are associated with an NLO antenna phase space mapping $(i,j,k)\to (I,K)$, followed by a second NLO antenna phase space mapping $(I, l, K)\to (I',K')$. 

\subsection*{Subtraction of angular correlations}
When using the antenna subtraction method to construct subtraction terms for higher order calculations, one encounters the problem of angular correlations in the collinear splitting of a gluon into massless partons. These 
angular correlations introduce non-factorizing terms which correlate the hard reduced matrix element with the splitting functions. As an example, for the $gg\to g$ splitting, the purely gluonic four 
parton antenna function factorises into the corresponding tensorial splitting functions and tensorial three parton antenna functions,
\begin{eqnarray}
F_4^0(g_1,g_2,g_3,g_4)&\stackrel{i_g \parallel j_g}{\longrightarrow}&\frac{1}{s_{ij}}
P_{gg\to G}^{\mu\nu}(z)(F_3^0)_{\mu\nu}((ij),k,l)\nonumber\\
&=&\frac{1}{s_{ij}}P_{gg\to G}(z)F_3^0((ij),k,l)+{\rm ang.}
\label{eq:angtermlimit}
\end{eqnarray}
$P_{ij\to(ij)}^{\mu\nu}$ stands for the spin dependent gluon splitting function given 
by~\cite{Catani:1996vz},
\begin{eqnarray}
P_{gg}^{\mu\nu}=2\left[-g^{\mu\nu}\left(\frac{z}{1-z}+\frac{1-z}{z}\right)-2(1-\epsilon)z(1-z)
\frac{k_{\perp}^\mu k_{\perp}^\nu}{k_{\perp}^2}\right],
\end{eqnarray}
while $P_{ij\to(ij)}$ stands for the spin averaged gluon splitting function:
\begin{equation}
P_{gg\to g}(z)=2\left(\frac{z}{1-z}+\frac{1-z}{z}+z(1-z)\right).
\end{equation}
The tensorial structure of the three-parton antenna function $(F_3^0)_{\mu\nu}$ is obtained by leaving the polarisation index of the gluon associated with momentum $P^\mu$ uncontracted and can be derived by analogy with the scalar three-parton antenna functions from physical matrix elements. 

Since we use spin-averaged scalar antenna functions to remove unresolved limits in QCD amplitudes, these do not subtract angular correlations in gluon-splittings. However, these angular terms vanish when the azimuthal 
variable of the collinear system is integrated out. This can be seen for the single collinear limits using the standard 
momentum parametrisation \cite{Catani:1996vz,Altarelli:1977zs} for the $i_g\parallel j_g$  limit:
\begin{eqnarray}
p_i^\mu = z p^\mu + k_\perp^\mu - \frac{k_\perp^2}{z}\frac{n^\mu}{2p\cdot n}
\;, &\qquad&
 p_j^\mu = (1-z) p^\mu - k_\perp^\mu 
- \frac{k_\perp^2}{1-z}\frac{n^\mu}{2p\cdot n}\;, \nonumber\\
\mbox{with } 2p_i\cdot p_j = -\frac{k_\perp^2}{z(1-z)}\;, &\qquad& p^2 
= n^2=k_\perp.p=k_\perp.n=0\;.\label{eq:colmap}
\end{eqnarray}
Here $p^\mu$ denotes the collinear momentum direction, and $n^\mu$ is an 
auxiliary vector. The collinear limit is approached as $k_\perp^2\to 0$. 

In the simple collinear $i\parallel j$ limit of the four-parton antenna function 
$F_4^0(l_g,i_g,j_g,k_g)$, 
one chooses $n=p_k$ to be one of the non-collinear momenta, such that the 
antenna function can be expressed in terms of $p$, $n$, $k_\perp$ and $p_l$. 
Expanding in $k_\perp^\mu$ yields only non-vanishing  scalar products of the 
form $p_l\cdot k_\perp$. Expressing the integral over the antenna phase 
space in the $(p,n)$ centre-of-mass frame, the angular average can be 
carried out as
\begin{equation}
\frac{1}{2\pi} \int_0^{2\pi} {\rm d} \phi\, (p_l\cdot k_\perp) =0 \;, \qquad 
 \frac{1}{2\pi} \int_0^{2\pi} {\rm d} \phi\, (p_l\cdot k_\perp)^2 = - 
k_\perp^2\, \frac{p\cdot p_l\, n\cdot p_l}{p\cdot n}\;.\label{eq:azymav}
\end{equation}

In this frame, the unsubtracted angular correlation in the gluon-gluon collinear limit of the four-parton purely gluonic antenna function is given by,
\begin{eqnarray}
\Theta_{F_3^0}(i,j,z,k_\perp)&=&\left[\frac{1}{s_{ij}}P^{\mu\nu}_{ij\to(ij)}(z,k_\perp)(F_3^0)_{\mu\nu}
-\frac{1}{s_{ij}}P_{ij\to(ij)}(z)F_3^0(1,(ij),2)\right]\nonumber\\
&=&\frac{4}{s_{ij}^2s_{1p2}^2}\left(\frac{s_{12}^2s_{1p2}^2+s_{1p}^2s_{p2}^2}{s_{12}^2s_{1p}^2s_{p2}^2}\right)
\Bigg[s_{12}s_{1p}s_{p2}\;k_\perp\cdot k_\perp\nonumber\\
&&-4p_1\cdot k_\perp p_2\cdot k_\perp s_{1p}s_{p2}
+2(p_1\cdot k_\perp)^2 s_{p2}^2+2(p_2\cdot k_\perp)^2 s_{1p}^2\Bigg],\nonumber\\
\label{eq:angfunction}
\end{eqnarray}
with $p$ and $k_{\perp}$ defined in (\ref{eq:colmap}). Using (\ref{eq:azymav}), we can easily see that (\ref{eq:angfunction}) integrates to zero. 

The same cancellation can be made to happen locally (before any integration), by deriving the azimuthal angular dependence of the angular correlation. In the $(p,n)$ centre-of-mass frame, 
it can be shown that 
\begin{equation}
\Theta_{F_3^0}(i,j,z,k_\perp) \sim A\cos(2\phi+\alpha)
\end{equation}
where $\phi$ is the same azimuthal angle as in \eqref{eq:azymav}.  Therefore, by combining two phase space points with azimuthal angles $\phi$ and $\phi+\pi/2$ and all other coordinates equal, 
the azimuthal correlations drop out. This strategy is implemented in the method and ensures a smooth cancellation of gluonic collinear splittings~\cite{Glover:2010im,Gehrmann:2018szu}.

The full double-real radiation subtraction term is given as a sum of all subtraction terms defined above:
\begin{equation}
{\rm d}\sigma_{NNLO}^{S}={\rm d}\sigma_{NNLO}^{S,a}+{\rm d}\sigma_{NNLO}^{S,b}+{\rm d}\sigma_{NNLO}^{S,c}+{\rm d}\sigma_{NNLO}^{S,d}+{\rm d}\sigma_{NNLO}^{S,e}\;,
\label{eq:RRsub}
\end{equation}
which correctly approximates the double real matrix element contribution in all double and single unresolved regions. Although individual terms in~\eqref{eq:RRsub} contain spurious
singularities in these limits, they cancel among each other in the sum.

\subsubsection{Real-virtual contribution}

As discussed in Section~\ref{sec:AntNNLO}, in order to carry out the numerical integration over the real-virtual matrix element, we need to introduce an
infrared subtraction term which removes the explicit infrared poles of the real-virtual one-loop matrix element and correctly describes its single unresolved limits. 
As in the previous section, we will explore the universal factorised form of the QCD amplitudes in the IR-singular regions to obtain the necessary antenna subtraction terms.

\subsubsection*{Subtraction of explicit poles}
It is a well known fact from NLO calculations, that the explicit infrared poles of one-loop matrix elements cancel with the corresponding infrared poles obtained by integrating out
all single unresolved configurations from the real radiation matrix elements contributing to the same (infrared safe) observable. We can therefore obtain an antenna subtraction
term to cancel the explicit poles of the one-loop real-virtual matrix element with the integrated counterpart of the single unresolved subtraction term introduced 
at the double real-level~${\rm d}\sigma_{NNLO}^{S,a}$. We obtain,
\begin{eqnarray}
{\rm d}\sigma_{NNLO}^{T,1}={\rm d}\sigma_{NNLO}^{T,a}=-\int_{1}{\rm d}\sigma_{NNLO}^{S,a}=-\sum_{ik}{\cal X}_{3}^{0}(s_{ik})|{\cal M}_{m+1}^{0}(p_{1},\hdots,p_i,p_k,\hdots,p_{m+1})|^2\;\nonumber
\end{eqnarray}
where the explicit $1/\epsilon$-poles in integrated antenna ${\cal X}_{3}^{0}(s_{ik})$ cancel analytically with the poles of the real-virtual matrix element as guarantee by the KLN theorem. 

\subsubsection*{Subtraction of soft and collinear phase space singularities at one-loop}
In single unresolved limits, the behaviour of the $(m + 1)$-parton real-virtual one-loop amplitude is described by the sum of two different 
contributions~\cite{Bern:1994zx,Kosower:1999xi,Kosower:1999rx,Bern:1998sc,Bern:1999ry}: a simple unresolved tree level
factor times a $m$-parton one-loop amplitude and a simple unresolved one-loop factor times a $m$-parton tree-level amplitude. Schematically the antenna subtraction
term reproduces this factorised form,
\begin{equation}
{\cal M}_{m+1}^{1}\to X_{3}^{0}\;{\cal M}_{m}^{1} + X_{3}^{1}\;{\cal M}_{m}^{0}\;,
\label{eq:RVfact}
\end{equation} 
where we have introduced a three-parton one-loop antenna function $X_{3}^{1}$ derived from properly normalised one-loop three-parton matrix elements, in an analogous way as for all other antennae. 

However, the factorised form on the right hand side of the limit in~\eqref{eq:RVfact}, contains new one-loop matrix elements in ${\cal M}_{m}^{1}$ and $X_{3}^{1}$ with explicit $1/\epsilon$-poles, 
whose cancellation needs to be fixed by the subtraction algorithm. The subtraction term for this contribution reads,
\begin{eqnarray}
{\rm d}\sigma_{NNLO}^{T,b}&=&\sum_{j}X_{ijk}^{0}\;\Big(|{\cal M}_{m}^{1}(p_{1},\hdots,p_{m})|^2+\sum_{IK}{\cal X}_{3}^{0}(s_{IK})|{\cal M}_{m}^{0}(p_{1},\hdots,p_{m})|^2\Big)\nonumber\\
&+&\sum_{j}\Big(X_{ijk}^{1} + \sum_{ik}{\cal X}_{3}^{0}(s_{ik})X_{ijk}^{0}\Big) |{\cal M}_{m}^{0}(p_{1},\hdots,p_{m})|^2\;.
\label{eq:RVTb}
\end{eqnarray}
In~\eqref{eq:RVTb} we have introduced terms of the type ${\cal X}_{3}^{0}X_{3}^{0}$ that cancel the explicit poles introduced by reduced one-loop $m$-parton matrix elements and 
one-loop antenna functions.

In particular, the subtraction of IR-poles from the $X_{3}^1$ antenna in~\eqref{eq:RVTb} is directly related to integrals of tree-level subtraction terms introduced at the double-real level 
$\int_{1}{\rm d}\sigma_{NNLO}^{S,b}$. The remaining integrals with $X_{3}^{1}$, ${\cal M}_{m}^{1}$ and $X_{3}^{0} {\cal X}_{3}^{0}(s_{IK})$ are genuine new contributions that 
can not be related to integrals of tree-level subtraction terms. Therefore, their contributions must cancel with parts of the
two-loop $m$-parton amplitude after analytic integration over the three-parton antenna phase space. 

We have therefore obtained in ${\rm d}\sigma_{NNLO}^{T,b}$, a universal antenna subtraction term, which is free from explicit $1/\epsilon$-poles by construction, and moreover,  it subtracts 
the phase space singularities of the physical real-virtual matrix element and simultaneously subtracts all phase space spurious singularities in~${\rm d}\sigma_{NNLO}^{T,a}$ defined above.

\subsubsection*{Subtraction of large angle soft emission}

For processes involving soft gluons the double-real channel has an additional subtraction contribution denoted by ${\rm d}\sigma_{NNLO}^{S,e}$ due to large angle soft gluon radiation. This term 
removed the remnant soft gluon behaviour associated with the phase space mappings of the iterated structures of the double-real subtraction contribution ${\rm d}\sigma_{NNLO}^{S,c}$. 
Both these subtracted contributions have an integrated counterpart, which can be obtained by integrating over the soft-eikonal factor in the former case, and over the outer antenna in the latter case. 

Both integrals are performed analytically over the factorised singly unresolved radiative phase space ${\rm d}\Phi_{X_{ijk}}$ making their IR-singularities explicit $1/\epsilon$-poles. 
This integration results in explicit $1/\epsilon$-poles whose cancellation needs to be fixed by the subtraction algorithm. The subtraction term for this contribution reads,
\begin{eqnarray}
{\rm d}\sigma_{NNLO}^{T,c}&=&\sum_{j}X_{ijk}^0\bigg[{\cal X}_{3}^{0}(s_{ik})-{\cal X}_{3}^{0}(s_{ai})-{\cal X}_{3}^{0}(s_{kb})
-{\cal X}_{3}^{0}(s_{IK})+{\cal X}_{3}^{0}(s_{aI})+{\cal X}_{3}^{0}(s_{Kb})\nonumber\\
&-&{\cal S}(s_{ik};s_{ik})+{\cal S}(s_{ai};s_{ik})+{\cal S}(s_{kb};s_{ik})+{\cal S}(s_{IK};s_{ik})-{\cal S}(s_{aI};s_{ik})-{\cal S}(s_{Kb};s_{ik}) \bigg]\nonumber\\
&\times&|{\cal M}_{m}^{0}(p_{1},\hdots,p_{m})|^2\;,
\label{eq:RVTc}
\end{eqnarray}
where ${\cal S}$ is the integrated soft-eikonal factor. With the analytic expressions for the integrated antennae and integrated soft-factors~\cite{GehrmannDeRidder:2007jk} we obtain by construction
a counterterm ${\rm d}\sigma_{NNLO}^{T,c}$ which is free from explicit $1/\epsilon$-poles and has no phase space soft or collinear singularities. In order to achieve this constraint it is necessary to add
genuine new terms of the type $X_{3}^{0} {\cal X}_{3}^{0}(s_{IK})$ to cancel the poles of the wide-angle soft term. Such contributions must be integrated analytically
over the three-parton antenna phase space and added in integrated form to the double-virtual $m$-parton contribution. 

The full real-virtual subtraction term is given as a sum of all subtraction terms constructed above:
\begin{equation}
{\rm d}\sigma_{NNLO}^{T}={\rm d}\sigma_{NNLO}^{T,a}+{\rm d}\sigma_{NNLO}^{T,b}+{\rm d}\sigma_{NNLO}^{T,c}\;,
\label{eq:RVsub}
\end{equation}
which correctly approximates the real-virtual one-loop matrix element in all single unresolved regions and simultaneously subtracts all of its $1/\epsilon$-explicit poles as guaranteed by the KLN theorem.

\subsubsection{Double-virtual contribution}
The double virtual contribution involves the two-loop $m$-parton matrix elements which have no implicit IR divergence in any regions of the
appropriate $m$-parton phase space. Therefore, to make this contribution finite, all that remains is to introduce the integrated forms
of the appropriate antenna subtraction terms such that the explicit IR-poles of the two-loop contribution are cancelled. We begin by reviewing the universal structure
of infrared singularities in on-shell QCD amplitudes at two-loop order in Catani's two-loop factorisation formula~\cite{Catani:1998bh},
\begin{eqnarray}
{\cal P}oles\big({\cal M}_{m}^{2}(1,\hdots,n)\big)&=&2\boldsymbol{I}_{m}^{(1)}(\epsilon;1,\hdots,m)\;{\cal M}_{m}^{1}(1,\hdots,m)\nonumber\\
&-&2\boldsymbol{I}_{m}^{(1)}(\epsilon;1,\hdots,m)^{2}{\cal M}_{m}^{0}(1,\hdots,m)\nonumber\\
&+&2e^{-\epsilon\gamma}\frac{\Gamma(1-2\epsilon)}{\Gamma(1-\epsilon)}\bigg(\frac{\beta_{0}}{\epsilon}+K\bigg)\boldsymbol{I}_{m}^{(1)}(2\epsilon;1\hdots,m)\,{\cal M}_{m}^{0}(1,\hdots,m)\nonumber\\
&+&2\boldsymbol{H}^{(2)}(\epsilon){\cal M}_{m}^{0}(1,\hdots,m).
\label{eq:catani2}
\end{eqnarray}
The poles of the two amplitude are organised according to the $\boldsymbol{I}_{m}^{(1)}$-operator given in~\cite{Catani:1998bh} and hard function, $\boldsymbol{H}^{(2)}$ and the
constant $K$, which depend on the particle content and order in $M$ under consideration. In the following we will obtain the integrated antenna subtraction terms in a form 
which is in one-to-one correspondence with~\eqref{eq:catani2} making the analytic cancellation of all explicit $1/\epsilon$-poles in the double-virtual contribution transparent. 

The first double virtual subtraction term is the integrated counterpart of the contribution introduced at the real-virtual level ${\rm d}\sigma_{NNLO}^{T,b}$. In that contribution,
we can perform the analytic integration over the factorised singly unresolved radiative phase space ${\rm d}\Phi_{X_{ijk}}$ of the antenna function proportional to the one-loop matrix element, obtaining,
\begin{eqnarray}
{\rm d}\sigma_{NNLO}^{U,a}=\sum_{ik}{\cal X}_{3}^{0}(s_{ik})\;|M_{m}^{1}(1,\hdots,m)|^2\;&=&\boldsymbol{J}_{m}^{(1)}(\epsilon;1,\hdots,m)\;|{\cal M}_{m}^{1}(1,\hdots,m)|^2\;,
\label{eq:Ua}
\end{eqnarray}
where in the last equality we have defined an IR-singular operator $\boldsymbol{J}_{m}^{(1)}$ containing a string of integrated three-parton antennae which contain the IR-poles of the integrated real-radiation
contribution, in analogy with the $\boldsymbol{I}_{m}^{(1)}$-operator which describes the IR-poles of the virtual matrix elements.

The second double virtual subtraction term is the integrated counterpart of the contributions introduced at the real-virtual level ${\rm d}\sigma_{NNLO}^{T,c}$ and double-real level 
${\rm d}\sigma_{NNLO}^{S,d}$, which combine and yield,
\begin{eqnarray}
{\rm d}\sigma_{NNLO}^{U,b}&=&\sum_{ik}\sum_{ml}{\cal X}_{3}^{0}(s_{ik})\otimes{\cal X}_{3}^{0}(s_{ml})\;|{\cal M}_{m}^{0}(1,\hdots,m)|^2\;\nonumber\\
&=&\frac{1}{2}\boldsymbol{J}_{m}^{(1)}(\epsilon;1,\hdots,m)\otimes \boldsymbol{J}_{m}^{(1)}(\epsilon;1,\hdots,m) \;|{\cal M}_{m}^{0}(1,\hdots,m)|^2\;,
\end{eqnarray}
where we explicitly introduced the square of the $\boldsymbol{J}_{m}^{(1)}$-operator introduced in~\eqref{eq:Ua}.

Finally, the third double virtual subtraction term is the integrated counterpart of the contributions introduced at the real-virtual level ${\rm d}\sigma_{NNLO}^{T,b}$ including the terms 
proportional to the one-loop $X_{3}^{1}$ antenna, and the contribution introduced at the double-real level ${\rm d}\sigma_{NNLO}^{S,b}$ involving the four-parton
antenna $X_{4}^{0}$,
\begin{eqnarray}
{\rm d}\sigma_{NNLO}^{U,c}&=&\sum_{ik}\Big({\cal X}_{4}^{0}(s_{ik})+{\cal X}_{3}^{1}(s_{ik})-\frac{1}{2}{\cal X}_{3}^{0}(s_{ik}){\cal X}_{3}^{0}(s_{ik})\Big)|{\cal M}_{m}^{0}(1,\hdots,m)|^2\;\nonumber\\
&=&\boldsymbol{J}_{m}^{(2)}(\epsilon;1,\hdots,m)\;|{\cal M}_{m}^{0}(1,\hdots,m)|^2\;.
\label{eq:Uc}
\end{eqnarray}
In Eq.~\eqref{eq:Uc} we introduced an IR-singular operator containing the double unresolved integrated antenna string $\boldsymbol{J}_{m}^{(2)}$.

The full double-virtual subtraction term is given as a sum of all subtraction terms constructed above:
\begin{equation}
{\rm d}\sigma_{NNLO}^{U}={\rm d}\sigma_{NNLO}^{U,a}+{\rm d}\sigma_{NNLO}^{U,b}+{\rm d}\sigma_{NNLO}^{U,c}\;,
\label{eq:VVsub}
\end{equation}
where in particular we can observe that ${\rm d}\sigma_{NNLO}^{U,a}$ and ${\rm d}\sigma_{NNLO}^{U,b}$ are in one-to-one correspondence with the first two lines in Eq.~\eqref{eq:catani2}, 
while the contribution ${\rm d}\sigma_{NNLO}^{U,c}$ subtracts the remaining IR singularities of the two-loop amplitude in the bottom two lines in~\eqref{eq:catani2}. 

\subsubsection*{Application: $N^2$ contribution to $q\bar{q}\to gg$ at NNLO}
In this section we present the double-virtual antenna subtraction term for the $N^2$ contribution to dijet production at hadron colliders at NNLO. Focusing on the 
$q_1\bar{q}_2\to g_3g_4$ channel, the subtraction term reads,
\begin{eqnarray}
{\rm d}\sigma_{q\bar{q},NNLO}^{U}&=\sum\limits_{\substack{P(i,j)}}&\bigg\{\boldsymbol{J}_{4}^{(1)}(\hat{\bar{1}}_{q},i_{g},j_{g},\hat{\bar{2}}_{\bar{q}})\ \biggl(B_{4}^{1}(\hat{\bar{1}}_{q},i_{g},j_{g},\hat{\bar{2}}_{\bar{q}})-\frac{b_{0}}{\epsilon}\ B_{4}^{0}(\hat{\bar{1}}_{q},i_{g},j_{g},\hat{\bar{2}}_{\bar{q}})\biggr)\nonumber\\
&+&\frac{1}{2}\ \boldsymbol{J}_{4}^{(1)}(\hat{\bar{1}}_{q},i_{g},j_{g},\hat{\bar{2}}_{\bar{q}})\otimes\boldsymbol{J}_{4}^{(1)}(\hat{\bar{1}}_{q},i_{g},j_{g},\hat{\bar{2}}_{\bar{q}})\ B_{4}^{0}(\hat{\bar{1}}_{q},i_{g},j_{g},\hat{\bar{2}}_{\bar{q}})\nonumber\\
&+&\boldsymbol{J}_{4}^{(2)}(\hat{\bar{1}}_{q},i_{g},j_{g},\hat{\bar{2}}_{\bar{q}})\ B_{4}^{0}(\hat{\bar{1}}_{q},i_{g},j_{g},\hat{\bar{2}}_{\bar{q}})\biggr\}\ J_{2}^{(2)}(p_{i},p_{j})\;,
\label{eq:Usub}
\end{eqnarray}
where $B_{4}^{1}$ and $B_{4}^{0}$ are the leading colour renormalised colour-ordered one-loop and tree level amplitudes for $q\bar{q}\to gg$ respectively. As demonstrated in the previous section, the IR-operators
$\boldsymbol{J}_{4}^{(1)}$ and $\boldsymbol{J}_{4}^{(2)}$ are built with integrated antenna strings that involve colour connected particles and that match the poles of the virtual amplitudes. 
For processes with coloured particles in the initial state these operators involve integrated antennae with hard radiators in the initial state that subtract radiation off incoming partons leading 
to initial-state collinear poles. These IR-singularities cancel with the redefinition (mass factorisation) of the parton distributions. In the example of this section we obtain, 
\begin{equation}
\boldsymbol{J}_{4}^{(1)}(\hat{\bar{1}}_{q},i_{g},j_{g},\hat{\bar{2}}_{\bar{q}})=\boldsymbol{J}_{2}^{(1)}(\hat{\bar{1}}_{q},i_{g})+\boldsymbol{J}_{2}^{(1)}(i_{g},j_{g})+\boldsymbol{J}_{2}^{(1)}(j_{g},\hat{\bar{2}}_{\bar{q}}),
\end{equation}
which when written in terms of integrated antennae and collinear splitting functions read~\cite{Currie:2013vh}:
\begin{eqnarray}
\boldsymbol{J}_{2}^{(1)}(\hat{\bar{1}}_{q},i_{g})&=&\frac{1}{2}{\cal D}_{3,q}^{0}(s_{\bar{1}i})-\Gamma_{qq}^{(1)}(x_{1})\;,\\
\boldsymbol{J}_{2}^{(1)}(i_{g},j_{g})&=&\frac{1}{3}{\cal F}_{3}^{0}(s_{ik})\;,\\
\boldsymbol{J}_{2}^{(1)}(j_{g},\hat{\bar{2}}_{q})&=&\boldsymbol{J}_{2}^{(1)}(\hat{\bar{2}}_{q},j_{g})=\frac{1}{2}{\cal D}_{3,q}^{0}(s_{\bar{2}j})-\Gamma_{qq}^{(1)}(x_{2})\;.
\end{eqnarray}
The analogous formula for $\boldsymbol{J}_{4}^{(2)}$ is given by,
\begin{eqnarray}
\boldsymbol{J}_{4}^{(2)}(\hat{\bar{1}}_{q},i_{g},j_{g},\hat{\bar{2}}_{\bar{q}})&=&\boldsymbol{J}_{2}^{(2)}(\hat{\bar{1}}_{q},i_{g})+\boldsymbol{J}_{2}^{(2)}(i_{g},j_{g})
+\boldsymbol{J}_{2}^{(2)}(j_{g},\hat{\bar{2}}_{q})-\overline{\boldsymbol{J}}_{2}^{(2)}(\hat{\bar{1}}_{q},\hat{\bar{2}}_{\bar{q}}),
\end{eqnarray}
where the renormalised two-parton double unresolved integrated antenna strings are given by~\cite{Currie:2013vh}:
\begin{eqnarray}
\boldsymbol{J}_{2}^{(2)}(\hat{\bar{1}}_{q},i_{g})&=&\frac{1}{2}{\cal{D}}_{4,q}^{0}(s_{\bar{1}i})+\frac{1}{2}{\cal{D}}_{3,q}^{1}(s_{\bar{1}i})+\frac{b_{0}}{2\epsilon}\bigg(\frac{|s_{\bar{1}i}|}{\mu^{2}}\bigg)^{-\epsilon}{\cal{D}}_{3,q}^{0}(s_{\bar{1}i})\nonumber\\
&-&\frac{1}{4}\big[{\cal{D}}_{3,q}^{0}(s_{\bar{1}i})\otimes{\cal{D}}_{3,q}^{0}(s_{\bar{1}i})\big](z_{1})-\overline{\Gamma}_{qq}^{(2)}(z_{1}),\\
\boldsymbol{J}_{2}^{(2)}(i_{g},j_{g})&=&\frac{1}{4}{\cal{F}}_{4}^{0}(s_{ij})+\frac{1}{3}{\cal{F}}_{3}^{1}(s_{ij})+\frac{b_{0}}{3\epsilon}\bigg(\frac{s_{ij}}{\mu^{2}}\bigg)^{-\epsilon}{\cal{F}}_{3}^{0}(s_{ij})\nonumber\\
&-&\frac{1}{9}\big[{\cal{F}}_{3}^{0}(s_{ij})\otimes{\cal{F}}_{3}^{0}(s_{ij})\big], \\
\boldsymbol{J}_{2}^{(2)}(i_{g},\hat{\bar{2}}_{\bar{q}})&=&\boldsymbol{J}_{2}^{(2)}(\hat{\bar{2}}_{q},i_{g}),\\
\overline{\boldsymbol{{J}}}_{2}^{(2)}(\hat{\bar{1}}_{q},\hat{\bar{2}}_{\bar{q}})&=&
\frac{1}{2}\tilde{\cal{A}}_{4,q\bar{q}}^{0}(s_{\bar{1}\bar{2}})+\tilde{\cal{A}}_{3,q\bar{q}}^{1}(s_{\bar{1}\bar{2}})
-\frac{1}{2}\big[{\cal{A}}_{3,q\bar{q}}^{0}(s_{\bar{1}\bar{2}})\otimes{\cal{A}}_{3,q\bar{q}}^{0}(s_{\bar{1}\bar{2}})\big].
\end{eqnarray}
Analytic expressions for the integrated antennae and collinear splitting functions introduced above can 
be found in~\cite{GehrmannDeRidder:2005cm,Daleo:2006xa,Daleo:2009yj,Boughezal:2010mc,Gehrmann:2011wi,GehrmannDeRidder:2012ja}. With these expressions
we can evaluate~\eqref{eq:Usub} and expand in powers of $\epsilon$ to obtain:
\begin{eqnarray}
&&{\rm d}\sigma_{q\bar{q},NNLO}^{U}=\sum\limits_{\substack{P(i,j)}}\Bigg[-\frac{9}{2\epsilon^{4}}+\frac{33}{18\epsilon^{3}}+\frac{1}{\epsilon^2}\Bigg(-\frac{5}{24}+\frac{\pi^2}{8}
+\frac{1}{2}\bigg(\log^2\left(\frac{s_{1i}}{\mu^2}\right)+\log^2\left(\frac{s_{ij}}{\mu^2}\right)\nonumber\\
&&+\log^2\left(\frac{s_{2j}}{\mu^2}\bigg)\right)+\log\left(\frac{s_{1i}}{\mu^2}\right)\log\left(\frac{s_{ij}}{\mu^2}\right)+\log\left(\frac{s_{2j}}{\mu^2}\right)\log\left(\frac{s_{ij}}{\mu^2}\right)+\log\left(\frac{s_{1i}}{\mu^2}\right)\log\left(\frac{s_{2j}}{\mu^2}\right)\nonumber\\
&&-\frac{37}{12}\bigg(\log\left(\frac{s_{1i}}{\mu^2}\right)+\log\left(\frac{s_{ij}}{\mu^2}\right)+\log\left(\frac{s_{2j}}{\mu^2}\right)\bigg)\Bigg)\nonumber\\
&&+\frac{1}{\epsilon}\Bigg(-\frac{10201}{288}+\frac{9}{4}\zeta_{3}+\frac{7\pi^2}{144}-\log^2\left(\frac{s_{1i}}{\mu^2}\right)-\log^2\left(\frac{s_{ij}}{\mu^2}\right)-\log^2\left(\frac{s_{2j}}{\mu^2}\right)\nonumber\\
&&-2\log\left(\frac{s_{1i}}{\mu^2}\right)\log\left(\frac{s_{ij}}{\mu^2}\right)-2\log\left(\frac{s_{2j}}{\mu^2}\right)\log\left(\frac{s_{ij}}{\mu^2}\right)
-2\log\left(\frac{s_{1i}}{\mu^2}\right)\log\left(\frac{s_{2j}}{\mu^2}\right)\nonumber\\
&&+\left(\frac{439}{36}-\frac{\pi^2}{12}\right)\left(\log\left(\frac{s_{1i}}{\mu^2}\right)+\log\left(\frac{s_{ij}}{\mu^2}\right)+\log\left(\frac{s_{2j}}{\mu^2}\right)\right)
\Bigg)\Bigg]\;B_{4}^{0}(\hat{\bar{1}}_{q},i_{g},j_{g},\hat{\bar{2}}_{\bar{q}})\nonumber\\
&&+\Bigg[-\frac{3}{\epsilon^2}+\frac{1}{\epsilon}\left(-\frac{31}{6}+\left(\log\left(\frac{s_{1i}}{\mu^2}\right)+\log\left(\frac{s_{ij}}{\mu^2}\right)
+\log\left(\frac{s_{2j}}{\mu^2}\right)\right)\right)\Bigg]\;B_{4}^{1}(\hat{\bar{1}}_{q},i_{g},j_{g},\hat{\bar{2}}_{\bar{q}})\;\nonumber\\
&&+ {\cal O}(\epsilon^0)
\label{eq:Uqqb}
\end{eqnarray} 
As expected, the initial-state collinear singularities in the integrated antennae with initial-state hard radiators cancelled against the PDF mass-factorisation collinear subtraction included
in the definitions of $\boldsymbol{J}_{2}^{(1)}$ and $\boldsymbol{J}_{2}^{(2)}$,
and all the remaining singularities in~\eqref{eq:Uqqb}, cancel explicitly and analytically with the IR-poles of the two-loop amplitude for $q\bar{q}\to gg$ as guaranteed by the KLN theorem.  

\subsection{Discussion}
In this report we have reviewed the main aspects of the antenna subtraction scheme for the subtraction of infrared singularities in
the calculation of jet observables at NNLO. We introduced subtraction terms for double real radiation at tree level and single real radiation at one 
loop based on antenna functions. These antenna functions at NLO and NNLO describe the colour-ordered radiation of unresolved partons between a pair of hard (radiator) partons,
and can be derived from physical matrix elements~\cite{GehrmannDeRidder:2005cm}.

We have shown how all singularities in intermediate steps of perturbative QCD calculations can be mapped to Born-like configurations exploiting the universal factorised 
structure of QCD amplitudes in the IR-limits. A key ingredient are the phase space mappings that smoothly interpolate between the various singular limits, and the 
factorisation of the real-radiation phase space, which allows for the analytic integration of the antenna functions, decoupling it from the integration over the Born configurations.
All the integrated counterterms that are necessary to have a fully general subtraction method for massless final-state~\cite{GehrmannDeRidder:2005cm} and
initial-state~\cite{Daleo:2006xa,Daleo:2009yj,Boughezal:2010mc,Gehrmann:2011wi,GehrmannDeRidder:2012ja} QCD have been computed. 

Phenomenological results for jet cross sections and transverse momentum distributions at NNLO at hadron colliders have been recently obtained within this approach. The results are obtained in the 
NNLOJET code framework~\cite{Gehrmann:2018szu} which is a parton-level event generator that provides the framework for the implementation of jet production processes to NNLO accuracy, 
using the antenna subtraction method.
It contains the event generator infrastructure (Monte Carlo phase-space integration, event handling and analysis routines) and provides the unintegrated and integrated antenna functions and
the phase-space mappings for all kinematical configurations. 

Processes included in NNLOJET up to now are $Z$ and $Z + j$ production~\cite{Ridder:2015dxa,Ridder:2016nkl,Gehrmann-DeRidder:2016jns}, 
$W$ and $W + j$ production~\cite{Gehrmann-DeRidder:2017mvr,Gehrmann-DeRidder:2019avi}, $WH+j$ production~\cite{Gauld:2020ced},
$H$ and $H + j$ production~\cite{Chen:2016zka,Chen:2019wxf}, $H+ 2j$ (VBF)~\cite{Cruz-Martinez:2018rod}, 
di-jet production in hadron-hadron collisions~\cite{Currie:2016bfm,Currie:2017eqf,Gehrmann-DeRidder:2019ibf} and in lepton-hadron collisions~\cite{Currie:2016ytq,Currie:2017tpe}, isolated
$\gamma$ and $\gamma+j$ production~\cite{Chen:2019zmr}, di-photon production~\cite{Gehrmann:2020oec} as well as three-jet 
production in electron-position annihilation~\cite{Gehrmann:2017xfb}. More recently, flavour sensitive observables at NNLO have been studied in~$pp\to HV$ with $H\to b\bar{b}$ 
and $V\to ll$~\cite{Gauld:2019yng}, and for $Z+b$ production~\cite{Gauld:2020deh}.

\section{Conclusions and Outlook }

The purpose of this section is twofold. In the first part, we briefly remark on the strengths ($+$)
and weaknesses ($-$) of the formalisms presented in this write-up. Whereas in the second part, we summarise the discussion in the closing session of the workshop.

\subsubsection*{FDH and DRED}

\begin{itemize}
    \item[$+$]
    The evaluation of the Lorentz algebra is significantly simpler than in conventional dimensional
    regularisation. For an NNLO computation in \DRED\ this is particularly true for double-real
    contributions and for integrated counterterms of subtraction methods since the
    $\mathcal{O}(\epsilon)$ terms of the matrix elements are not required.

    \item[$+$]
    \FDH\ is more amenable to methods that rely on strictly four-dimensional objects like the spinor-helicity
    formalism and unitarity. Similar to completely four-dimensional regularisation approaches, however,
    this is not true for the treatment of $\gamma_5$.

    \item[$+$]
    As $\mathcal{O}(\epsilon)$ terms cannot contain any physical information, \FDH\ and \DRED\ might help
    to improve the conceptual understanding of regularisation and of subtraction methods. Both schemes
    constitute the most promising candidates to find links between dimensional regularisation and strictly
    four-dimensional approaches like \FDU, \FDR, and \IReg.

    \item[$-$]
    The evaluation of (master) integrals is not affected. Compared to \CDR\ we still need the same loop
    and phase-space integrals.

    \item[$-$]
    The UV renormalisation is slightly more complicated than in \CDR. The procedure, however, is standardised
    and well understood. For an NNLO computation in \FDH\ or \DRED, the evanescent renormalisation constants
    at most have to be known at one-loop order.
\end{itemize}

\subsubsection*{FDR}
\begin{itemize}
\item[$+$]Both UV and IR divergences are regularised strictly in four dimensions.
\item[$+$]No UV counterterms need to be computed.
\item[$+$]Lowest order parts of the calculation remain the same even when embedded in higher order computations. For instance, no ${\cal O} (\epsilon)$ terms need to be included at two loops.
\item[$+$]Being four dimensional, \FDR\ is suitable for a fully numerical treatment.
\item[$-$] One cannot rely on existing libraries and/or reduction methods. In particular, loop and phase-space integrals have to be computed.
\item[$-$] At the moment, a local cancellation algorithm for final-state IR singularities has been implemented at NLO only.
\end{itemize}

\subsubsection*{FDU}
\begin{itemize}
\item[$+$]Direct connection between multi-loop and tree-level amplitudes within
a common phase-space by means of the loop-tree duality formalism.
In other words, real contributions are mapped onto virtual ones.
\item[$+$]Local cancellation of IR and UV singularities at integrand level.
Namely, no need of ad-hoc and integrated counterterms. The local cancellation
allows to perform loop and phase-space integration in four dimensions.
\item[$+$]Local UV renormalisation at one- and two-loop level has been tested
with proof-of-concept calculations.
\item[$+$]The integrand representation of multi-loop Feynman integrals and scattering
amplitudes in the loop-tree duality is manifestly causal, i.e. it
displays only physical information. Absence of spurious singularities. 
\item[$-$]The subleading UV local counterterms that implement the renormalisation
scheme are still evaluated in $d$ dimensions.
\item[$-$]Contrary to calculations at NLO, the treatment of IR singularities at NNLO
is not yet fully developed. 
\item[$-$]Processes including initial state radiation have not yet been studied
within \FDU. This is because an integrand and local representation
of the Altarelli-Parisi kernels is currently missing.
\item[$-$]\FDU\ cannot profit from current techniques for the calculation of multi-loop
Feynman integrals, such as integration-by-parts identities and differential/difference
equation methods. This is because the latter modify the local IR and
UV behaviour. 
\end{itemize}

\subsubsection*{IREG}

\begin{itemize}

\item[$+$]Fully four-dimensional scheme for momentum integration and Clifford algebra. Although implicit four dimensional schemes such as \IReg\ share the same problems with the $\gamma_5$ matrix a well-defined procedure can avoid inconsistencies as shown in~\cite{Bruque:2018bmy, Viglioni:2016nqc}.

\item[$+$] The UV content of a given Feynman integral can be cast in terms of a well-defined set of basic divergent integrals which do not need to be evaluated. From the viewpoint of anomalies in perturbation theory it is a useful scheme.

\item[$+$] Generalisation to $L$-loops is straightforward and compatible with local subtraction theorems such as the BHPZ scheme and the Bogolyubov recursion formula.

\item[$-$] Although IR and UV divergences are clearly separated in a gauge invariant way, and no extra fields are needed in the Lagrangian (such as epsilon-scalar fields) compatibility with factorisation theorems are yet to be studied beyond leading order.

\item[$-$] Although a diagrammatic all order proof of gauge invariance in abelian models can be constructed for \IReg, a general all order proof for the non-abelian case need to be constructed based on quantum action principles.

\end{itemize}

\subsubsection*{Local analytic Sector Subtraction}

\begin{itemize}
\item[$+$] IR singularities are locally cancelled at NLO (validated for generic processes with 
QCD partons in the final state) and NNLO (ongoing validation in the general case).

\item[$+$] Subtraction counterterms feature a minimal structure, thanks to the radiative phase-space 
partition with sector functions (inspired by FKS).

\item[$+$] An optimal phase-space parametrisation (via multiple CS mappings) enables the analytic
integration of the counterterms by means of standard techniques.

\item[$+$] The scheme is valid for an arbitrary number of QCD partons in the final state. It has been tested 
at NNLO with a proof-of-concept calculation of the $T_R C_F$ contribution to  $e^+e^-\rightarrow 2j$.

\item[$-$] Several checks still need to be performed to test the cancellation of IR poles 
in the general case at NNLO.

\item[$-$] In order to stick to a minimal structure, a delicate tuning is needed
for the counterterm definition and for the corresponding phase-space mappings.

\item[$-$] A complete implementation in a Monte Carlo code is still missing.

\item[$-$] At NNLO, the scheme is currently designed only for FSR and for massless partons. 
Preliminary investigation is ongoing for the extension to ISR and to the massive case.
\end{itemize}

\subsubsection*{qt-subtraction}

\begin{itemize}

\item[$+$]
The method benefits from any existing calculation for ``F+jet'' production at LO, NLO,
NNLO, etc, to produce results for ``F'' production at one corresponding higher order
of the perturbative expansion: NLO, NNLO, N3LO, etc. 
In fact, the singular behaviour of ``F+jet'' as $\qt \to 0$ is well-known from the resummation
program of logarithmically-enhanced contributions to $\qt$ distributions.

\item[$+$]The universality of the logarithmic contributions to the $q_T$ distributions
  allows to construct counterterms which require minimal information about
  the process, such as the Born subprocess and the finite remainder of the
  multi-loop scattering amplitudes (at any corresponding order in the
  coupling).

\item[$+$] $q_T$-subtraction is fully developed for production processes of colourless particles and massive quarks.
The general structure is the same for any number of colourless particles in the final state,
any number of massive quarks in the final state, or any combination between colourless
particles and massive quarks.

\item[$-$] The subtraction is non-local and the control on large cancellations is an
  issue. A small resolution variable, i.e. a cutoff, leads to numerical
  difficulties and slower integrations. Conversely, a greater value for the
  cutoff enhances the contributions from power corrections in the resolution
  variable, which are actually neglected in the general formulation of the
  subtraction. 

\item[$-$] Real scattering amplitudes and counterterms are integrated in different
  phase spaces, with the counterterm always evaluated in the corresponding
  Born phase space.

\item[$-$] $q_T$-subtraction is not fully developed for production processes of detected massless
coloured particles and/or jets.

\end{itemize}

\subsubsection*{Antenna subtraction}

\begin{itemize}
\item[$+$] Local subtraction scheme with phase-space averaging. Good control on the numerical accuracy of the final result with double-real, real-virtual and double-virtual contributions separately finite.
\item[$+$] No need to introduce phase-space slicing parameters in the calculation.
\item[$+$] IR singularities are cancelled analytically, i.e., the explicit $\epsilon$-poles in the dimension regularisation parameter of one- and 
two-loop matrix elements are cancelled in analytic and local form against the $\epsilon$-poles of the integrated antenna subtraction terms. Good control on the correctness of the pole cancellation.
\item[$+$] Fully general subtraction scheme at NNLO for massless final-state and initial state QCD for any  jet multiplicity. 
\item[$-$] The antenna functions are scalar objects and do not subtract angular correlations in gluon-splittings, which vanish when the azimuthal variable of the collinear system 
(with respect to the collinear axis, defined by the collinear momentum and a light-like recoil momentum) is integrated out. Cancellation is accomplished with the method locally, by
combining phase space points correlated by a $\pi/2$ rotation around the collinear. axis%~\cite{Glover:2010im}.
\item[$-$] Involves many re-mappings and subtraction terms as expected for a local method. Can be improved by introducing a caching system to store the evaluation of the phase space
mappings applied to the real contributions. 
\end{itemize}

\subsection*{Closing discussion} 
%\label{sec:closing}
%
The closing event of the workshop was a discussion session 
%led by Dr. Stefano Catani and Dr. Germ\'an Rodrigo, 
in which all the attendants participated and shared opinions. Here, we present a summary of the main topics and questions that were mentioned in that session, preserving the 
%natural 
original 
ordering in which the discussion took place.

As stated in the general introduction, the purpose of the present workshop was to deepen into technical aspects of modern high-precision computations for QFTs. This implies covering several topics, several techniques and subtleties, that might hide conceptual and/or computational issues. These issues range from subtle definitions (e.g., \emph{what does locality mean?}) to deeper conceptual problems (e.g., \emph{does it make sense going till N$^k$LO in perturbation theory
without having quantitative control on the size of various corrections of
non-perturbative origin?}). Since 
%this is a very wide field, and 
the time was limited, the discussion focused on three points: theoretical errors, factorisation breaking and $\gamma^5$ issues.

\subsubsection*{About theoretical errors}
\label{ssec:theoreticalerrors}
It is a fact that experiments are reaching an impressive level of accuracy, due to the increase in the data collection and its treatment. So, one question is: \emph{how can theory keep the pace and produce predictions within the required precision? 
%In other words: 
How can we control the theoretical errors in a 
%credible 
reliable
way? 
}

%At this point, 
Nowadays there is a shared opinion within the HEP community that more legs and more loops will lead to an error reduction, but there are not yet established precise 
%formal (i.e. mathematical) statement 
procedures to properly quantify the error estimation (see, e.g., 
Refs.~\cite{Cacciari:2011ze, Bonvini:2020xeo}
and related references therein).

Moreover, regarding theoretical errors in collider observables, we need to include the non-negligible impact of PDFs. Essentially, the predictions are being affected by perturbative and non-perturbative contributions, and both of them are potential sources of errors that need to be kept under control. Thus, another important question is: \emph{how does the theoretical framework affect our skills to extract predictions from QFTs?} 
PDF extractions relies not only on highly-accurate experimental data but also
on highly-precise partonic predictions (coming from multi-loop multi-leg computations). 
A non-trivial interplay 
%competition 
between perturbative uncertainties at partonic level and ensuing PDF errors is always present. So, it is not possible to claim a certain accuracy if there is not a rigorous control on the errors present in all the ingredients of the calculation
(see. e.g., Refs.~\cite{AbdulKhalek:2019bux, AbdulKhalek:2019ihb} and related references therein).

\subsubsection*{Factorization breaking}
\label{ssec:factorizationbreaking}
Most of the time, we focus our methods on trying to compute NLO, NNLO and even higher-order corrections to hard-scattering processes at hadron colliders. We 
%strongly 
rely on the validity of the factorisation theorem, but only a few times we are (fully) aware of the potential limitations. 

Our ability to compute predictions for high-energy colliders strongly relies on the parton model and factorisation formulae, which isolate the dominant
(i.e., `leading twist') non-perturbative nature of the colliding proton inside 
process-independent PDFs. If collinear factorisation is spoiled at some perturbative orders, then PDFs would carry an implicit dependence on the process. Correlations among initial- and final-state partons will survive, and this will break the possibility of using the factorisation theorem. 

We recall that a general (process and observable independent) proof of the factorisation theorem to all perturbative orders is still lacking.
In Ref.~\cite{Catani:2011st}, 
the violation of strict collinear factorisation at the scattering amplitude level
was pointed out.
%a serious and deep discussion about strict factorization breaking was carried out.
%In a naive picture, it is expected that matrix elements factorizes 
In the collinear limit, the scattering amplitude ${\cal M}$ factorises    
according to
\beq
{\cal M}_n(p_1,\ldots,p_l,p_{l+1},\ldots,p_n) \to {\rm Sp}(p_1,\ldots,p_l,\tilde{P}) \times {\cal M}_{n-l+1}(\tilde{P},p_{l+1},\ldots,p_n) \, ,
\eeq
where $\{p_1,\ldots,p_l\}$ ($\{p_{l+1},\ldots,p_n\}$) are the collinear (non-collinear) momenta and $\tilde{P}$ denotes the light-like vector carrying the total momenta of the collinear partons. The factor ${\rm Sp}$ embodies the contributions that are singular in the collinear region. In a naive picture, the singular factor
${\rm Sp}$ is expected to be universal (process-independent), namely it can only depend on the momenta and flavours of the collinear partons.
This picture is valid at the tree level and, more generally, in the time-like region, but, including loop corrections, it was proven 
\cite{Catani:2011st}
that color and momentum correlations among collinear and non-collinear partons are present in 
${\rm Sp}$ in the space-like region, i.e., in the case of collinear emission from
initial-state colliding partons. The contributions that break strict collinear factorisation originate from absorptive interactions that takes place in the 
far past (long before the occurrence of the hard scattering) between the 
initial-state colliding partons. Owing to their absorptive origin, these contributions are imaginary at the lowest perturbative order
and, therefore, they cancel at the squared amplitude level up to NNLO
(and also N$^3$LO in pure QCD processes \cite{Catani:2011st,Forshaw:2012bi}).  

The occurrence of such cancellation mechanism of factorization breaking terms is not guaranteed at high perturbative orders. Further studies and investigations 
\cite{Catani:2011st, Forshaw:2012bi, Rothstein:2016bsq, Schwartz:2017nmr,
Dixon:2019lnw} of these factorization breaking terms (and of their precise structure) are certainly relevant in view of the conceptual and computational importance of the factorisation theorem (`assumption') for extracting predictions within QFT.
Even if collinear-factorisation breaking contributions eventually do not spoil the validity of the factorisation theorem, their presence definitely introduce technical complications in the cancellation mechanism of IR divergences for multileg
hard-scattering observables in hadron collisions. Such complications have to be overcome to extend IR-subtraction methods beyond the NNLO.

\subsubsection*{$\gamma^5$ problems}
\label{ssec:gamma5}
One of the main topics of this workshop regarded the subtleties that appear when extending a theory to $d$ dimensions, whatever $d$ means (as in the context of \DREG). However, there are problems that also arise when $d=4$: this is the case of $\gamma^5$. Of course, in the standard four-dimensional space-time there is a well established recipe to mathematically define $\gamma^5$. Moreover, in any even-dimensional Minkowskian manifold analogous objects to $\gamma^5$ are properly defined.

Regularisation involving $\gamma_5$ is problematic. In dimensional schemes the problems
are well-known (see e.g. the review~\cite{Jegerlehner:2000dz}), and recent references 
have focused on comparing different $\gamma_5$-prescriptions up to the two-loop 
level~\cite{Gnendiger:2017rfh} and on determining gauge invariance-restoring counterterms 
for the Breitenlohner/Maison/'t Hooft/Veltman prescription of $\gamma_5$~\cite{Belusca-Maito:2020ala}. 
%But even non-dimensional schemes cannot be defined in a gauge 
%invariant way without leading to very similar $\gamma_5$-problems \cite{Bruque:2018bmy}.
Quite surprisingly, non-dimensional schemes are not exempted of issues in the presence of 
$\gamma_{5}$~\cite{Viglioni:2016nqc,Porto:2017asd}. The reason boils down to requiring very basic properties such as shift invariance and numerator-denominator consistency to be respected, showing that virtually any regularization scheme will need to deal with 
$\gamma_{5}$-problems~\cite{Bruque:2018bmy}. 

Therefore, consistent definition of $\gamma_5$, together with full understanding of its 
properties with respect to symmetries, gauge invariance and anomaly cancellation, is crucial 
for higher-order calculations. This is especially important in the context of 
high-precision predictions taking into account electroweak corrections.

%In spite of these properties, having a clear definition of $\gamma^5$ does not imply that we can unambiguously perform any calculation involving it. In fact, it is necessary to set additional prescriptions to obtain well-defined results. In the case of gauge theories, this is required to preserve gauge invariance and allow the cancellation of anomalies.
%
%Thus, it is crucial to define a consistent \emph{standard} prescription to deal with $\gamma^5$, specially at higher-orders. This is an important point in the context of high-precision particle physics, because several observables depend of this object and the predictions must be unambiguously defined (i.e. physically consistent).

\subsubsection*{Further open questions}
\label{ssec:openQ}
After the exciting discussion session, many issues and questions remained 
%unsolved. 
opened. In particular, we would like to highlight: 
\begin{itemize}
 \item\emph{How to re-define a QFT in such a way that no distinction among real and virtual corrections is done?} 
\item Even if we manage to combine the real and virtual contributions from the very beginning, still threshold singularities might survive. \emph{How to tackle them and develop efficient techniques to integrate through thresholds?}
\end{itemize}
Recent studies at NNLO point to computational frameworks in which IR, UV and threshold singularities are treated in purely four dimensions.
In fact, NLO calculations in a four-dimensional framework started to be carried out long time ago by Soper~\cite{Soper:1999xk} and subsequent related works.
More recently, studies that aim at achieving a complete cancellation of singularities
at the integrand level were presented in Refs.~\cite{Hernandez-Pinto:2015ysa,Sborlini:2016gbr,Sborlini:2016hat,Driencourt-Mangin:2019yhu} 
at NLO, and
preliminary results that involve two-loop
scattering amplitudes were presented in Ref.~\cite{Driencourt-Mangin:2019aix}.
The studies in Ref.~\cite{Capatti:2020xjc}, based on the knowledge of the infrared structure of scattering
amplitudes~\cite{Anastasiou:2018rib,Anastasiou:2020sdt}, 
point towards the same direction.
Furthermore, novel techniques for the evaluation of multi-loop Feynman integrals, inspired by the
loop-tree duality approach~\cite{Bierenbaum:2010cy,Catani:2008xa}, 
have shown to display a causal representation depending
only on physical singularities (see Refs.~\cite{Capatti:2019edf, Verdugo:2020kzh, Aguilera-Verdugo:2020kzc, Capatti:2020ytd, Aguilera-Verdugo:2020nrp,Ramirez-Uribe:2020hes,Capatti:2019ypt} and related
references therein).

Alternatives approaches based on analytic and semi-numerical
techniques for NNLO calculations are summarised in the recent review~\cite{Heinrich:2020ybq}.

%%%%%%%%%%%%%%%%%%%%%%%%%%%%%%%%%%%%%%%%%%%%%%%%%%%
%%%%%%%%%%%%%%%%%%%%%%%%%%%%%%%%%%%%%%%%%%%%%%%%%%%

\section*{Acknowledgments}

In this work, we summarise all discussions originated as a result of the WorkStop/ThinkStart~3.0:
\textit{paving the way to alternative NNLO strategies}
that took place on 4.--6.~November~2019 at the Galileo Galilei Institute for 
Theoretical Physics (GGI). 
We gratefully acknowledge the support of GGI
and the COST Action CA16201 PARTICLEFACE.
We wish to thank to W.M.~Marroqu\'in and M.~Morandini for their help in organising the workshop.

\noindent P.~Banerjee acknowledges support by the European Union's Horizon 2020 research and innovation programme under the Marie Sk\l{}odowska-Curie grant agreement No 701647.

\noindent A.L.~Cherchiglia, B.~Hiller and M.Sampaio
acknowledge support from Fundação para a Ciência e Tecnologia (FCT) through the projects  UID/FIS/04564/2020 and  CERN/FIS-COM/0035/2019. 

\noindent The work of L.~Cieri has received funding from the 
European Union's Horizon 2020 research and innovation programme under the 
Marie Sklodowska-Curie grant agreement No 754496.

\noindent The work of F.~Driencourt-Mangin, G.~Rodrigo, G.~Sborlini and W.J.~Torres~Bobadilla is supported by the Spanish Government (Agencia Estatal de Investigaci\'on), 
ERDF funds from European Commission (Grant No. FPA2017-84445-P), 
Generalitat Valenciana (Grant No. PROMETEO/2017/053) 
and from the Spanish Government (FJCI-2017-32128). 

\noindent T.~Engel acknowledges support by the Swiss National Science Foundation (SNF) under contract 200021\_178967.

\noindent C.~Gnendiger, R.~Pittau, A.~Signer and D.~St\"ockinger wish to thank B. Page for his help in establishing~\eqref{eq:EEIb}. 

\noindent The work of R.J.~Hern\'andez-Pinto is supported by CONACyT through the Project No. A1-S-33202 (Ciencia Basica) and Sistema Nacional de Investigadores.

\noindent G.Pelliccioli.~was supported by the Bundesministerium f\"ur Bildung
und Forschung (BMBF, German Federal Ministry for Education and Research) under
contract no.~05H18WWCA1.

\noindent J.~Pires was supported by Fundação para a Ciência e Tecnologia (FCT, Portugal) 
through the contract UIDP/50007/2020 and project CERN/FIS-PAR/0024/2019.

\noindent The work of R.~Pittau has been supported by the Spanish Government grant PID2019-106087GB-C21 and by the Junta de Andalucía project P18-FR-4314 (fondos FEDER).

\noindent M.~Sampaio acknowledges a research grant from CNPq (Conselho Nacional de Desenvolvimento Cient\'\i fico e Tecnol\'ogico - 303482/2017-6). 

\noindent C.~Signorile-Signorile was supported by the Deutsche
Forschungsgemeinschaft (DFG, German Research Foundation) under grant
no.~396021762 - TRR 257. 

%\newpage
\bibliography{refs}
\bibliographystyle{JHEP}

\end{document}